\documentclass[a4paper,11pt]{article}
\pdfoutput=1
\usepackage{jheppub}
\usepackage[utf8]{inputenc}
\usepackage{url}
\usepackage{mathrsfs}  
\usepackage{pdflscape}
\usepackage{refcount}
\usepackage{booktabs}
\usepackage{todonotes}
\usepackage{enumitem}
\usepackage{adjustbox}
\usepackage{afterpage}
\usepackage{placeins}
\usepackage{mdframed}

\usepackage{pdfpages}
\usepackage{caption}
\usepackage{subcaption}

\usepackage{tikz}
\usetikzlibrary{positioning}
\usetikzlibrary{arrows}
\usetikzlibrary{decorations.pathreplacing}
\usetikzlibrary{shapes.geometric}
\usetikzlibrary{calc}
\usetikzlibrary{decorations.pathmorphing}
\usetikzlibrary{shapes.misc}

\newcommand{\convexpath}[2]{
  [   
  create hullcoords/.code={
    \global\edef\namelist{#1}
    \foreach [count=\counter] \nodename in \namelist {
      \global\edef\numberofnodes{\counter}
      \coordinate (hullcoord\counter) at (\nodename);
    }
    \coordinate (hullcoord0) at (hullcoord\numberofnodes);
    \pgfmathtruncatemacro\lastnumber{\numberofnodes+1}
    \coordinate (hullcoord\lastnumber) at (hullcoord1);
  },
  create hullcoords
  ]
  ($(hullcoord1)!#2!-90:(hullcoord0)$)
  \foreach [
  evaluate=\currentnode as \previousnode using \currentnode-1,
  evaluate=\currentnode as \nextnode using \currentnode+1
  ] \currentnode in {1,...,\numberofnodes} {
    let \p1 = ($(hullcoord\currentnode) - (hullcoord\previousnode)$),
    \n1 = {atan2(\y1,\x1) + 90},
    \p2 = ($(hullcoord\nextnode) - (hullcoord\currentnode)$),
    \n2 = {atan2(\y2,\x2) + 90},
    \n{delta} = {Mod(\n2-\n1,360) - 360}
    in 
    {arc [start angle=\n1, delta angle=\n{delta}, radius=#2]}
    -- ($(hullcoord\nextnode)!#2!-90:(hullcoord\currentnode)$) 
  }
}

\def\ns#1#2{
	\node[circle, draw, fill=white] (#2) at (#1){};
	\node[cross out, draw] at (#1){};
}

\tikzset{flavour/.style={draw=none,minimum size=0.3mm,fill=white, regular polygon,regular polygon sides=4,draw}}
\tikzset{gaugeBig/.style={inner sep=1mm,draw=none,fill=white,minimum size=2mm,circle, draw}}
\tikzset{bd/.style={circle, draw=black, inner sep=0pt, fill=black, minimum size=2mm}}
\tikzset{wd/.style={circle, draw=black, inner sep=0pt, fill=white, minimum size=2mm}}
\tikzset{Dynkin/.style={circle, draw=black, inner sep=0pt, fill=white, minimum size=2mm}}
\tikzstyle{ligne}=[draw, very thick] 
\tikzstyle{gridline}=[draw, gray] 
\tikzset{gauge/.style={circle, draw,inner sep=2.5pt}}
\tikzset{gaugeo/.style={circle, draw,inner sep=2.5pt,fill=orange}}
\tikzset{gaugeol/.style={circle, draw,inner sep=2.5pt,fill=olive}}
\tikzset{gauger/.style={circle, draw,inner sep=2.5pt,fill=red}}
\tikzset{gaugeb/.style={circle, draw,inner sep=2.5pt,fill=blue}}
\tikzset{gaugeg/.style={circle, draw,inner sep=2.5pt,fill=green}}
\tikzset{gaugem/.style={circle, draw,inner sep=2.5pt,fill=magenta}}
\tikzset{gaugec/.style={circle, draw,inner sep=2.5pt,fill=cyan}}
\tikzset{hasse/.style={circle, fill,inner sep=2pt}}
\tikzset{shrinky/.style={circle, fill,inner sep=1pt}}
\tikzset{sized/.style={circle, draw, inner sep=1.5pt}}
\tikzset{seven/.style={circle, draw,inner sep=3pt}}

\definecolor{goodgreen}{RGB}{55,169,49}

\definecolor{darkyellow}{RGB}{204,204,10}

 \newcommand{\za}{
\begin{tikzpicture}
\def\x{1cm};
\draw (.5,-\x)--(.5,\x);
\draw (1,-\x)--(1,\x);
\draw (4,-\x)--(4,\x);
\draw (4.5,-\x)--(4.5,\x);
\draw (.5,-.25)--(1,-.25);
\draw (1,0)--(4,0);
\draw (1,-.5)--(4,-.5);
\draw (4,-.25)--(4.5,-.25);
\ns{2,.5};
\ns{2.5,.75};
\ns{3,1};
\end{tikzpicture}}
 \newcommand{\zba}{
\begin{tikzpicture}
\def\x{1cm};
\draw (.5,-\x)--(.5,\x);
\draw (1,-\x)--(1,\x);
\draw (4,-\x)--(4,\x);
\draw (4.5,-\x)--(4.5,\x);
\draw (.5,-.5)--(1,-.5);
\draw (1,-.05)--(4,-.05);
\draw (1,.05)--(4,.05);
\draw (4,0)--(4.5,0);
\ns{2,0};
\ns{2.5,0};
\ns{3,0};
\end{tikzpicture}}
 \newcommand{\zbb}{
\begin{tikzpicture}
\def\x{1cm};
\draw (.5,-\x)--(.5,\x);
\draw (1,-\x)--(1,\x);
\draw (4,-\x)--(4,\x);
\draw (4.5,-\x)--(4.5,\x);
\draw (.5,-0)--(1,0);
\draw (1,-.05)--(4,-.05);
\draw (1,.05)--(4,.05);
\draw (4,0)--(4.5,0);
\ns{2,.5};
\ns{2.5,0};
\ns{3,0};
\end{tikzpicture}}
 \newcommand{\zbc}{
\begin{tikzpicture}
\def\x{1cm};
\draw (.5,-\x)--(.5,\x);
\draw (1,-\x)--(1,\x);
\draw (4,-\x)--(4,\x);
\draw (4.5,-\x)--(4.5,\x);
\draw (.5,-0)--(1,0);
\draw (1,-.05)--(4,-.05);
\draw (1,.05)--(4,.05);
\draw (4,0)--(4.5,0);
\ns{2,0};
\ns{2.5,.5};
\ns{3,0};
\end{tikzpicture}}
 \newcommand{\zbd}{
\begin{tikzpicture}
\def\x{1cm};
\draw (.5,-\x)--(.5,\x);
\draw (1,-\x)--(1,\x);
\draw (4,-\x)--(4,\x);
\draw (4.5,-\x)--(4.5,\x);
\draw (.5,-0)--(1,0);
\draw (1,-.05)--(4,-.05);
\draw (1,.05)--(4,.05);
\draw (4,0)--(4.5,0);
\ns{2,0};
\ns{2.5,0};
\ns{3,.5};
\end{tikzpicture}}
 \newcommand{\zbe}{
\begin{tikzpicture}
\def\x{1cm};
\draw (.5,-\x)--(.5,\x);
\draw (1,-\x)--(1,\x);
\draw (4,-\x)--(4,\x);
\draw (4.5,-\x)--(4.5,\x);
\draw (.5,-0)--(1,0);
\draw (1,-.05)--(4,-.05);
\draw (1,.05)--(4,.05);
\draw (4,-.5)--(4.5,-.5);
\ns{2,0};
\ns{2.5,0};
\ns{3,0};
\end{tikzpicture}}
 \newcommand{\zc}{
\begin{tikzpicture}
\def\x{1cm};
\draw (.5,-\x)--(.5,\x);
\draw (1,-\x)--(1,\x);
\draw (4,-\x)--(4,\x);
\draw (4.5,-\x)--(4.5,\x);
\draw (.5,-0)--(1,0);
\draw (1,-.05)--(4,-.05);
\draw (1,.05)--(4,.05);
\draw (4,0)--(4.5,0);
\ns{2,0};
\ns{2.5,0};
\ns{3,0};
\end{tikzpicture}}
 \newcommand{\zd}{
\begin{tikzpicture}
\def\x{1cm};
\draw (.5,-\x)--(.5,\x);
\draw (1,-\x)--(1,\x);
\draw (4,-\x)--(4,\x);
\draw (4.5,-\x)--(4.5,\x);
\draw (.5,-.25)--(1,-.25);
\draw (1,0)--(4,0);
\draw (1,-.5)--(4,-.5);
\draw (4,-.25)--(4.5,-.25);
\ns{2,.5};
\ns{2,1};
\ns{3,.75};
\end{tikzpicture}}
 \newcommand{\ze}{
\begin{tikzpicture}
\def\x{1cm};
\draw (.5,-\x)--(.5,\x);
\draw (1,-\x)--(1,\x);
\draw (4,-\x)--(4,\x);
\draw (4.5,-\x)--(4.5,\x);
\draw (.5,-.25)--(1,-.25);
\draw (1,0)--(4,0);
\draw (1,-.5)--(4,-.5);
\draw (4,-.25)--(4.5,-.25);
\ns{2,.75};
\ns{3,.75};
\node at (2.25,1) {2};
\end{tikzpicture}}
 \newcommand{\zfa}{
\begin{tikzpicture}
\def\x{1cm};
\draw (.5,-\x)--(.5,\x);
\draw (1,-\x)--(1,\x);
\draw (4,-\x)--(4,\x);
\draw (4.5,-\x)--(4.5,\x);
\draw (.5,-0)--(1,0);
\draw (1,-.05)--(4,-.05);
\draw (1,.05)--(4,.05);
\draw (4,-.5)--(4.5,-.5);
\ns{2,0};
\ns{3,0};
\node at (2.25,.25) {2};
\end{tikzpicture}}
 \newcommand{\zfb}{
\begin{tikzpicture}
\def\x{1cm};
\draw (.5,-\x)--(.5,\x);
\draw (1,-\x)--(1,\x);
\draw (4,-\x)--(4,\x);
\draw (4.5,-\x)--(4.5,\x);
\draw (.5,-0)--(1,0);
\draw (1,-.05)--(4,-.05);
\draw (1,.05)--(4,.05);
\draw (4,0)--(4.5,0);
\ns{2,0};
\ns{3,.5};
\node at (2.25,.25) {2};
\end{tikzpicture}}
 \newcommand{\zfc}{
\begin{tikzpicture}
\def\x{1cm};
\draw (.5,-\x)--(.5,\x);
\draw (1,-\x)--(1,\x);
\draw (4,-\x)--(4,\x);
\draw (4.5,-\x)--(4.5,\x);
\draw (.5,-.5)--(1,-.5);
\draw (1,-.05)--(4,-.05);
\draw (1,.05)--(4,.05);
\draw (4,0)--(4.5,0);
\ns{2,0};
\ns{3,0};
\node at (2.25,.25) {2};
\end{tikzpicture}}
 \newcommand{\zg}{
\begin{tikzpicture}
\def\x{1cm};
\draw (.5,-\x)--(.5,\x);
\draw (1,-\x)--(1,\x);
\draw (4,-\x)--(4,\x);
\draw (4.5,-\x)--(4.5,\x);
\draw (.5,0)--(1,0);
\draw (1,-.05)--(4,-.05);
\draw (1,.05)--(4,.05);
\draw (4,0)--(4.5,0);
\ns{2,0};
\ns{3,0};
\node at (2.25,.25) {2};
\end{tikzpicture}} 
\newcommand{\zh}{
\begin{tikzpicture}
\def\x{1cm};
\draw (.5,-\x)--(.5,\x);
\draw (1,-\x)--(1,\x);
\draw (4,-\x)--(4,\x);
\draw (4.5,-\x)--(4.5,\x);
\draw (.5,-.25)--(1,-.25);
\draw (1,0)--(4,0);
\draw (1,-.5)--(4,-.5);
\draw (4,-.25)--(4.5,-.25);
\ns{3,.5};
\ns{3,1};
\ns{2,.75};
\end{tikzpicture}}
 \newcommand{\zi}{
\begin{tikzpicture}
\def\x{1cm};
\draw (.5,-\x)--(.5,\x);
\draw (1,-\x)--(1,\x);
\draw (4,-\x)--(4,\x);
\draw (4.5,-\x)--(4.5,\x);
\draw (.5,-.25)--(1,-.25);
\draw (1,0)--(4,0);
\draw (1,-.5)--(4,-.5);
\draw (4,-.25)--(4.5,-.25);
\ns{2,.75};
\ns{3,.75};
\node at (3.25,1) {2};
\end{tikzpicture}}
 \newcommand{\zja}{
\begin{tikzpicture}
\def\x{1cm};
\draw (.5,-\x)--(.5,\x);
\draw (1,-\x)--(1,\x);
\draw (4,-\x)--(4,\x);
\draw (4.5,-\x)--(4.5,\x);
\draw (.5,-0)--(1,0);
\draw (1,-.05)--(4,-.05);
\draw (1,.05)--(4,.05);
\draw (4,-.5)--(4.5,-.5);
\ns{2,0};
\ns{3,0};
\node at (3.25,.25) {2};
\end{tikzpicture}}
 \newcommand{\zjb}{
\begin{tikzpicture}
\def\x{1cm};
\draw (.5,-\x)--(.5,\x);
\draw (1,-\x)--(1,\x);
\draw (4,-\x)--(4,\x);
\draw (4.5,-\x)--(4.5,\x);
\draw (.5,-0)--(1,0);
\draw (1,-.05)--(4,-.05);
\draw (1,.05)--(4,.05);
\draw (4,0)--(4.5,0);
\ns{3,0};
\ns{2,.5};
\node at (3.25,.25) {2};
\end{tikzpicture}}
 \newcommand{\zjc}{
\begin{tikzpicture}
\def\x{1cm};
\draw (.5,-\x)--(.5,\x);
\draw (1,-\x)--(1,\x);
\draw (4,-\x)--(4,\x);
\draw (4.5,-\x)--(4.5,\x);
\draw (.5,-.5)--(1,-.5);
\draw (1,-.05)--(4,-.05);
\draw (1,.05)--(4,.05);
\draw (4,0)--(4.5,0);
\ns{2,0};
\ns{3,0};
\node at (3.25,.25) {2};
\end{tikzpicture}}
 \newcommand{\zk}{
\begin{tikzpicture}
\def\x{1cm};
\draw (.5,-\x)--(.5,\x);
\draw (1,-\x)--(1,\x);
\draw (4,-\x)--(4,\x);
\draw (4.5,-\x)--(4.5,\x);
\draw (.5,0)--(1,0);
\draw (1,-.05)--(4,-.05);
\draw (1,.05)--(4,.05);
\draw (4,0)--(4.5,0);
\ns{2,0};
\ns{3,0};
\node at (3.25,.25) {2};
\end{tikzpicture}}
 \newcommand{\zl}{
\begin{tikzpicture}
\def\x{1cm};
\draw (.5,-\x)--(.5,\x);
\draw (1,-\x)--(1,\x);
\draw (4,-\x)--(4,\x);
\draw (4.5,-\x)--(4.5,\x);
\draw (.5,-.25)--(1,-.25);
\draw (1,0)--(4,0);
\draw (1,-.5)--(4,-.5);
\draw (4,-.25)--(4.5,-.25);
\ns{2.5,.25};
\ns{2.5,.75};
\ns{2.5,1.25};
\end{tikzpicture}}
 \newcommand{\zm}{
\begin{tikzpicture}
\def\x{1cm};
\draw (.5,-\x)--(.5,\x);
\draw (1,-\x)--(1,\x);
\draw (4,-\x)--(4,\x);
\draw (4.5,-\x)--(4.5,\x);
\draw (.5,-.25)--(1,-.25);
\draw (1,0)--(4,0);
\draw (1,-.5)--(4,-.5);
\draw (4,-.25)--(4.5,-.25);
\ns{2.5,.5};
\ns{2.5,1};
\node at (2.75,.75) {2};
\end{tikzpicture}}
 \newcommand{\zn}{
\begin{tikzpicture}
\def\x{1cm};
\draw (.5,-\x)--(.5,\x);
\draw (1,-\x)--(1,\x);
\draw (4,-\x)--(4,\x);
\draw (4.5,-\x)--(4.5,\x);
\draw (.5,0)--(1,0);
\draw (1,-.05)--(4,-.05);
\draw (1,.05)--(4,.05);
\draw (4,0)--(4.5,0);
\ns{2.5,0};
\ns{2.5,1};
\node at (2.75,0.25) {2};
\end{tikzpicture}}
 \newcommand{\zo}{
\begin{tikzpicture}
\def\x{1cm};
\draw (.5,-\x)--(.5,\x);
\draw (1,-\x)--(1,\x);
\draw (4,-\x)--(4,\x);
\draw (4.5,-\x)--(4.5,\x);
\draw (.5,-.25)--(1,-.25);
\draw (1,0)--(4,0);
\draw (1,-.5)--(4,-.5);
\draw (4,-.25)--(4.5,-.25);
\ns{2.5,.5};
\node at (2.75,.75) {3};
\end{tikzpicture}}
 \newcommand{\zp}{
\begin{tikzpicture}
\def\x{1cm};
\draw (.5,-\x)--(.5,\x);
\draw (1,-\x)--(1,\x);
\draw (4,-\x)--(4,\x);
\draw (4.5,-\x)--(4.5,\x);
\draw (.5,0)--(1,0);
\draw (1,-.05)--(4,-.05);
\draw (1,.05)--(4,.05);
\draw (4,-.5)--(4.5,-.5);
\ns{2.5,0};
\node at (2.75,0.25) {3};
\end{tikzpicture}}
 \newcommand{\zq}{
\begin{tikzpicture}
\def\x{1cm};
\draw (.5,-\x)--(.5,\x);
\draw (1,-\x)--(1,\x);
\draw (4,-\x)--(4,\x);
\draw (4.5,-\x)--(4.5,\x);
\draw (.5,-.5)--(1,-.5);
\draw (1,-.05)--(4,-.05);
\draw (1,.05)--(4,.05);
\draw (4,0)--(4.5,0);
\ns{2.5,0};
\node at (2.75,0.25) {3};
\end{tikzpicture}}
 \newcommand{\zr}{
\begin{tikzpicture}
\def\x{1cm};
\draw (.5,-\x)--(.5,\x);
\draw (1,-\x)--(1,\x);
\draw (4,-\x)--(4,\x);
\draw (4.5,-\x)--(4.5,\x);
\draw (.5,0)--(1,0);
\draw (1,-.05)--(4,-.05);
\draw (1,.05)--(4,.05);
\draw (4,0)--(4.5,0);
\ns{2.5,0};
\node at (2.75,0.25) {3};
\end{tikzpicture}}

\preprint{\hspace{1cm}}
\title{Fibrations and Hasse diagrams for 6d SCFTs}
\author[I,II]{Antoine Bourget}
\author[III]{and Julius F. Grimminger}
\affiliation[I]{Laboratoire de Physique de l'\'Ecole Normale Sup\'erieure, PSL University, \\ 24 rue Lhomond, 75005 Paris, France}
\affiliation[II]{Université Paris-Saclay, CNRS, CEA, Institut de physique théorique, 91191, Gif-sur-Yvette, France}
\affiliation[III]{Theoretical Physics Group, The Blackett Laboratory, Imperial College London, Prince Consort Road
London, SW7 2AZ, UK}
\emailAdd{antoine.bourget@polytechnique.org}
\emailAdd{julius.grimminger17@imperial.ac.uk}

\abstract{We study the full moduli space of vacua of 6d worldvolume SCFTs on M5 branes probing an $A$-type singularity, focusing on the geometric incarnation of the discrete gauging mechanism which acts as a discrete quotient on the Higgs branch fibered over the tensor branch. We combine insights from brane constructions and magnetic quiver techniques, in which discrete gauging is implemented through the concept of decoration introduced in \cite{Bourget:2022ehw}. We discover and characterize new transverse slices between phases of 6d SCFTs, identifying some of them with a family of isolated symplectic singularities recently discovered in \cite{2021arXiv211215494B}, and conjecturing the existence of two new isolated symplectic singularities. }

\begin{document}
\maketitle

\section{Introduction and Summary}

Supersymmetric quantum field theories with 8 supercharges can be characterized by their moduli space of vacua $\mathcal{M}$, which possesses an intricate geometry. This geometry contains a lot of crucial physical content, which has been successfully used in the past decades to learn about QFTs. The work by Seiberg and Witten \cite{Seiberg:1994rs,Seiberg:1994aj} used the structure of the Coulomb branch of 4d $\mathcal{N}=2$ theories to understand the strong coupling dynamics, in particular showing that certain singularities correspond to non-perturbative states becoming massless, thereby giving an explanation to confinement. 
It has been realized that restrictions imposed by physical requirements on the types of singularities can lead to partial classifications of QFTs, in particular in the conformal case. In 4d, the classification has been attempted at low rank \cite{Caorsi:2018ahl,Martone:2020nsy,Argyres:2020wmq} (the rank is the complex dimension of the Coulomb branch), yielding impressive results at rank 1 \cite{Argyres:2015ffa,Argyres:2015gha,Argyres:2016xmc,Caorsi:2019vex} and rank 2 \cite{Argyres:2005pp,Argyres:2005wx,Argyres:2018zay,Argyres:2019ngz,Argyres:2022lah,Argyres:2022puv,Argyres:2022fwy}.

A feature of the geometry of the moduli spaces of vacua that is central in all the works mentioned above is the singularity structure. This naturally defines a stratification which can be depicted in a phase diagram encoding a partial order (making it a Hasse diagram) \cite{Bourget:2019aer}. Dots in the diagram represent the locus in $\mathcal{M}$ corresponding to a given phase, and edges are labeled with the local singular geometry which encodes phase transitions. These ideas are for instance reflected in flavor symmetry breaking patterns in 6d SCFTs \cite{Heckman:2016ssk,Heckman:2018pqx,Cremonesi:2015bld,Mekareeya:2016yal,Hassler:2019eso,Baume:2021qho}, and in 5d SCFTs \cite{Apruzzi:2019vpe,Apruzzi:2019opn,Apruzzi:2019enx,Martone:2021drm}. 

The Higgs branch $\mathcal{M}_H \subset \mathcal{M}$ is a hyperK\"ahler singular space \cite{Hitchin:1986ea,Argyres:1996eh,Antoniadis:1996ra}, i.e. a symplectic singularity \cite{beauville2000symplectic,kaledin2006symplectic,fu2006survey}. The study of the singular stratification of $\mathcal{M}_H$, which coincides with its symplectic leaves stratification, was put forward in connection to generalized Higgsing in \cite{Bourget:2019aer}. This is the physical incarnation of the stratification of symplectic singularities \cite{kaledin2006symplectic}, which can be seen as an extension of the partial order on nilpotent orbits of simple algebras \cite{kraft1980minimal,Kraft1982,2015arXiv150205770F} which are realized explicitly in brane setups \cite{Gaiotto:2013bwa,Cabrera:2016vvv,Cabrera:2017njm}. A powerful tool to compute this hyperK\"ahler singular structure is to realize it as the 3d $\mathcal{N}=4$ Coulomb branch \cite{Cremonesi:2013lqa,Bullimore:2015lsa,Nakajima:2015txa,Braverman:2016wma,Dedushenko:2017avn,Dedushenko:2018icp} of a quiver gauge theory, which is then called a \emph{magnetic quiver} for that singularity
\cite{Hanany:1996ie,DelZotto:2014kka,Ferlito:2017xdq,Hanany:2018uhm,Cabrera:2018jxt,Cabrera:2019izd,Cabrera:2019dob,Bourget:2019rtl,Bourget:2020asf,Bourget:2020gzi,Beratto:2020wmn,Closset:2020scj,Akhond:2020vhc,Bourget:2020mez,vanBeest:2020kou,Giacomelli:2020gee,VanBeest:2020kxw,Closset:2020afy,Akhond:2021knl,Martone:2021ixp,Arias-Tamargo:2021ppf,Bourget:2021xex,vanBeest:2021xyt,Carta:2021dyx,Xie:2021ewm,Sperling:2021fcf,Nawata:2021nse,Closset:2021lwy,Bhardwaj:2021mzl,Akhond:2022jts,Kang:2022zsl,Bertolini:2022osy,Hanany:2022itc}. This notion extends the notion of 3d mirror symmetry \cite{Intriligator:1996ex,deBoer:1996mp,deBoer:1996ck,Feng:2000eq,Benini:2010uu,Nanopoulos:2010bv}, its scope encompassing much beyond 3d Higgs branches. The stratification is obtained using an algorithm dubbed \emph{quiver subtraction}, which is derived from brane manipulations \cite{Cabrera:2018ann,Bourget:2019aer}. In certain situations, it turns out to be necessary to extend the standard notion of quiver : non simply laced quivers \cite{Cremonesi:2014xha,Dey:2014tka,Dey:2016qqp,Kimura:2017hez,Nakajima:2019olw}, wreathed quivers \cite{Bourget:2020bxh,Arias-Tamargo:2021ppf}, decorated quivers \cite{Bourget:2022ehw}. A Hasse diagram, or phase diagram, not only is a highly discriminant invariant one can associate to a QFT with 8 supercharges, but it also shows how the theory is related to others under various kinds of deformations or flows (see  \cite{Eckhard:2020jyr,Santilli:2021eon,Gledhill:2021cbe,Nawata:2021nse,Santilli:2021rlf,DeMarco:2021try,Giacomelli:2022drw,Carta:2022fxc,Fazzi:2022hal} for a selection of recent works along these lines).  

The stratification of all of $\mathcal{M}$ was studied in \cite{Assel:2017jgo,Assel:2018exy,Grimminger:2020dmg,Argyres:2019yyb,Argyres:2020wmq}.
The general picture is as follows: $\mathcal{M}$ can be described as a fibration with a global section,\footnote{In this paper, all fibrations are implicitly assumed to be fibrations with a global section.} where the base is the Coulomb/tensor branch, while the fiber is the local Higgs branch. The fiber geometry stays the same as long as it lies above the same leaf in the base space, but can jump when going from one leaf to the next -- this statement can be seen as a refinement of the classical non-renormalization theorems for the Higgs branch. The nature of these jumps depends on the theory, but can be severe in general, including a change of the dimension of the fiber, which is the usually observed effect of extra Higgs branch directions opening up at singularities of the Coulomb branch, implying the fibration is non-flat; or a discrete quotient of the fiber which is the case we will describe here. It is also possible to change the point of view and see the Higgs branch as the base, and the local Coulomb/tensor branch as the fiber. 

\begin{figure}
\begin{center}
\begin{tikzpicture}
\node at (-3,0) {\begin{tikzpicture}
\draw[draw=white,fill=blue!20] (0,0)--(1,2)--(-1,2);
\draw[draw=white,fill=blue!20] (0,0)--(1,-2)--(-1,-2);
\draw (-1,2)--(1,-2);
\draw (1,2)--(-1,-2);
\draw[fill=blue!20] (.99,2) arc(0:360:.99 and .4);
\draw[dotted] (.99,-2) arc(0:180:.99 and .4);
\draw[fill=blue!20] (-.99,-1.98) arc(180:360:.99 and .4);
\draw[<->] (0,1.8)--(0,2.2);
\node at (.4,2) {$\mathbb{Z}_2$};
\end{tikzpicture}};
\node at (0,0) {
\begin{tikzpicture}
\draw[draw=white,fill=blue!20] (-1,2)--(1,-2)--(-1,-2)--(1,2)--(-1,2);
\draw[very thick,olive] (-1,2)--(1,-2) (-1,-2)--(1,2);
\end{tikzpicture}};
\draw[purple] (-4,0)--(0,0);
\node[hasse] at (0,0) {};
\node at (3.5,0) {\begin{tikzpicture}
        \node[hasse,blue!50,label=right:{\tiny $10$}] (t) at (0,1) {};
        \node[hasse,olive,label=right:{\tiny $6$}] (m) at (0,0) {};
        \node[hasse,label=right:{\tiny $0$}] (b) at (0,-1) {};
        \draw (t)--(m)--(b);
        \node at (0.3,0.5) {$A_1$};
        \node at (0.3,-0.5) {$b_3$};
        \node[hasse,blue!50,label=left:{\tiny $11$}] (tl) at (-1,1.5) {};
        \node[hasse,purple,label=left:{\tiny $1$}] (bl) at (-1,-0.5) {};
        \draw (bl)--(tl);
        \node at (-1.3,0.5) {$d_4$};
        \draw[red] (bl)--(b);
        \draw[red,dashed] (tl)--(t);
        \node at (-0.2,1.5) {\scriptsize$(\mathbb{R},\mathbb{Z}_2)$};
        \node at (-0.6,-1) {\scriptsize$\mathbb{R}/\mathbb{Z}_2$};
    \end{tikzpicture}};
    \node at (7,0) {\begin{tikzpicture}
        \node[hasse,olive,label=right:{\tiny $6$}] (m) at (0,0) {};
        \node[hasse,label=right:{\tiny $0$}] (b) at (0,-1) {};
        \draw (m)--(b);
        \node at (0.3,-0.5) {$b_3$};
        
        \node[hasse,blue!50,label=left:{\tiny $11$}] (tl) at (-1,1.5) {};
        \node[hasse,purple,label=left:{\tiny $1$}] (bl) at (-1,-0.5) {};
        \draw (bl)--(tl);
        \node at (-1.3,0.5) {$d_4$};
        \draw[red] (bl)--(b);
        \draw[orange] (tl)--(m);
        \node at (0,1) {\scriptsize$\mathbb{R}^5/\mathbb{Z}_2$};
        \node at (-0.6,-1) {\scriptsize$\mathbb{R}/\mathbb{Z}_2$};
    \end{tikzpicture}};
    \node at (-1.5,-3) {$(a)$};
    \node at (3.5,-3) {$(b)$};
    \node at (7,-3) {$(c)$};
\node at (-3,1.2) {\scriptsize$(d_4,\mathbb{Z}_2)$};
\node at (0,1.2) {\scriptsize$d_4/\mathbb{Z}_2$};
\node at (-1.5,-0.3) {\scriptsize$\mathbb{R}/\mathbb{Z}_2$};
\end{tikzpicture}
\end{center}
\caption{$(a)$ Full moduli space $\mathcal{M}$ of the theory \eqref{eq:linearquiver} with $n=k=2$, viewed as a fibration by the Higgs branch (blue) over the tensor branch (purple). At the origin of the tensor branch, the fiber develops a singularity (olive) due to quotienting by the $\mathbb{Z}_2$ symmetry on the generic $d_4$ fiber (indicated by the notation $(d_4,\mathbb{Z}_2)$ borrowed from  \cite{slodowy1980simple,2015arXiv150205770F}) producing the next-to-minimal nilpotent orbit closure of $\mathfrak{so}(7)$. The flavor symmetry is reduced from $\mathfrak{so}(8)$ to $\mathfrak{so}(7)$ at the origin of the tensor branch \cite{Hanany:2018vph}. $(b)$ Hasse diagram of $\mathcal{M}$ with respect to the fibration depicted in $(a)$, with the real dimension of each locus. The red lines denote $\mathcal{N}=(1,0)$ tensor branch transitions; when it is dashed, the transition is actually smooth (no phase transition). Indeed the top two blue vertices represent the same phase of the theory, they are distinct only because the diagram respects the fibration. The real codimension 1 locus is selected by the $\mathbb{Z}_2$ symmetry acting on the smooth tensor branch $\mathbb{R}$. $(c)$ Hasse diagram of $\mathcal{M}$. There are only four phases. The orange line denotes an $\mathcal{N}=(2,0)$ tensor branch transition, the transverse geometry being $\mathbb{R}^5/\mathbb{Z}_2$.  }
 \label{fig:coneFibrationZ2}
\end{figure}
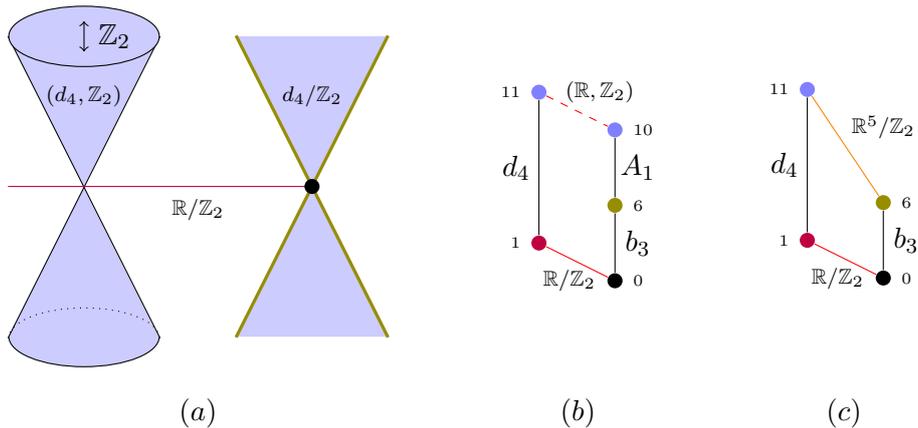

In this paper, we leverage these tools to explore the full moduli space of 6d $\mathcal{N}=(1,0)$ theories. We focus on a simple yet very rich family of 6d theories, which can be realized in M-theory by $n$ M5 branes probing a $\mathbb{C}^2 / \mathbb{Z}_k$ singularity \cite{Intriligator:1997kq,Brunner:1997gk,Hanany:1997gh}.  When the M5 branes are separated, the theory is on its tensor branch, it can be described by the quiver 
\begin{equation}
  \raisebox{-.5\height}{  \begin{tikzpicture}
        \node[flavour,label=below:{$k$}] (0) at (0,0) {};
        \node[gaugeBig,label=below:{SU$(k)$}] (1) at (1,0) {};
        \node (2) at (2,0) {$\cdots$};
        \node[gaugeBig,label=below:{SU$(k)$}] (3) at (3,0) {};
        \node[flavour,label=below:{$k$}] (4) at (4,0) {};
        \draw (0)--(1)--(2)--(3)--(4);
		\draw [decorate,decoration={brace,amplitude=5pt}] (3.2,-0.8)--(0.8,-0.8);
		\node at (2,-1.2) {$n-1$ nodes};
    \end{tikzpicture}}\; 
    \label{eq:linearquiver}
\end{equation}
and the Higgs branch can be studied classically.  When some M5 branes coincide, we move on to more singular loci on the tensor branch, and the theory experiences so-called discrete gauging \cite{Hanany:2018vph}. One remarkable and somewhat counter-intuitive consequence of this phenomenon is the fact that the flavor symmetry of SU(2) SQCD with $N_f = 4$ (i.e. the case $n=k=2$), which is classically $\mathfrak{so}(8)$, reduces to $\mathfrak{so}(7)$ at the SCFT point at the origin of the tensor branch \cite{Ohmori:2015pia,Mekareeya:2017jgc,Bah:2017gph,Hanany:2018vph}. Geometrically, this can be interpreted in the context of the fibration of $\mathcal{M}$ over the tensor branch: while the fiber has a fixed dimension everywhere on the tensor branch, the singularity structure of the fiber changes at the origin of the tensor branch, see Figure \ref{fig:coneFibrationZ2}$(a)$. This physical effect and its generalizations can be understood from the magnetic quiver perspective, thus motivating an in-depth study of Hasse diagram of symmetric products of symplectic singularities. A central role is played by \emph{decorated quivers}, recently introduced in \cite{Bourget:2022ehw}. A summary of why these are needed is given in Appendix \ref{app:decorated}.

While the description of $\mathcal{M}$ as a fibration provides a powerful organizing principle that allows to apply the decorated quiver techniques to compute the singularity structure of the fibers, it can also introduce spurious loci in these fibers, which breaks the one-to-one correspondence between dots of the Hasse diagrams and phases of the theory. This is illustrated in Figure \ref{fig:coneFibrationZ2}$(b)$ and $(c)$. 

In view of this, it is useful to draw two general lessons:
\begin{enumerate}
    \item[$(i)$] Describing a space as a fibration can change the Hasse diagram.  
    \item[$(ii)$] The Hasse diagram of a transverse slice in a space $X$ might not be a sub-diagram of the Hasse diagram of $X$. 
\end{enumerate}

The main motivation of this note is the full phase diagram for the 6d theory described above, derived independently from string theory and magnetic quiver technology. In the steps needed to reach this result, we uncover several new features: 
\begin{itemize}
    \item We identify the recently discovered family of new symplectic singularities $\mathcal{Y}(d)$ \cite{2021arXiv211215494B} in 6d physics. 
    \item We conjecture the existence of two new isolated symplectic singularities, that we call $\mathcal{J}_{32}$ and $\mathcal{J}_{33}$, unknown in both the mathematics and physics literature. Their existence is derived from brane systems, and their Hilbert series is computed using magnetic quivers. Their isometry algebra is $\mathfrak{so}(4) = \mathfrak{su}(2) \oplus \mathfrak{su}(2)$. 
    \item We point out that the transverse slice between two adjacent leaves in a Higgs branch Hasse diagram can be a union of arbitrary many irreducible singularities, a known fact from geometry of which we give an interpretation in string theory. 
    \item We give a general recipe to compute the Hasse diagram of symmetric products of symplectic singularities. 
\end{itemize}

The paper is organized as follows. We start in Section \ref{sec:unionCones} by a review and extension of the decorated quiver technology of \cite{Bourget:2022ehw}, emphasizing the possibility of having multiple cones in a transverse slice. We apply these tools to symmetric products of symplectic singularities in Section \ref{sec:symmetricProducts} as a necessary step before constructing the full moduli space of the 6d theory described above in Section \ref{sec:6d}. The text is complemented by appendices which review the geometry of the singularities $\mathcal{Y}(d)$ (Appendix \ref{app:slice}) and $\mathcal{J}_{32}$ and $\mathcal{J}_{33}$ (Appendix \ref{app:Jslices}). Appendix \ref{app:decorated} gathers basic facts about decorated quivers, explaining why they are necessary and illustrating with an example. In appendix \ref{app:16susy} we show how non-singular transitions can arise in Hasse diagrams of the full moduli space on the example of theories with 16 supercharges as well as 3d $\mathcal{N}=4$ SO$(3)$ SQCD.

\section{Unions of Cones in Quiver Subtraction}
\label{sec:unionCones}

In this section we briefly review the algorithm of \cite{Bourget:2022ehw} to compute Hasse diagrams of unitary quivers with adjoint loops and the possibility of subtracting the same slice multiple times through the use of decorated quivers. We add an important ingredient to the algorithm, to deal with elementary slices which are unions of cones. Throughout the paper we use the notation
\begin{equation}
    nS=\underbrace{S\cup S \cup...\cup S}_{n \, \textnormal{times}}\;.
\end{equation}
When such a slice appears in a diagram, we draw $n$ parallel lines. 

\subsection{Decorated Node Merging}
As pointed out in \cite{Bourget:2022ehw} one can merge U$(1)$ nodes which are decorated in the same color as a generalised quiver subtraction. Let us consider as an example the 4-th symmetric product of $\mathbb{C}^2$. We can realize this as the Higgs branch of a brane system with 4 parallel D$p$ branes moving inside a single D$(p+4)$ brane. We draw a D$p$ brane as $\begin{tikzpicture}
\node[shrinky] at (0,0) {};
\end{tikzpicture}$, and denote coincident D$p$ branes by drawing a circle around them, e.g.  $\begin{tikzpicture}
\node[shrinky] (1) at (0,0) {};
\node[shrinky] (2) at (0.3,0) {};
\draw \convexpath{1,2}{3pt};
\end{tikzpicture}$. We can draw the Hasse diagram of brane phases and associate partitions to each brane phase:
\begin{equation}
    \raisebox{-.5\height}{\begin{tikzpicture}
                \node (4) at (-0.3,0) {$\begin{tikzpicture}
                    \node[shrinky] (4x1) at (0,0) {};
                    \node[shrinky] (4x2) at (0.3,0) {};
                    \node[shrinky] (4x3) at (0.6,0) {};
                    \node[shrinky] (4x4) at (0.9,0) {};
                    \draw \convexpath{4x1,4x4}{3pt};
                \end{tikzpicture}
                $};
                \node (31) at (-0.7-0.3,1) {$\begin{tikzpicture}
                    \node[shrinky] (31x1) at (0,0) {};
                    \node[shrinky] (31x2) at (0.3,0) {};
                    \node[shrinky] (31x3) at (0.6,0) {};
                    \node[shrinky] (31x4) at (0.9,0) {};
                    \draw \convexpath{31x1,31x3}{3pt};
                \end{tikzpicture}
                $};
                \node (22) at (0.7-0.3,1) {$\begin{tikzpicture}
                    \node[shrinky] (22x1) at (0,0) {};
                    \node[shrinky] (22x2) at (0.3,0) {};
                    \node[shrinky] (22x3) at (0.6,0) {};
                    \node[shrinky] (22x4) at (0.9,0) {};
                    \draw \convexpath{22x1,22x2}{3pt};
                    \draw \convexpath{22x3,22x4}{3pt};
                \end{tikzpicture}
                $};
                \node (211) at (0-0.3,2) {$\begin{tikzpicture}
                    \node[shrinky] (211x1) at (0,0) {};
                    \node[shrinky] (211x2) at (0.3,0) {};
                    \node[shrinky] (211x3) at (0.6,0) {};
                    \node[shrinky] (211x4) at (0.9,0) {};
                    \draw \convexpath{211x1,211x2}{3pt};
                \end{tikzpicture}
                $};
                \node (1111) at (0-0.3,3) {$\begin{tikzpicture}
                    \node[shrinky] (1111x1) at (0,0) {};
                    \node[shrinky] (1111x2) at (0.3,0) {};
                    \node[shrinky] (1111x3) at (0.6,0) {};
                    \node[shrinky] (1111x4) at (0.9,0) {};
                \end{tikzpicture}
                $};
                \draw (4)--(31)--(211)--(1111) (4)--(22)--(211);
    \end{tikzpicture} } \qquad\leftrightarrow\qquad   \raisebox{-.5\height}{\begin{tikzpicture}
                \node (4) at (-0.3,0) {$[4]$};
                \node (31) at (-0.7-0.3,1) {$[3,1]$};
                \node (22) at (0.7-0.3,1) {$[2^2]$};
                \node (211) at (0-0.3,2) {$[2,1^2]$};
                \node (1111) at (0-0.3,3) {$[1^4]$};
                \draw (4)--(31)--(211)--(1111) (4)--(22)--(211);
    \end{tikzpicture}}
    \label{eq:Sym4Branes}
\end{equation}
We argued in \cite{Bourget:2022ehw} that a transition where two stacks of $n_1$ and $n_2$ coincident branes are made to coincide as $n_1+n_2$ coincident branes is
\begin{itemize}
    \item $A_1$ if $n_1=n_2$
    \item $m$ if $n_1\neq n_2$
\end{itemize}
This statement needs to be amended however. Let us focus on the transition between the partitions $[2,2]$ and $[2,1^2]$. Going upwards in the Hasse diagram, there are two possibilities, one could split the `left' or the `right' 2-stack of branes:
\begin{equation}
    \raisebox{-.5 \height}{\begin{tikzpicture}
    \node (a) at (0,0) {$\begin{tikzpicture}
                    \node[shrinky] (22x1) at (0,0) {};
                    \node[shrinky] (22x2) at (0.3,0) {};
                    \node[shrinky] (22x3) at (0.6,0) {};
                    \node[shrinky] (22x4) at (0.9,0) {};
                    \draw \convexpath{22x1,22x2}{3pt};
                    \draw \convexpath{22x3,22x4}{3pt};
                \end{tikzpicture}
                $};
    \node (b) at (-1,1) {$\begin{tikzpicture}
                    \node[shrinky] (211x1) at (0,0) {};
                    \node[shrinky] (211x2) at (0.3,0) {};
                    \node[shrinky] (211x3) at (0.6,0) {};
                    \node[shrinky] (211x4) at (0.9,0) {};
                    \draw \convexpath{211x1,211x2}{3pt};
                \end{tikzpicture}
                $};
    \node (c) at (1,1) {$\begin{tikzpicture}
                    \node[shrinky] (211x1) at (0,0) {};
                    \node[shrinky] (211x2) at (0.3,0) {};
                    \node[shrinky] (211x3) at (0.6,0) {};
                    \node[shrinky] (211x4) at (0.9,0) {};
                    \draw \convexpath{211x3,211x4}{3pt};
                \end{tikzpicture}
                $};
    \draw[->] (a)--(b);
    \draw[->] (a)--(c);
    \end{tikzpicture}}\;.
    \label{eq:A1unionA1TransBranes}
\end{equation}
Both transitions are $A_1$. The two brane systems on top of \eqref{eq:A1unionA1TransBranes} are related by moving branes around without going through any singular transitions. Hence the two configurations are actually the same leaf. This is no surprise, and it is how the Hasse diagram was already drawn in \eqref{eq:Sym4Branes}. However, there are in fact two $A_1$ transitions from the leaf $[2,2]$ to the leaf $[2,1^2]$. The transverse slice between the two leaves is therefore not $A_1$ but rather $2A_1=A_1\cup A_1$.\footnote{This can be seen in the algebraic description of $\mathrm{Sym}^4(\mathbb{C}^2)$. Denote the elements of $\mathrm{Sym}^4(\mathbb{C}^2)$ by $\mathcal{O}_{(x_1 , x_2 , x_3 , x_4)}$, the orbit of $(x_1 , x_2 , x_3 , x_4) \in \mathbb{C}^4$ under the permutation group $S_4$. The $[2^2]$ leaf is $\{ \mathcal{O}_{(x_1 , x_1 , x_2 , x_2)} | x_1 \neq x_2 \}$, and given a point $\mathcal{O}_{(x_1 , x_1 , x_2 , x_2)}$ in this leaf, the transverse slice in the leaf $[2,1^2]$ is  $\{ \mathcal{O}_{(x_1 +u , x_1 -u , x_2 +v , x_2 -v)} | (u,v) \in (\mathbb{C}^2 / \mathbb{Z}_2 )\times \{0\} \cup \{0\}  \times (\mathbb{C}^2 / \mathbb{Z}_2) \} $, which we denote as $A_1 \cup A_1$.  } $\mathrm{Sym}^4(\mathbb{C}^2)$ can be realized as the Coulomb branch of
\begin{equation}
    \raisebox{-.5 \height}{\begin{tikzpicture}
        \node[gauge,label=below:{$1$}] (a) at (0,0) {};
        \node[gauge,label=left:{$4$}] (u) at (0,1) {};
        \draw (a)--(u);
        \draw (u) to [out=45,in=135,looseness=8] (u);
    \end{tikzpicture}} \cong \raisebox{-.5 \height}{ \begin{tikzpicture}
        \node[gauge,label=below:{$1$}] (a) at (0,0) {};
        \node[gauge,label=left:{$1$}] (u1) at (-1.5,1) {};
        \draw (a)--(u1);
        \draw (u1) to [out=45,in=135,looseness=8] (u1);
        \node[gauge,label=left:{$1$}] (u2) at (-0.5,1) {};
        \draw (a)--(u2);
        \draw (u2) to [out=45,in=135,looseness=8] (u2);
        \node[gauge,label=left:{$1$}] (u3) at (0.5,1) {};
        \draw (a)--(u3);
        \draw (u3) to [out=45,in=135,looseness=8] (u3);
        \node[gauge,label=left:{$1$}] (u4) at (1.5,1) {};
        \draw (a)--(u4);
        \draw (u4) to [out=45,in=135,looseness=8] (u4);
        \draw[purple] (u1) circle (0.6cm);
        \draw[purple] (u2) circle (0.6cm);
        \draw[purple] (u3) circle (0.6cm);
        \draw[purple] (u4) circle (0.6cm);
    \end{tikzpicture}}\;.
\end{equation}
The node merging procedure is as follows:
\begin{equation}
\raisebox{-.5 \height}{\scalebox{0.7}{\begin{tikzpicture}
                \node (4) at (0,0) {
    $\begin{tikzpicture}
        \node[gauge,label=below:{$1$}] (a) at (-1,0) {};
        \node[gauge,label=left:{$1$}] (u) at (-1,1) {};
        \draw (u) to [out=45,in=135,looseness=8] (u);
        \draw[transform canvas={xshift=3pt}] (a)--(u);
        \draw[transform canvas={xshift=1pt}] (a)--(u);
        \draw[transform canvas={xshift=-1pt}] (a)--(u);
        \draw[transform canvas={xshift=-3pt}] (a)--(u);
        \draw (-1-0.2,0.5-0.1)--(-1,0.5+0.1)--(-1+0.2,0.5-0.1);
        \draw[purple] (u) circle (0.6cm);
    \end{tikzpicture}$};
                \node (22) at (2,3.5) {
    $\begin{tikzpicture}
        \node[gauge,label=below:{$1$}] (a) at (-1,0) {};
        \begin{scope}[rotate around={45:(-1,0)}]
        \node[gauge,label=left:{$1$}] (u) at (-1,1) {};
        \draw (u) to [out=45,in=135,looseness=8] (u);
        \draw[transform canvas={xshift=1.3pt,yshift=1.3pt}] (a)--(u);
        \draw[transform canvas={xshift=-1.3pt,yshift=-1.3pt}] (a)--(u);
        \draw (-1-0.2,0.5-0.1)--(-1,0.5+0.1)--(-1+0.2,0.5-0.1);
        \end{scope}
        \begin{scope}[rotate around={-45:(-1,0)}]
        \node[gauge,label=left:{$1$}] (u2) at (-1,1) {};
        \draw (u2) to [out=45,in=135,looseness=8] (u2);
        \draw[transform canvas={xshift=1.3pt,yshift=-1.3pt}] (a)--(u2);
        \draw[transform canvas={xshift=-1.3pt,yshift=1.3pt}] (a)--(u2);
        \draw (-1-0.2,0.5-0.1)--(-1,0.5+0.1)--(-1+0.2,0.5-0.1);
        \end{scope}
        \draw[purple] (u) circle (0.6cm);
        \draw[purple] (u2) circle (0.6cm);
    \end{tikzpicture}$};
                \node (31) at (-2,3.5) {
    $\begin{tikzpicture}
        \node[gauge,label=below:{$1$}] (a) at (-1,0) {};
        \begin{scope}[rotate around={45:(-1,0)}]
        \node[gauge,label=left:{$1$}] (u) at (-1,1) {};
        \draw (u) to [out=45,in=135,looseness=8] (u);
        \draw[transform canvas={xshift=1.5pt,yshift=1.5pt}] (a)--(u);
        \draw (a)--(u);
        \draw[transform canvas={xshift=-1.5pt,yshift=-1.5pt}] (a)--(u);
        \draw (-1-0.2,0.5-0.1)--(-1,0.5+0.1)--(-1+0.2,0.5-0.1);
        \end{scope}
        \begin{scope}[rotate around={-45:(-1,0)}]
        \node[gauge,label=left:{$1$}] (u2) at (-1,1) {};
        \draw (u2) to [out=45,in=135,looseness=8] (u2);
        \draw (a)--(u2);
        \end{scope}
        \draw[purple] (u) circle (0.6cm);
        \draw[purple] (u2) circle (0.6cm);
    \end{tikzpicture}$};
                \node (211) at (0,7) {
    $\begin{tikzpicture}
        \node[gauge,label=below:{$1$}] (a) at (-1,0) {};
        \begin{scope}[rotate around={45:(-1,0)}]
        \node[gauge,label=left:{$1$}] (u) at (-1,1) {};
        \draw (u) to [out=45,in=135,looseness=8] (u);
        \draw[transform canvas={xshift=1.3pt,yshift=1.3pt}] (a)--(u);
        \draw[transform canvas={xshift=-1.3pt,yshift=-1.3pt}] (a)--(u);
        \draw (-1-0.2,0.5-0.1)--(-1,0.5+0.1)--(-1+0.2,0.5-0.1);
        \end{scope}
        \node[gauge,label=left:{$1$}] (u2) at (-0.5,1) {};
        \draw (a)--(u2);
        \node[gauge,label=left:{$1$}] (u3) at (0.5,1) {};
        \draw (a)--(u3);
        \draw[purple] (u) circle (0.6cm);
        \draw[purple] (u2) circle (0.6cm);
        \draw[purple] (u3) circle (0.6cm);
        \draw (u2) to [out=45,in=135,looseness=8] (u2);
        \draw (u3) to [out=45,in=135,looseness=8] (u3);
    \end{tikzpicture}$};
                \node (1111) at (0,10.5) {
    $    \begin{tikzpicture}
        \node[gauge,label=below:{$1$}] (a) at (0,0) {};
        \node[gauge,label=left:{$1$}] (u1) at (-1.5,1) {};
        \draw (a)--(u1);
        \draw (u1) to [out=45,in=135,looseness=8] (u1);
        \node[gauge,label=left:{$1$}] (u2) at (-0.5,1) {};
        \draw (a)--(u2);
        \draw (u2) to [out=45,in=135,looseness=8] (u2);
        \node[gauge,label=left:{$1$}] (u3) at (0.5,1) {};
        \draw (a)--(u3);
        \draw (u3) to [out=45,in=135,looseness=8] (u3);
        \node[gauge,label=left:{$1$}] (u4) at (1.5,1) {};
        \draw (a)--(u4);
        \draw (u4) to [out=45,in=135,looseness=8] (u4);
        \draw[purple] (u1) circle (0.6cm);
        \draw[purple] (u2) circle (0.6cm);
        \draw[purple] (u3) circle (0.6cm);
        \draw[purple] (u4) circle (0.6cm);
    \end{tikzpicture}$};
                \draw (4)--(31)--(211)--(1111) (4)--(22)--(211);
                \draw[transform canvas={xshift=3pt}] (211)--(22);
    \node at (-0.5,8.75) {$A_1$};
    \node at (2,5.25) {$2A_1$};
    \node at (-2.25,5.25) {$m$};
    \node at (2.25,1.75) {$A_1$};
    \node at (-2.25,1.75) {$m$};
\end{tikzpicture}}}
\end{equation}
From this example it is straight forward to read of a general rule when unions appear as elementary slices in the quiver merging procedure. We modify the rule of decorated node merging of \cite{Bourget:2022ehw} as follows:

\begin{mdframed}
\paragraph{Decorated Node Merging:}
When merging two U$(1)$s which are decorated in the same color and have shortness $1/n_1$ and $1/n_2$ respectively, we have the following elementary transition
\begin{itemize}
    \item $(k+1)A_1$, if $n_1=n_2$
    \item $(k+1)m$, if $n_1\neq n_2$
\end{itemize}
Where $k$ is the number U$(1)$s decorated in the same color of shortness $1/(n_1+n_2)$.\\

Schematically we can represent this subtraction as follows:
\begin{equation}
\label{eq:rule1}
   \raisebox{-.5 \height}{ \begin{tikzpicture}
        \node (a) at (0,0) {$\raisebox{-.5 \height}{\begin{tikzpicture}
            \node (1) at (0,0) {$\mathsf{Q}$};
            \node[gauge,label=below:{$1$}] (2u) at (1,0) {};
            \draw (2u) to [out=315,in=45,looseness=8] (2u);
            \node[gauge,label=right:{$1$}] (2d) at (0,-1) {};
            \draw (2d) to [out=225,in=315,looseness=8] (2d);
            \node[gauge,label=left:{$k$}] (3) at (0,1) {};
            \draw (3) to [out=45,in=135,looseness=8] (3);
            	\draw[transform canvas={xshift=1.5pt}] (1)--(3);
            	\draw[transform canvas={xshift=-1.5pt}] (1)--(3);
            	\draw (-0.2,0.5-0.1)--(0,0.5+0.1)--(0.2,0.5-0.1);
            	\draw[transform canvas={xshift=1.5pt}] (1)--(2d);
            	\draw[transform canvas={xshift=-1.5pt}] (1)--(2d);
            	\draw (-0.2,-0.5+0.1)--(0,-0.5-0.1)--(0.2,-0.5+0.1);
            	\draw[transform canvas={yshift=1.5pt}] (1)--(2u);
            	\draw[transform canvas={yshift=-1.5pt}] (1)--(2u);
            	\draw (0.5-0.1,0.2)--(0.5+0.1,0)--(0.5-0.1,-0.2);
            \draw[purple] (2u) circle (0.3cm);
            \draw[purple] (2d) circle (0.3cm);
            \draw[purple] (3) circle (0.25cm);
            \node at (-0.8,0.5) {\scriptsize$n_1+n_2$};
            \node at (0.5,0.3) {\scriptsize$n_1$};
            \node at (-0.3,-0.5) {\scriptsize$n_2$};
        \end{tikzpicture}}=\raisebox{-.5 \height}{\begin{tikzpicture}
            \node (1) at (0,0) {$\mathsf{Q}$};
            \node[gauge,label=below:{$1$}] (2u) at (1,0) {};
            \draw (2u) to [out=315,in=45,looseness=8] (2u);
            \node[gauge,label=right:{$1$}] (2d) at (0,-1) {};
            \draw (2d) to [out=225,in=315,looseness=8] (2d);
            \node[gauge,label=left:{$1$}] (31) at (-0.5,1) {};
            \draw (31) to [out=45,in=135,looseness=8] (31);
            \node at (0,1) {$\cdots$};
            \node[gauge,label=right:{$1$}] (33) at (0.5,1) {};
            \draw (33) to [out=45,in=135,looseness=8] (33);
            	\draw[transform canvas={xshift=1.5pt}] (1)--(31);
            	\draw[transform canvas={xshift=-1.5pt}] (1)--(31);
            	\draw (-0.2-0.3,0.5-0.1)--(0-0.3,0.5+0.1)--(0.2-0.3,0.5-0.1);
            	\draw[transform canvas={xshift=1.5pt}] (1)--(33);
            	\draw[transform canvas={xshift=-1.5pt}] (1)--(33);
            	\draw (-0.2+0.3,0.5-0.1)--(0+0.3,0.5+0.1)--(0.2+0.3,0.5-0.1);
            	\draw[transform canvas={xshift=1.5pt}] (1)--(2d);
            	\draw[transform canvas={xshift=-1.5pt}] (1)--(2d);
            	\draw (-0.2,-0.5+0.1)--(0,-0.5-0.1)--(0.2,-0.5+0.1);
            	\draw[transform canvas={yshift=1.5pt}] (1)--(2u);
            	\draw[transform canvas={yshift=-1.5pt}] (1)--(2u);
            	\draw (0.5-0.1,0.2)--(0.5+0.1,0)--(0.5-0.1,-0.2);
		    \draw [decorate,decoration={brace,amplitude=5pt}] (-0.7,1.5)--(0.7,1.5);
		    \node at (0,2) {$k$};
            \draw[purple] (2u) circle (0.3cm);
            \draw[purple] (2d) circle (0.3cm);
            \draw[purple] (31) circle (0.25cm);
            \draw[purple] (33) circle (0.25cm);
        \end{tikzpicture}}$};
        \node (b) at (0,-5) {$\raisebox{-.5 \height}{\begin{tikzpicture}
            \node (1) at (0,0) {$\mathsf{Q}$};
            \node[gauge,label=left:{$k+1$}] (3) at (0,1) {};
            \draw (3) to [out=45,in=135,looseness=8] (3);
            	\draw[transform canvas={xshift=1.5pt}] (1)--(3);
            	\draw[transform canvas={xshift=-1.5pt}] (1)--(3);
            	\draw (-0.2,0.5-0.1)--(0,0.5+0.1)--(0.2,0.5-0.1);
            \draw[purple] (3) circle (0.25cm);
            \node at (-0.8,0.5) {\scriptsize$n_1+n_2$};
        \end{tikzpicture}}=\raisebox{-.5 \height}{\begin{tikzpicture}
            \node (1) at (0,0) {$\mathsf{Q}$};
            \node[gauge,label=left:{$1$}] (31) at (-0.5,1) {};
            \draw (31) to [out=45,in=135,looseness=8] (31);
            \node at (0,1) {$\cdots$};
            \node[gauge,label=right:{$1$}] (33) at (0.5,1) {};
            \draw (33) to [out=45,in=135,looseness=8] (33);
            	\draw[transform canvas={xshift=1.5pt}] (1)--(31);
            	\draw[transform canvas={xshift=-1.5pt}] (1)--(31);
            	\draw (-0.2-0.3,0.5-0.1)--(0-0.3,0.5+0.1)--(0.2-0.3,0.5-0.1);
            	\draw[transform canvas={xshift=1.5pt}] (1)--(33);
            	\draw[transform canvas={xshift=-1.5pt}] (1)--(33);
            	\draw (-0.2+0.3,0.5-0.1)--(0+0.3,0.5+0.1)--(0.2+0.3,0.5-0.1);
		    \draw [decorate,decoration={brace,amplitude=5pt}] (-0.7,1.5)--(0.7,1.5);
		    \node at (0,2) {$k+1$};
		    \draw[purple] (31) circle (0.25cm);
            \draw[purple] (33) circle (0.25cm);
        \end{tikzpicture}}$};
        \draw (a)--(b);
        \node at (-2,-3) {$(k+1)A_1$ or $(k+1)m$};
    \end{tikzpicture}}
\end{equation}
\end{mdframed}

\subsection{Repeated Identical Quiver Subtraction}
We can now turn to quiver subtraction in the case where the same slice can be subtracted multiple times. As we showed in \cite{Bourget:2022ehw} when there is the possibility of subtracting the same slice multiple times on the same set of nodes, there is a need to decorate the U$(1)$ rebalancing nodes together with the subtracted part of the quiver. Let us consider $\mathcal{M}_{2,\mathrm{SU}(3)}$, the moduli space of two SU$(3)$ instantons on $\mathbb{C}^2$. We can realize this moduli space as the Higgs branch of two parallel D$p$ branes inside a stack of three coincident D$(p+2)$ branes. We draw D$p$ branes just like before, and D$p$ branes which are dissolved into the D$(p+4)$ branes -- realizing instanton moduli -- as $\begin{tikzpicture}
\node[sized] at (0,0) {};
\end{tikzpicture}$. We can draw the Hasse diagram of brane phases and associate partitions to each brane phase:
\begin{equation}
   \raisebox{-.5 \height}{ \begin{tikzpicture}
        \node (2) at (0,0) {$\begin{tikzpicture}
            \node[shrinky] (a) at (0,0) {};
            \node[shrinky] (b) at (0.3,0) {};
            \draw \convexpath{a,b}{3pt};
        \end{tikzpicture}$};
        \node (11) at (0,1) {$\begin{tikzpicture}
            \node[shrinky] at (0,0) {};
            \node[shrinky] at (0.3,0) {};
        \end{tikzpicture}$};
        \node (1) at (1,1.5) {$\begin{tikzpicture}
            \node[shrinky] at (0,0) {};
            \node[sized] at (0.3,0) {};
        \end{tikzpicture}$};
        \node (0) at (2,2) {$\begin{tikzpicture}
            \node[sized] at (0,0) {};
            \node[sized] at (0.3,0) {};
        \end{tikzpicture}$};
        \draw (2)--(11)--(1)--(0);
    \end{tikzpicture}}\qquad\leftrightarrow\qquad  \raisebox{-.5 \height}{\begin{tikzpicture}
        \node (2) at (0,0) {$[2]$};
        \node (11) at (0,1) {$[1^2]$};
        \node (1) at (1,1.5) {$[1]$};
        \node (0) at (2,2) {$[0]$};
        \draw (2)--(11)--(1)--(0);
    \end{tikzpicture}}
    \label{eq:2instHasseBranes}
\end{equation}
Let us focus on the transition $[1^2]$ to $[1]$. Similar to what we discussed before there is a choice to make. There are two branes of which we should pick one to dissolve:
\begin{equation}
   \raisebox{-.5 \height}{ \begin{tikzpicture}
        \node (11) at (0,0) {$\begin{tikzpicture}
            \node[shrinky] (1) at (0,0) {};
            \node[shrinky] (2) at (0.3,0) {};
        \end{tikzpicture}$};
        \node (1a) at (-1,1) {$\begin{tikzpicture}
            \node[sized] (1) at (0,0) {};
            \node[shrinky] (2) at (0.3,0) {};
        \end{tikzpicture}$};
        \node (1b) at (1,1) {$\begin{tikzpicture}
            \node[shrinky] (1) at (0,0) {};
            \node[sized] (2) at (0.3,0) {};
        \end{tikzpicture}$};
    \draw[->] (11)--(1a);
    \draw[->] (11)--(1b);
    \end{tikzpicture}}
\end{equation}
Both transitions are $a_2$ transitions, since we are dealing with SU$(3)$ instantons. For $G$-instantons it would be the minimal nilpotent orbit closure of the corresponding algebra. However the two brane systems are related by a non-singular transition, and therefore are part of the same leaf. Just like before, this is no surprise, and we already drew our Hasse diagram like this in \eqref{eq:2instHasseBranes}. The transverse slice between $[11]$ and $[1]$ is the union $2a_2=a_2\cup a_2$.\\

This is also visible from partial Higgsings. $\mathcal{M}_{2,\mathrm{SU}(3)}$ can be realized through the ADHM construction \cite{atiyah1994construction} as the Higgs branch of
\begin{equation}
   \raisebox{-.5 \height}{ \begin{tikzpicture}
        \node[gauge,label=below:{U$(2)$}] (g) at (0,0) {};
        \draw (g) to [out=135,in=225,looseness=8] (g);
        \node[flavour,label=below:{$3$}] (f) at (1,0) {};
        \draw (g)--(f);
    \end{tikzpicture}}\;.
\end{equation}
Turning on a scalar VEV in the adjoint hyper one Higgses the theory to a product
\begin{equation}
   \raisebox{-.5 \height}{ \begin{tikzpicture}
        \node[gauge,label=below:{U$(1)$}] (g) at (0,0) {};
        \draw (g) to [out=135,in=225,looseness=8] (g);
        \node[flavour,label=below:{$3$}] (f) at (1,0) {};
        \draw (g)--(f);
        \node[gauge,label=below:{U$(1)$}] (g2) at (0,-1) {};
        \draw (g2) to [out=135,in=225,looseness=8] (g2);
        \node[flavour,label=below:{$3$}] (f2) at (1,-1) {};
        \draw (g2)--(f2);
    \end{tikzpicture}}\;.
\end{equation}
While there is a $\mathbb{Z}_2\subset\mathrm{U}(2)$ which exchanges the two U$(1)$s, this $\mathbb{Z}_2$ is broken by the Higgs mechanism.\footnote{We did not appreciate this point in \cite{Bourget:2022ehw} and we thank Carlo Meneghelli for correcting us.} One can now turn on a scalar VEV in the fundamental hyper of either of the U$(1)$ gauge groups -- one has two choices -- to trigger an additional partial Higgsing to a single U$(1)$ gauge theory. However the two choices are not distinguished from the point of view of the U$(2)$ gauge theory where the $\mathbb{Z}_2$ exchanging the U$(1)$s is gauged. Therefore there is a single leaf in the Higgs branch of the U$(2)$ theory on which it is broken to a U$(1)$ theory, while there are two distinct leaves in the Higgs branch U$(1)\times$U$(1)$ theory on which it is broken to a U$(1)$ theory, explaining the elementary slice being a union of cones.\\

$\mathcal{M}_{2,\mathrm{SU}(3)}$ can also be realized as the Coulomb branch of \cite{deBoer:1996mp}
\begin{equation}
   \raisebox{-.5\height}{ \begin{tikzpicture}
            \node[gauge,label=left:{$1$}] (a) at (0,0) {};
            \node[gauge,label=below:{$2$}] (c1) at (1,0) {};
            \node[gauge,label=right:{$2$}] (c2) at (2,0.5) {};
            \node[gauge,label=right:{$2$}] (c3) at (2,-0.5) {};
            \draw (a)--(c1)--(c2)--(c3)--(c1);
            \end{tikzpicture}}\;.
\end{equation}
On the level of quiver subtraction we have the following:
\begin{equation}
   \raisebox{-.5 \height}{ \begin{tikzpicture}
        \node (2) at (0,0) {$\begin{tikzpicture}
            \node[gauge,label=left:{$1$}] (a) at (0,0) {};
            \node[gauge,label=left:{$1$}] (b) at (0,1) {};
            \draw (b) to [out=45,in=135,looseness=8] (b);
            \draw[transform canvas={xshift=1.5pt}] (a)--(b);
            \draw[transform canvas={xshift=-1.5pt}] (a)--(b);
            \draw (0-0.2,0.5-0.1)--(0,0.5+0.1)--(0+0.2,0.5-0.1);
            \end{tikzpicture}$};
        \node (11) at (0,3) {$\begin{tikzpicture}
            \node[gauge,label=left:{$1$}] (a) at (0,0) {};
            \node[gauge,label=left:{$1$}] (b1) at (-0.5,1) {};
            \draw (b1) to [out=45,in=135,looseness=8] (b1);
            \node[gauge,label=left:{$1$}] (b2) at (0.5,1) {};
            \draw (b2) to [out=45,in=135,looseness=8] (b2);
            \draw(a)--(b1);
            \draw(a)--(b2);
            \draw[purple] (b1) circle (0.25cm);
            \draw[purple] (b2) circle (0.25cm);
            \end{tikzpicture}$};
        \node (1) at (4,4) {$\begin{tikzpicture}
            \node[gauge,label=left:{$1$}] (a) at (0,0) {};
            \node[gauge,label=left:{$1$}] (b1) at (0,1) {};
            \draw (b1) to [out=45,in=135,looseness=8] (b1);
            \draw(a)--(b1);
            \draw[purple] (b1) circle (0.25cm);
            \node[gauge,label=below:{$1$}] (c1) at (1,0) {};
            \node[gauge,label=right:{$1$}] (c2) at (2,0.5) {};
            \node[gauge,label=right:{$1$}] (c3) at (2,-0.5) {};
            \draw (a)--(c1)--(c2)--(c3)--(c1);
            \draw[purple] \convexpath{c1,c2,c3}{0.25cm};
            \end{tikzpicture}$};
        \node (0) at (8,5) {$\begin{tikzpicture}
            \node[gauge,label=left:{$1$}] (a) at (0,0) {};
            \node[gauge,label=below:{$2$}] (c1) at (1,0) {};
            \node[gauge,label=right:{$2$}] (c2) at (2,0.5) {};
            \node[gauge,label=right:{$2$}] (c3) at (2,-0.5) {};
            \draw (a)--(c1)--(c2)--(c3)--(c1);
            \end{tikzpicture}$};
        \draw (2)--(11)--(1)--(0);
        \draw[transform canvas={yshift=-3pt}] (11)--(1);
        \node at (-0.5,1.5) {$A_1$};
        \node at (1.5,3.8) {$2a_2$};
        \node at (5.8,4.8) {$a_2$};
    \end{tikzpicture}} \;.
\end{equation}
It is straight forward to modify the rule of repeatedly subtracting the same slice given in \cite{Bourget:2022ehw} to take into account unions of cones:

\begin{mdframed}
\paragraph{Repeated Identical Quiver Subtraction:}
\label{rule2}
When subtracting a quiver $\mathsf{D}$ from a set of nodes, such that one can subtract the same quiver again from the same nodes, one must add a decoration between this set of nodes and the rebalancing U$(1)$, and every rebalancing U$(1)$ coming from successive subtractions. The slice associated to the subtraction is\footnote{Here and everywhere in the paper, the notation $\mathcal{C}$ stands for the 3d $\mathcal{N}=4$ Coulomb branch. } $(k+1)\mathcal{C}(\mathsf{D})$ where $k$ is the number of rebalancing U$(1)$ nodes present in the quiver before subtraction decorated in the same color. Only U$(1)$ nodes of shortness $1$ are counted, if two or more such U$(1)$s have already been merged (i.e.\ have shortness $<1$) they are not counted.\\

Schematically we can represent this subtraction as follows:
\begin{equation}
\label{eq:rule2}
   \raisebox{-.5 \height}{  \begin{tikzpicture}
        \node (a) at (0,0) {$\raisebox{-.5 \height}{ \begin{tikzpicture}
            \node (1) at (0,0) {$\mathsf{Q}$};
            \node (2) at (1,0) {$n\mathsf{D}$};
            \node[gauge,label=left:{$k$}] (3) at (0,1) {};
            \draw (3) to [out=45,in=135,looseness=8] (3);
            \draw (3)--(1)--(2);
            \draw[purple] (2) circle (0.3cm);
            \draw[purple] (3) circle (0.25cm);
        \end{tikzpicture}}=\raisebox{-.5 \height}{ \begin{tikzpicture}
            \node (1) at (0,0) {$\mathsf{Q}$};
            \node (2) at (1,0) {$n\mathsf{D}$};
            \node[gauge,label=left:{$1$}] (31) at (-0.5,1) {};
            \draw (31) to [out=45,in=135,looseness=8] (31);
            \node at (0,1) {$\cdots$};
            \node[gauge,label=right:{$1$}] (33) at (0.5,1) {};
            \draw (33) to [out=45,in=135,looseness=8] (33);
            \draw (31)--(1)--(2) (33)--(1);
		    \draw [decorate,decoration={brace,amplitude=5pt}] (-0.7,1.5)--(0.7,1.5);
		    \node at (0,2) {$k$};
            \draw[purple] (2) circle (0.3cm);
            \draw[purple] (31) circle (0.25cm);
            \draw[purple] (33) circle (0.25cm);
        \end{tikzpicture}}$};
        \node (b) at (0,-5) {$\raisebox{-.5 \height}{ \begin{tikzpicture}
            \node (1) at (0,0) {$\mathsf{Q}$};
            \node (2) at (1,0) {$(n-1)\mathsf{D}$};
            \node[gauge,label=left:{$k+1$}] (3) at (0,1) {};
            \draw (3) to [out=45,in=135,looseness=8] (3);
            \draw (3)--(1)--(2);
            \draw[purple] (2) circle (0.7cm);
            \draw[purple] (3) circle (0.25cm);
        \end{tikzpicture}}=\raisebox{-.5 \height}{ \begin{tikzpicture}
            \node (1) at (0,0) {$\mathsf{Q}$};
            \node (2) at (1,0) {$(n-1)\mathsf{D}$};
            \node[gauge,label=left:{$1$}] (31) at (-0.5,1) {};
            \draw (31) to [out=45,in=135,looseness=8] (31);
            \node at (0,1) {$\cdots$};
            \node[gauge,label=right:{$1$}] (33) at (0.5,1) {};
            \draw (33) to [out=45,in=135,looseness=8] (33);
            \draw (31)--(1)--(2) (33)--(1);
		    \draw [decorate,decoration={brace,amplitude=5pt}] (-0.7,1.5)--(0.7,1.5);
		    \node at (0,2) {$k+1$};
            \draw[purple] (2) circle (0.7cm);
            \draw[purple] (31) circle (0.25cm);
            \draw[purple] (33) circle (0.25cm);
        \end{tikzpicture}}$};
        \draw (a)--(b);
        \node at (-1,-2.5) {$(k+1)\mathcal{C}(\mathsf{D})$};
    \end{tikzpicture}}
\end{equation}
\end{mdframed}

\FloatBarrier

\section{Symmetric Products}
\label{sec:symmetricProducts}

\subsection{\texorpdfstring{Symmetric product of $\mathbb{C}^2$}{Symmetric product of C2}}

Before we turn to symmetric products of a singular space we start by analyzing $\mathrm{Sym}^k (\mathbb{C}^2)$ taking into account elementary slices which are unions of cones using what we discussed in the last section. The results for $k \leq 6$ are shown in Figure \ref{fig:Higgsing0}.

\afterpage{
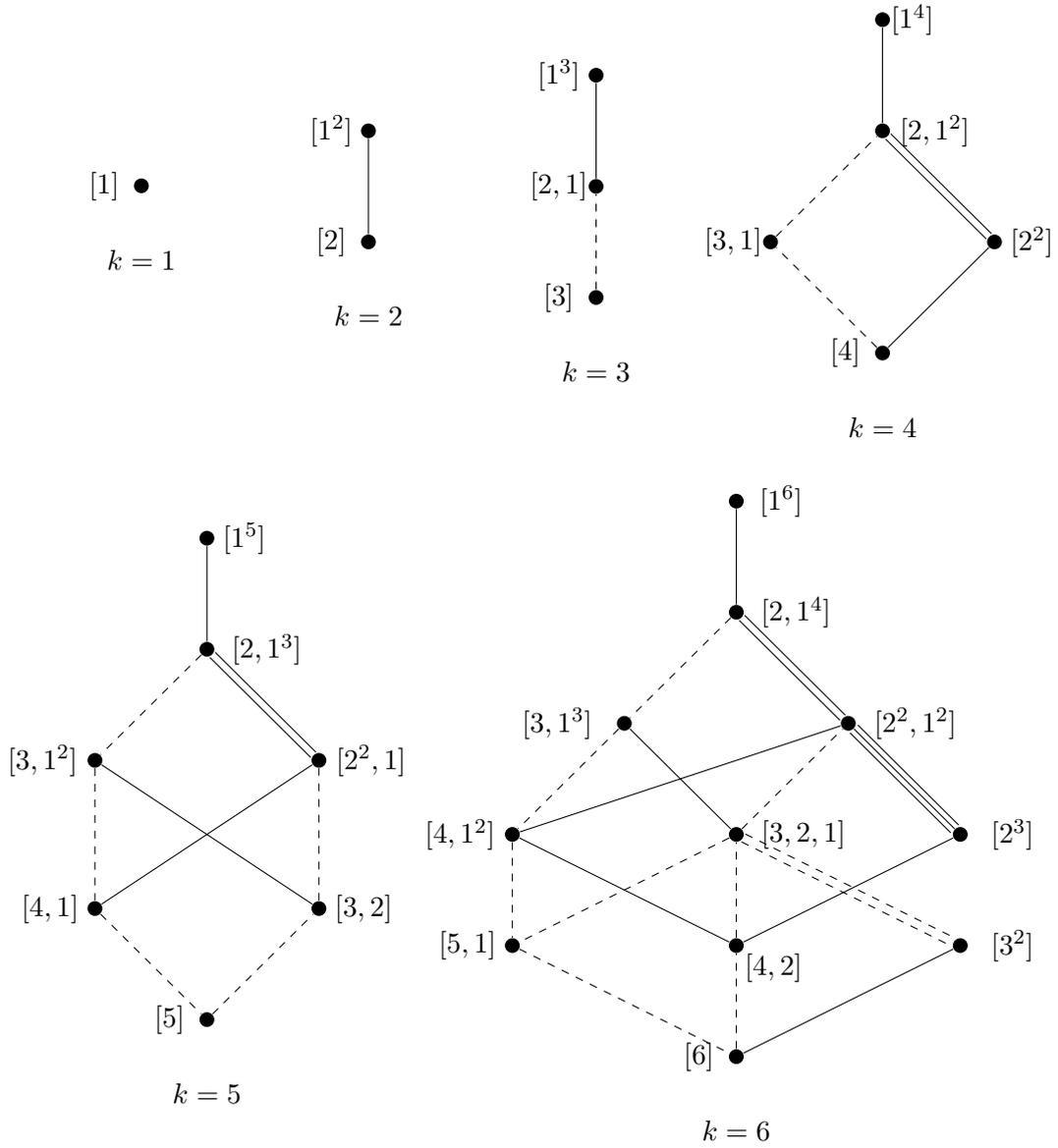
\begin{figure}
    \centering
  \begin{tikzpicture}
\node at (0,8) {  \begin{tikzpicture}
                \node[hasse] (1) at (0,0) {};
                \node at (-.5,0) {$[1]$};
                \node at (0,-1) {$k=1$};
        \end{tikzpicture}};
\node at (3,8) {  \begin{tikzpicture}
                \node[hasse] (1) at (0,0) {};
                \node[hasse] (2) at (0,1.5) {};
                \draw (1)--(2);
                \node at (-.5,1.5) {$[1^2]$};
                \node at (-.5,0) {$[2]$};
                \node at (0,-1) {$k=2$};
        \end{tikzpicture}};
        \node at (6,8) {\begin{tikzpicture}
                \node[hasse] (1) at (0,0) {};
                \node[hasse] (2) at (0,1.5) {};
                \node[hasse] (3) at (0,3) {};
                \draw (3)--(2);
                \draw[dashed] (1)--(2);
                \node at (-.5,3) {$[1^3]$};
                \node at (-.5,1.5) {$[2,1]$};
                \node at (-.5,0) {$[3]$};
                \node at (0,-1) {$k=3$};
        \end{tikzpicture}};
        \node at (10,8) {\begin{tikzpicture}
                \node[hasse] (1) at (0,0) {};
                \node[hasse] (2) at (1.5,1.5) {};
                \node[hasse] (3) at (-1.5,1.5) {};
                \node[hasse] (4) at (0,3) {};
                \node[hasse] (5) at (0,4.5) {};
                \draw (5)--(4) (2)--(1);
                \draw[transform canvas={xshift=-1pt,yshift=-1pt}] (2)--(4);
                \draw[transform canvas={xshift=1pt,yshift=1pt}] (2)--(4);
                \draw[dashed] (1)--(3)--(4);
                \node at (.4,4.5) {$[1^4]$};
                \node at (.7,3) {$[2,1^2]$};
                \node at (-2,1.5) {$[3,1]$};
                \node at (2,1.5) {$[2^2]$};
                \node at (-.5,0) {$[4]$};
                \node at (0,-1) {$k=4$};
        \end{tikzpicture}};
        \node at (1,0) {\begin{tikzpicture}
                \node[hasse] (1) at (0,0) {};
                \node[hasse] (2) at (-1.5,1.5) {};
                \node[hasse] (3) at (1.5,1.5) {};
                \node[hasse] (4) at (-1.5,3.5) {};
                \node[hasse] (5) at (1.5,3.5) {};
                \node[hasse] (6) at (0,5) {};
                \node[hasse] (7) at (0,6.5) {};
                \draw (7)--(6) (5)--(2) (3)--(4);
                \draw[transform canvas={xshift=-1pt,yshift=-1pt}] (5)--(6);
                \draw[transform canvas={xshift=1pt,yshift=1pt}] (5)--(6);
                \draw[dashed] (6)--(4)--(2)--(1)--(3)--(5);
                \node at (.5,6.5) {$[1^5]$};
                \node at (.8,5) {$[2,1^3]$};
                \node at (-2.2,3.5) {$[3,1^2]$};
                \node at (-2.1,1.5) {$[4,1]$};
                \node at (2.1,1.5) {$[3,2]$};
                \node at (2.2,3.5) {$[2^2,1]$};
                \node at (-.5,0) {$[5]$};
                \node at (0,-1) {$k=5$};
        \end{tikzpicture}};
        \node at (8,-0) { \begin{tikzpicture}
                \node[hasse] (1) at (0,0) {};
                \node[hasse] (2) at (-3,1.5) {};
                \node[hasse] (3) at (0,1.5) {};
                \node[hasse] (4) at (-3,3) {};
                \node[hasse] (5) at (0,3) {};
                \node[hasse] (6) at (3,3) {};
                \node[hasse] (7) at (-1.5,4.5) {};
                \node[hasse] (8) at (1.5,4.5) {};
                \node[hasse] (9) at (0,6) {};
                \node[hasse] (10) at (0,7.5) {};
                \node[hasse] (11) at (3,1.5) {};
                \draw[dashed] (9)--(7)--(4)--(2)--(1)--(3)--(5)--(2) (5)--(8);
                \draw (10)--(9) (8)--(6)--(3)--(4)--(8) (7)--(5) (1)--(11);
                \draw[transform canvas={xshift=-1pt,yshift=-1pt}] (8)--(9);
                \draw[transform canvas={xshift=1pt,yshift=1pt}] (8)--(9);
                \draw[transform canvas={xshift=-1.5pt,yshift=-1.5pt}] (6)--(8);
                \draw[transform canvas={xshift=1.5pt,yshift=1.5pt}] (6)--(8);
                \draw[dashed,transform canvas={xshift=0.75pt,yshift=1.5pt}] (5)--(11);
                \draw[dashed,transform canvas={xshift=-0.75pt,yshift=-1.5pt}] (5)--(11);
                \node at (.6,7.5) {$[1^6]$};
                \node at (.8,6) {$[2,1^4]$};
                \node at (-2.4,4.5) {$[3,1^3]$};
                \node at (2.4,4.5) {$[2^2,1^2]$};
                \node at (-3.7,3) {$[4,1^2]$};
                \node at (0.9,3) {$[3,2,1]$};
                \node at (3.7,3) {$[2^3]$};
                \node at (-3.6,1.5) {$[5,1]$};
                \node at (.5,1.2) {$[4,2]$};
                \node at (3.7,1.5) {$[3^2]$};
                \node at (-.5,0) {$[6]$};
                \node at (0,-1) {$k=6$};
        \end{tikzpicture}  };
        \end{tikzpicture}
    \caption{Diagrams for the $k$-th symmetric products of $\mathbb{C}^2$ for $k=1,2,3,4,5,6$. The full lines denote $A_1$-type transitions, the dashed lines denote $m$-type transition. $n$ parallel lines denote a union of $n$ cones.} 
    \label{fig:Higgsing0}
\end{figure}

}

\paragraph{Transverse slices in $\mathrm{Sym}^k(\mathbb{C}^2)$.} Transverse slices from any leaf in $\mathrm{Sym}^k(\mathbb{C}^2)$ to the top leaf are products $\prod_i \mathrm{Sym}^{k_i}_0(\mathbb{C}^2)$\footnote{We thank Travis Schedler for pointing this out to us.}, $\sum_i k_i\leq k$, where $\mathrm{Sym}^{k_i}_0(\mathbb{C}^2)=\mathrm{Sym}^{k_i}(\mathbb{C}^2)/\mathbb{C}^2$ is the symmetric product with the free part (center of mass) removed. However the Hasse diagrams of these product spaces are not a subdiagram of the bigger $\mathrm{Sym}^k(\mathbb{C}^2)$ spaces in which they are a transverse slices. For example we have
\begin{equation}
    \mathrm{Sym}^{2} (\mathbb{C}^2) \times \mathrm{Sym}^{2} (\mathbb{C}^2) = \raisebox{-.5 \height}{ \begin{tikzpicture}
                \node[hasse] (1) at (0,0) {};
                \node[hasse] (2) at (0,1.5) {};
                \draw (1)--(2);
                \node at (-.5,1.5) {$[1^2]$};
                \node at (-.5,0) {$[2]$};
                \node at (-.5,.75) {$A_1$};
        \end{tikzpicture}} \times \raisebox{-.5 \height}{ \begin{tikzpicture}
                \node[hasse] (1) at (0,0) {};
                \node[hasse] (2) at (0,1.5) {};
                \draw (1)--(2);
                \node at (-.5,1.5) {$[1^2]$};
                \node at (-.5,0) {$[2]$};
                \node at (-.5,.75) {$A_1$};
        \end{tikzpicture}}
        = \raisebox{-.5 \height}{\begin{tikzpicture}
                \node[hasse] (1) at (0,0) {};
                \node[hasse] (2) at (-1.5,1.5) {};
                \node[hasse] (3) at (1.5,1.5) {};
                \node[hasse] (4) at (0,3) {};
                \draw (1)--(2)--(4)--(3)--(1);
                \node at (-.9,3) {$[1^2][1^2]$};
                \node at (-2.2,1.5) {$[2][1^2]$};
                \node at (2.2,1.5) {$[1^2][2]$};
                \node at (-.7,0) {$[2][2]$};
                \node at (-1.2,.75) {$A_1$};
                \node at (1.2,.75) {$A_1$};
                \node at (-1.2,2.25) {$A_1$};
                \node at (1.2,2.25) {$A_1$};
        \end{tikzpicture}}
        \label{eq:sym2sym2Hasse}
\end{equation}

\afterpage{

\begin{figure} 
\begin{center}
\scalebox{.8}{
\begin{tabular}{|c|c|c|}
\hline 
Partition & Subdiagram of bigger Hasse diagram & Hasse diagram of transverse slice \\ \hline 
(2,2) & 
\raisebox{-.5 \height}{\begin{tikzpicture}
                \node[hasse,label=left:{$[2^2]$}] (1) at (0,0) {};
                \node[hasse,label=left:{$[2,1^2]$}] (2) at (0,1.5) {};
                \node[hasse,label=left:{$[1^4]$}] (4) at (0,3) {};
                \draw (2)--(4);
                \draw[transform canvas={xshift=-1.5pt,yshift=0pt}] (1)--(2);
                \draw[transform canvas={xshift=1.5pt,yshift=0pt}] (1)--(2);
        \end{tikzpicture}} &\raisebox{-.5 \height}{\begin{tikzpicture}
                \node[hasse,label=left:{$[2][2]$}] (1) at (0,0) {};
                \node[hasse,label=left:{$[2][1^2]$}] (2) at (-1.5,1.5) {};
                \node[hasse,label=right:{$[1^2][2]$}] (3) at (1.5,1.5) {};
                \node[hasse,label=left:{$[1^2][1^2]$}] (4) at (0,3) {};
                \draw (1)--(2)--(4)--(3)--(1);
                \node at (0,3.5) {};
                \node at (0,-.5) {};
        \end{tikzpicture}} \\  \hline 
(2,2,2) & 
        \raisebox{-.5 \height}{\begin{tikzpicture}
                \node[hasse,label=left:{$[2^3]$}] (0) at (0,0) {};
                \node[hasse,label=left:{$[2^2,1^2]$}] (1) at (0,1.5) {};
                \node[hasse,label=left:{$[2,1^4]$}] (2) at (0,3) {};
                \node[hasse,label=left:{$[1^6]$}] (3) at (0,4.5) {};
                \draw (0)--(1) (2)--(3);
                \draw[transform canvas={xshift=-1.5pt,yshift=0pt}] (1)--(2);
                \draw[transform canvas={xshift=1.5pt,yshift=0pt}] (1)--(2);
                \draw[transform canvas={xshift=-2.5pt,yshift=0pt}] (0)--(1);
                \draw[transform canvas={xshift=2.5pt,yshift=0pt}] (0)--(1);
        \end{tikzpicture}} & \raisebox{-.5 \height}{\begin{tikzpicture}
                \node[hasse,label=left:{$[2][2][2]$}] (000) at (0,0) {};
                \node[hasse,label=left:{$[2][2][1^2]$}] (100) at (-2.5,1.5) {};
                \node[hasse,label=left:{$[2][1^2][2]$}] (010) at (0,1.5) {};
                \node[hasse,label=right:{$[1^2][2][2]$}] (001) at (2.5,1.5) {};
                \node[hasse,label=left:{$[2][1^2][1^2]$}] (110) at (-2.5,3) {};
                \node[hasse,label=left:{$[1^2][2][1^2]$}] (101) at (0,3) {};
                \node[hasse,label=right:{$[1^2][1^2][2]$}] (011) at (2.5,3) {};
                \node[hasse,label=left:{$[1^2][1^2][1^2]$}] (111) at (0,4.5) {};
                \draw (000)--(100)--(110)--(111)--(011)--(001)--(000)--(010)--(110) (100)--(101)--(001) (010)--(011) (101)--(111);
                \node at (0,5) {};
                \node at (0,-.5) {};
        \end{tikzpicture}} \\  \hline 
        (2,3) & 
        \raisebox{-.5 \height}{\begin{tikzpicture}
                \node[hasse,label=left:{$[3,2]$}] (00) at (0,0) {};
                \node[hasse,label=left:{$[3,1^2]$}] (10) at (-1.5,1.5) {};
                \node[hasse,label=right:{$[2^2,1]$}] (01) at (1.5,1.5) {};
                \node[hasse,label=left:{$[2,1^3]$}] (11) at (0,3) {};
                \node[hasse,label=left:{$[1^5]$}] (12) at (0,4.5) {};
                \draw (00)--(10) (11)--(12);
                \draw[transform canvas={xshift=1.5pt,yshift=1.5pt}] (01)--(11);
                \draw[transform canvas={xshift=-1.5pt,yshift=-1.5pt}] (01)--(11);
                \draw[dashed] (10)--(11) (00)--(01);
        \end{tikzpicture}} & \raisebox{-.5 \height}{\begin{tikzpicture}
                \node[hasse,label=left:{$[3][2]$}] (00) at (0,0) {};
                \node[hasse,label=left:{$[3][1^2]$}] (10) at (-1.5,1.5) {};
                \node[hasse,label=right:{$[2,1][2]$}] (01) at (1.5,1.5) {};
                \node[hasse,label=left:{$[2,1][1^2]$}] (11) at (0,3) {};
                \node[hasse,label=right:{$[1^3][2]$}] (02) at (3,3) {};
                \node[hasse,label=left:{$[1^3][1^2]$}] (12) at (1.5,4.5) {};
                \draw (00)--(10) (11)--(12)--(02)--(01) (00) (11)--(01);
                \draw[dashed] (10)--(11) (00)--(01);
                \node at (0,5) {};
                \node at (0,-.5) {};
        \end{tikzpicture}} \\  \hline 
        (3,3) & 
        \raisebox{-.5 \height}{\begin{tikzpicture}
                \node[hasse,label=left:{$[3^2]$}] (00) at (0,0) {};
                \node[hasse,label=left:{$[3,2,1]$}] (10) at (0,1.5) {};
                \node[hasse,label=left:{$[3,1^3]$}] (20) at (-1.5,3) {};
                \node[hasse,label=right:{$[2^2,1^2]$}] (11) at (1.5,3) {};
                \node[hasse,label=left:{$[2,1^4]$}] (21) at (0,4.5) {};
                \node[hasse,label=left:{$[1^6]$}] (22) at (0,6) {};
                \draw (10)--(20) (21)--(22);
                \draw[transform canvas={xshift=1.5pt,yshift=1.5pt}] (11)--(21);
                \draw[transform canvas={xshift=-1.5pt,yshift=-1.5pt}] (11)--(21);
                \draw[dashed] (10)--(11) (20)--(21);
                \draw[dashed,transform canvas={xshift=-1.5pt,yshift=0pt}] (00)--(10);
                \draw[dashed,transform canvas={xshift=1.5pt,yshift=0pt}] (00)--(10);
        \end{tikzpicture}} & \raisebox{-.5 \height}{\begin{tikzpicture}
                \node[hasse,label=left:{$[3][3]$}] (00) at (0,0) {};
                \node[hasse,label=left:{$[3][2,1]$}] (10) at (-1.5,1.5) {};
                \node[hasse,label=left:{$[3][1^3]$}] (20) at (-3,3) {};
                \node[hasse,label=left:{$[2,1][1^3]$}] (21) at (-1.5,4.5) {};
                \node[hasse,label=right:{$[1^3][1^3]$}] (22) at (0,6) {};
                \node[hasse,label=right:{$[2,1][3]$}] (01) at (1.5,1.5) {};
                \node[hasse,label=left:{$[2,1][2,1]$}] (11) at (0,3) {};
                \node[hasse,label=right:{$[1^3][3]$}] (02) at (3,3) {};
                \node[hasse,label=right:{$[1^3][2,1]$}] (12) at (1.5,4.5) {};
                \draw (22)--(21)--(11)--(12)--(22) (10)--(20) (01)--(02);
                \draw[dashed] (01)--(11)--(10)--(00)--(01) (20)--(21) (02)--(12);
                \node at (0,6.5) {};
                \node at (0,-.5) {};
        \end{tikzpicture}} \\  \hline 
\end{tabular}}
\end{center}
\caption{Transverse slices in $\mathrm{Sym}^k(\mathbb{C}^2)$. On the left we display the corresponding subdiagram in the Hasse diagram of $\mathrm{Sym}^k(\mathbb{C}^2)$. On the right we display the Hasse diagram of the transverse slice. This illustrates that the Hasse diagram of a transverse slice in a space $X$ need not be a subdiagram of the Hasse diagram of $X$. The Hasse diagram conventions are the same as in Figure \ref{fig:Higgsing0}.}
\label{fig:productsPartitions}
\end{figure}
}

This product space can be identified with the transverse slice from $[2^2]$ to $[1^4]$ in $\mathrm{Sym}^4(\mathbb{C}^2)$. The relevant subdiagram of the Hasse diagram is
\begin{equation}
    \raisebox{-.5 \height}{\begin{tikzpicture}
                \node[hasse] (1) at (0,0) {};
                \node[hasse] (2) at (0,1.5) {};
                \node[hasse] (4) at (0,3) {};
                \draw (2)--(4);
                \draw[transform canvas={xshift=-1.5pt}] (1)--(2);
                \draw[transform canvas={xshift=+1.5pt}] (1)--(2);
                \node at (-.6,3) {$[1^4]$};
                \node at (-1.,1.5) {$[2,1^2]$};
                \node at (-.7,0) {$[2,2]$};
                \node at (.7,.75) {$2A_1$};
                \node at (.7,2.25) {$A_1$};
        \end{tikzpicture}}\;.
        \label{eq:sym2sym2sub}
\end{equation}
This is clearly not the same as \eqref{eq:sym2sym2Hasse}, however the $2A_1$ at the bottom of \eqref{eq:sym2sym2sub} tell us that there are really two minimal leaves whose closure is $A_1$, and we can infer the correct Hasse diagram of symplectic leaves. This shows that it is important to be sensitive to unions of cones as elementary slices.
Other examples of similar computations are shown in Figure \ref{fig:productsPartitions}, which shows the diagrams for the $k$-th symmetric products of $\mathbb{C}^2$ for $k=1,2,3,4,5,6$. Going further in $k$ presents no difficulty. Note that the cases $k=3,4,5$ can be checked against computations in the math literature. Indeed, \cite[Theorem 1.3]{2015arXiv150205770F} shows that these three diagrams should appear in the diagrams for nilpotent orbits in $G_2$, $F_4$ and $E_8$ respectively, namely as the transverse slices between $G_2 (a_1)$ and $A_1$, between $F_4 (a_3)$ and $A_2 + \tilde{A}_1$, and between $E_8 (a_7)$ and $A_4 + A_3$. This can be checked directly on the diagrams at the end of \cite{2015arXiv150205770F}.

\FloatBarrier

\subsection{Symmmetric product of Klein singularities}

We now consider the symmetric product of an ADE singularity, $\mathrm{Sym}^k(\mathbb{C}^2/\Gamma_G)$, for $G=ADE$.

\paragraph{The leaves of $\mathrm{Sym}^k(X)$.}
The moduli space of $k$ points on $X$, i.e.\ $\mathrm{Sym}^k(X)$, has a much more intricate Hasse diagram. Let us first review how to count the number of leaves, following \cite[9.4(vii)]{Webster1407}. We can have $0\leq k'\leq k$ points on $\mathcal{L}$, while $k-k'$ points remain at the singularity $\{0\}$. For a given $k'$ there is the possibility for several of the $k'$ points to coincide. Each such possibility is captured by a partition of $k'$, where the parts count the multiplicities of points. Therefore we have
\begin{equation}
\label{eq:numberLeaves11}
    \#\textnormal{ leaves of }\mathrm{Sym}^k(X)=\sum_{k'=0}^k p(k'), 
\end{equation}
where $p(k')$ denotes the number of partitions of $k'$. 
The leaf associated to a partition $\lambda_{k'}=[n_1,\dots,n_l]$ has quaternionic dimension $l$, the number of parts.

\paragraph{The Hasse diagram of $\mathrm{Sym}^k(X)$.}

Before constructing Hasse diagrams of $\mathrm{Sym}^k(X)$, let us note that the subdiagram for a fixed $k'\leq k$ is the Hasse diagram of $\mathrm{Sym}^{k'}(\mathbb{C}^2)$ studied in the previous subsection. 
The Hasse diagram of $\mathrm{Sym}^k(X)$ is obtained in the following way:
\begin{enumerate}
    \item Take the disjoint union of the Hasse diagrams of $\mathrm{Sym}^{k'}(\mathbb{C}^2)$ for $0\leq k'\leq k$.
 \item For each partition $\lambda = [n_1,\dots,n_l]$ of $k'$, with $0 \leq k' <k$, add an elementary transition from $\lambda$ to the partition $[n_1,\dots,n_l ,k-k']$ of $k$. This transition has multiplicity 1.
\end{enumerate}
The transitions added in the second step are justified in the following way. From the $k-k'$ points at the origin, we may move $n\leq k-k'$ coincident points onto $\mathcal{L}$. This is expected to be an elementary transition.

The diagrams for $1\leq k\leq 4$ are shown in Figure \ref{fig:growth_sym}. We now show how these can be derived from quiver subtraction. It is necessary to introduce some notations first.

\paragraph{Notation.} In the following we will encounter quivers with non-simply laced edges with multiplicity. We will denote such edges by two numbers -- $m,l$ -- where $l$ denotes the non-simply lacedness and $m$ denotes the multiplicity of the link.
A link $1,l$ denotes a standard non-simply laced edge of order $l$, and a link $m,1$ denotes $m$ simply laced links.
We have
\begin{equation}
    \mathcal{C}\left(
    \vcenter{\hbox{\scalebox{1}{
	\begin{tikzpicture}
		\node[gauge,label=below:{$1$}] (0) at (0,0) {};
		\node[gauge,label=below:{$1$}] (1) at (1,0) {};
		\draw[double,thick] (0)--(1);
		\draw (0.4,0.2)--(0.6,0)--(0.4,-0.2);
		\node at (0.5,0.3) {\scriptsize$m,l$};
	\end{tikzpicture}}}}
	\right)=\mathbb{C}^2/\mathbb{Z}_m
\end{equation}
independent of $l$.

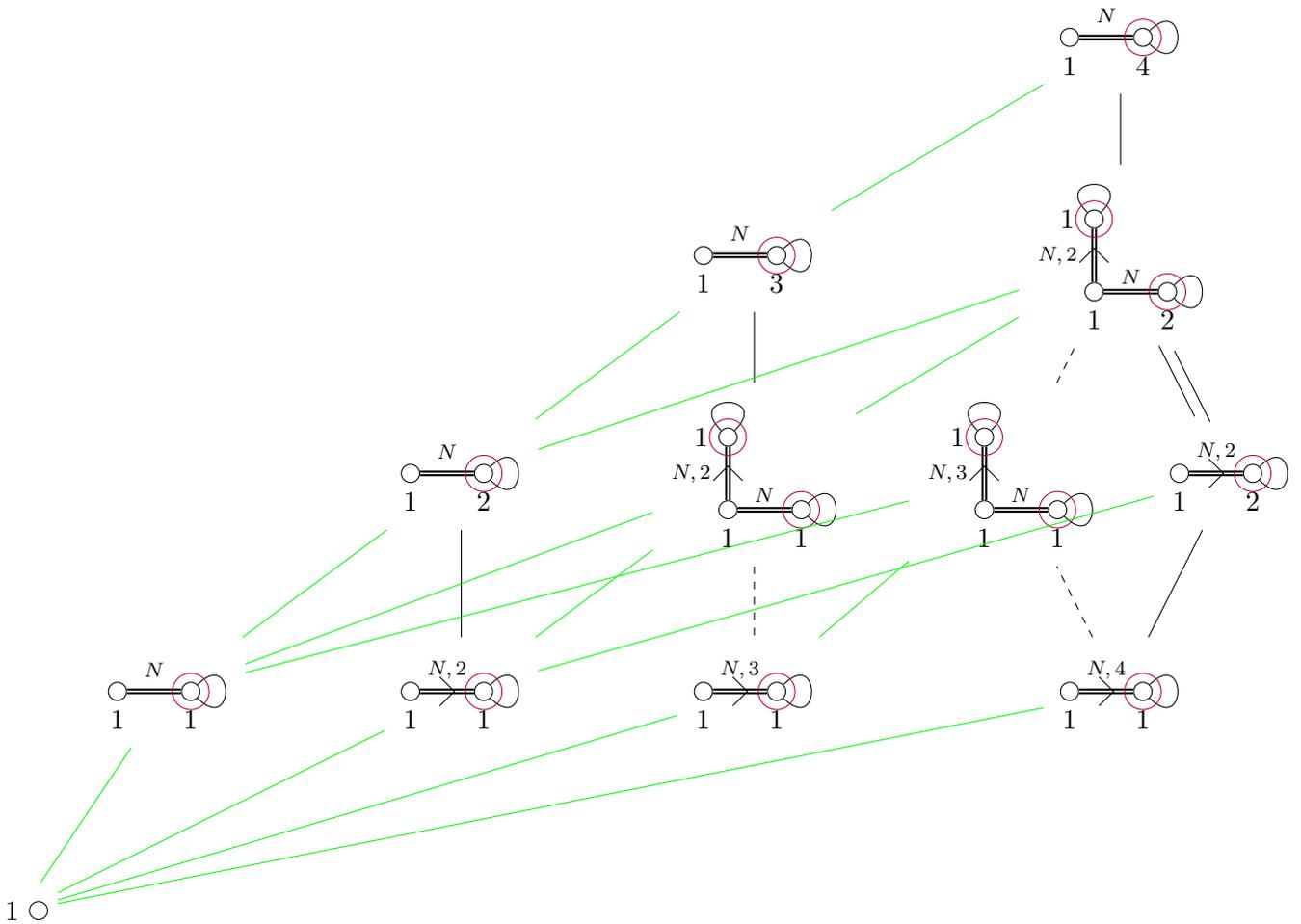
\begin{figure}
\newcommand{\quivoneoneoneone}{\begin{tikzpicture}
		\node[gauge,label=below:{$1$}] (0) at (0,0) {};
		\node[gauge,label=below:{$4$}] (1) at (1,0) {};
		\draw[double,thick] (0)--(1);
		\node at (0.5,0.3) {\scriptsize $N$};
        \draw (1) to [out=315,in=45,looseness=10] (1);
        \draw[purple] (1) circle (0.25cm);
	\end{tikzpicture}}\newcommand{\quivoneoneone}{\begin{tikzpicture}
		\node[gauge,label=below:{$1$}] (0) at (0,0) {};
		\node[gauge,label=below:{$3$}] (1) at (1,0) {};
		\draw[double,thick] (0)--(1);
		\node at (0.5,0.3) {\scriptsize $N$};
        \draw (1) to [out=315,in=45,looseness=10] (1);
        \draw[purple] (1) circle (0.25cm);
	\end{tikzpicture}}
	\newcommand{\quivoneone}{\begin{tikzpicture}
		\node[gauge,label=below:{$1$}] (0) at (0,0) {};
		\node[gauge,label=below:{$2$}] (1) at (1,0) {};
		\draw[double,thick] (0)--(1);
		\node at (0.5,0.3) {\scriptsize $N$};
        \draw (1) to [out=315,in=45,looseness=10] (1);
        \draw[purple] (1) circle (0.25cm);
	\end{tikzpicture}}
		\newcommand{\quivone}{\begin{tikzpicture}
		\node[gauge,label=below:{$1$}] (0) at (0,0) {};
		\node[gauge,label=below:{$1$}] (1) at (1,0) {};
		\draw[double,thick] (0)--(1);
		\node at (0.5,0.3) {\scriptsize $N$};
        \draw (1) to [out=315,in=45,looseness=10] (1);
        \draw[purple] (1) circle (0.25cm);
	\end{tikzpicture}}
\newcommand{\quivtwooneone}{\begin{tikzpicture}
		\node[gauge,label=below:{$1$}] (0) at (0,0) {};
		\node[gauge,label=below:{$2$}] (1) at (1,0) {};
		\draw[double,thick] (0)--(1);
		\node at (0.5,0.2) {\scriptsize $N$};
        \draw (1) to [out=315,in=45,looseness=10] (1);
        \node[gauge,label=left:{$1$}] (l) at (0,1) {};
        \draw (l) to [out=45,in=45+90,looseness=10] (l);
        \draw[double,thick] (l)--(0);
        \draw (0-0.2,0.5-0.1)--(0,0.5+0.1)--(0+0.2,0.5-0.1);
        \draw[purple] (l) circle (0.25cm);
        \draw[purple] (1) circle (0.25cm);
        \node at (-0.5,0.5) {\scriptsize $N,2$};
	\end{tikzpicture}}
\newcommand{\quivtwoone}{\begin{tikzpicture}
		\node[gauge,label=below:{$1$}] (0) at (0,0) {};
		\node[gauge,label=below:{$1$}] (1) at (1,0) {};
		\draw[double,thick] (0)--(1);
		\node at (0.5,0.2) {\scriptsize $N$};
        \draw (1) to [out=315,in=45,looseness=10] (1);
        \node[gauge,label=left:{$1$}] (l) at (0,1) {};
        \draw (l) to [out=45,in=45+90,looseness=10] (l);
        \draw[double,thick] (l)--(0);
        \draw (0-0.2,0.5-0.1)--(0,0.5+0.1)--(0+0.2,0.5-0.1);
        \draw[purple] (l) circle (0.25cm);
        \draw[purple] (1) circle (0.25cm);
        \node at (-0.5,0.5) {\scriptsize $N,2$};
	\end{tikzpicture}}
\newcommand{\quivtwotwo}{\begin{tikzpicture}
		\node[gauge,label=below:{$1$}] (0) at (0,0) {};
		\node[gauge,label=below:{$2$}] (1) at (1,0) {};
		\draw[double,thick] (0)--(1);
		\node at (0.5,0.3) {\scriptsize $N,2$};
        \draw (1) to [out=315,in=45,looseness=10] (1);
        \draw[purple] (1) circle (0.25cm);
        \draw (.4,-.2)--(.6,0)--(.4,.2);
	\end{tikzpicture}  }
\newcommand{\quivthreeone}{\begin{tikzpicture}
		\node[gauge,label=below:{$1$}] (0) at (0,0) {};
		\node[gauge,label=below:{$1$}] (1) at (1,0) {};
		\draw[double,thick] (0)--(1);
		\node at (0.5,0.2) {\scriptsize $N$};
        \draw (1) to [out=315,in=45,looseness=10] (1);
        \node[gauge,label=left:{$1$}] (l) at (0,1) {};
        \draw (l) to [out=45,in=45+90,looseness=10] (l);
        \draw[double,thick] (l)--(0);
        \draw (0-0.2,0.5-0.1)--(0,0.5+0.1)--(0+0.2,0.5-0.1);
        \draw[purple] (l) circle (0.25cm);
        \draw[purple] (1) circle (0.25cm);
        \node at (-0.5,0.5) {\scriptsize $N,3$};
	\end{tikzpicture}}
\newcommand{\quivfour}{\begin{tikzpicture}
		\node[gauge,label=below:{$1$}] (0) at (0,0) {};
		\node[gauge,label=below:{$1$}] (1) at (1,0) {};
		\draw[double,thick] (0)--(1);
		\node at (0.5,0.3) {\scriptsize $N,4$};
        \draw (1) to [out=315,in=45,looseness=10] (1);
        \draw[purple] (1) circle (0.25cm);
        \draw (.4,-.2)--(.6,0)--(.4,.2);
	\end{tikzpicture}  }
\newcommand{\quivthree}{\begin{tikzpicture}
		\node[gauge,label=below:{$1$}] (0) at (0,0) {};
		\node[gauge,label=below:{$1$}] (1) at (1,0) {};
		\draw[double,thick] (0)--(1);
		\node at (0.5,0.3) {\scriptsize $N,3$};
        \draw (1) to [out=315,in=45,looseness=10] (1);
        \draw[purple] (1) circle (0.25cm);
        \draw (.4,-.2)--(.6,0)--(.4,.2);
	\end{tikzpicture}  }
\newcommand{\quivtwo}{\begin{tikzpicture}
		\node[gauge,label=below:{$1$}] (0) at (0,0) {};
		\node[gauge,label=below:{$1$}] (1) at (1,0) {};
		\draw[double,thick] (0)--(1);
		\node at (0.5,0.3) {\scriptsize $N,2$};
        \draw (1) to [out=315,in=45,looseness=10] (1);
        \draw[purple] (1) circle (0.25cm);
        \draw (.4,-.2)--(.6,0)--(.4,.2);
	\end{tikzpicture}  }
\makebox[\textwidth][c]{	
\begin{tikzpicture}
\node(1111) at (17,12) {\quivoneoneoneone};
\node(211) at (17,9) {\quivtwooneone};
\node(22) at (18.5,6) {\quivtwotwo};
\node(31) at (15.5,6) {\quivthreeone};
\node(4) at (17,3) {\quivfour};
\node(111) at (12,9) {\quivoneoneone};
\node(21) at (12,6) {\quivtwoone};
\node(3) at (12,3) {\quivthree};
\node(11) at (8,6) {\quivoneone};
\node(2) at (8,3) {\quivtwo};
\node(1) at (4,3) {\quivone};
\node(0) at (2,0) {$\begin{tikzpicture}
\node[gauge,label=left:{$1$}] at (0,0) {};
\end{tikzpicture}$};
\draw (1111)--(211) (22)--(4);
\draw[transform canvas={xshift=3pt,yshift=-1pt}] (22)--(211);
\draw[transform canvas={xshift=-3pt,yshift=1pt}] (22)--(211);
\draw[dashed] (4)--(31)--(211);
\draw (111)--(21);
\draw[dashed] (21)--(3);
\draw (11)--(2);
\draw[green] (1111)--(111)--(11)--(1)--(0);
\draw[green] (211)--(11) (211)--(21);
\draw[green] (22)--(2);
\draw[green] (31)--(3);
\draw[green] (31)--(1);
\draw[green] (21)--(2);
\draw[green] (21)--(1);
\draw[green] (4)--(0) (3)--(0) (2)--(0);
\end{tikzpicture}}
    \caption{Hasse diagram for the space $\mathrm{Sym}^4 (\mathbb{C}^2 / \mathbb{Z}_N)$ computed from the magnetic quiver using the rules of Section \ref{sec:unionCones}. The green transitions denote $A_{N-1}$ while the other transitions are those shown in Figure \ref{fig:Higgsing0}.}
    \label{fig:hasseSym4}
\end{figure}

Let's get to quiver subtraction. We have
\begin{equation}
	\mathrm{Sym}^k(\mathbb{C}^2/\mathbb{Z}_N)=
	\mathcal{C}\left(\vcenter{\hbox{\scalebox{1}{
	\begin{tikzpicture}
		\node[gauge,label=below:{$1$}] (0) at (0,0) {};
		\node[gauge,label=below:{$k$}] (1) at (1,0) {};
		\draw[double,thick] (0)--(1);
		\node at (0.5,0.2) {\scriptsize$N$};
        \draw (1) to [out=315,in=45,looseness=10] (1);
	\end{tikzpicture}}}}
	\right).
\end{equation}
There are two possible subtractions: either $A_1$, using the rule \eqref{eq:rule1}, or the standard subtraction of $A_{N-1}$: 
\begin{equation}
   \raisebox{-.5\height}{ \begin{tikzpicture}
        \node (a) at (0,0) {
    $\begin{tikzpicture}
		\node[gauge,label=below:{$1$}] (0) at (0,0) {};
		\node[gauge,label=below:{$k$}] (1) at (1,0) {};
		\draw[double,thick] (0)--(1);
		\node at (0.5,0.2) {\scriptsize$N$};
        \draw (1) to [out=315,in=45,looseness=10] (1);
	\end{tikzpicture}$};
        \node (b) at (3,-3) {
    $\begin{tikzpicture}
		\node[gauge,label=below:{$1$}] (0) at (0,0) {};
		\node[gauge,label=below:{$k-2$}] (1) at (1,0) {};
		\draw[double,thick] (0)--(1);
		\node at (0.5,0.2) {\scriptsize$N$};
        \draw (1) to [out=315,in=45,looseness=10] (1);
        \node[gauge,label=left:{$1$}] (l) at (-0.5,1) {};
        \draw (l) to [out=45,in=45+90,looseness=10] (l);
        \node[gauge,label=right:{$1$}] (r) at (0.5,1) {};
        \draw (r) to [out=45,in=45+90,looseness=10] (r);
        \draw[double,thick] (l)--(0)--(r);
        \draw[purple] (l) circle (0.25cm);
        \draw[purple] (r) circle (0.25cm);
        \draw[purple] (1) circle (0.25cm);
        \draw[dashed] \convexpath{l,r}{0.6cm};
	\end{tikzpicture}$};
        \node (c) at (0,-6) {
    $\begin{tikzpicture}
		\node[gauge,label=below:{$1$}] (0) at (0,0) {};
		\node[gauge,label=below:{$k-2$}] (1) at (1,0) {};
		\draw[double,thick] (0)--(1);
		\node at (0.5,0.2) {\scriptsize$N$};
        \draw (1) to [out=315,in=45,looseness=10] (1);
        \node[gauge,label=left:{$1$}] (l) at (0,1) {};
        \draw (l) to [out=45,in=45+90,looseness=10] (l);
        \draw[double,thick] (l)--(0);
        \draw (0-0.2,0.5-0.1)--(0,0.5+0.1)--(0+0.2,0.5-0.1);
        \draw[purple] (l) circle (0.25cm);
        \draw[purple] (1) circle (0.25cm);
        \node at (-0.5,0.5) {\scriptsize$N,2$};
	\end{tikzpicture}$};
    \draw[->] (a) .. controls (1,-3) .. (c);
    \node at (0,-3) {$-A_1$};
    \end{tikzpicture}}
    \qquad \textrm{and}  \qquad
     \raisebox{-.5\height}{    \begin{tikzpicture}
        \node (a) at (0,0) {
    $\begin{tikzpicture}
		\node[gauge,label=below:{$1$}] (0) at (0,0) {};
		\node[gauge,label=below:{$k$}] (1) at (1,0) {};
		\draw[double,thick] (0)--(1);
		\node at (0.5,0.2) {\scriptsize$N$};
        \draw (1) to [out=315,in=45,looseness=10] (1);
	\end{tikzpicture}$};
        \node (b) at (3,-3) {
    $\begin{tikzpicture}
    	\node at (-0.4,0) {$-$};
		\node[gauge,label=below:{$1$}] (0) at (0,0) {};
		\node[gauge,label=below:{$1$}] (1) at (1,0) {};
		\draw[double,thick] (0)--(1);
		\node at (0.5,0.2) {\scriptsize$N$};
        \draw (1) to [out=315,in=45,looseness=10] (1);
	\end{tikzpicture}$};
        \node (c) at (0,-6) {
    $\begin{tikzpicture}
		\node[gauge,label=below:{$1$}] (0) at (0,0) {};
		\node[gauge,label=below:{$k-1$}] (1) at (1,0) {};
		\draw[double,thick] (0)--(1);
		\node at (0.5,0.2) {\scriptsize$N$};
        \draw (1) to [out=315,in=45,looseness=10] (1);
	\end{tikzpicture}$};
    \draw[green,->] (a) .. controls (1,-3) .. (c);
    \node at (0,-3) {$-A_{N-1}$};
    \end{tikzpicture}}
\end{equation}
Using these two types of subtractions repeatedly, making sure that rule \eqref{eq:rule1} is applied properly to account for multiplicities, one can compute the Hasse diagram for an arbitrary symmetric product of $A_{N-1}$. An example is given in Figure \ref{fig:hasseSym4} for $k=4$. 

\subsection{Comparison with the Moduli Space of Instantons} 
\begin{figure}
    \centering
     \begin{subfigure}[b]{1\textwidth}
         \centering
    \begin{tikzpicture}
                \node (4) at (0,0) {$4$};
                \node (31) at (-0.5,1) {$31$};
                \node (22) at (0.5,1) {$22$};
                \node (211) at (0,2) {$211$};
                \node (1111) at (0,3) {$1111$};
                \node (3) at (2,1.5) {$3$};
                \node (21) at (2,2.5) {$21$};
                \node (111) at (2,3.5) {$111$};
                \node (2) at (4,3) {$2$};
                \node (11) at (4,4) {$11$};
                \node (1) at (6,4.5) {$1$};
                \node (0) at (8,5) {$0$};
                    \draw (1111)--(211) (22)--(4);
                    \draw[transform canvas={xshift=2pt,yshift=1pt}] (22)--(211);
                    \draw[transform canvas={xshift=-2pt,yshift=-1pt}] (22)--(211);
                    \draw[dashed] (4)--(31)--(211);
                    \draw (111)--(21);
                    \draw[dashed] (21)--(3);
                    \draw (11)--(2);
                \draw[red] (31)--(3) (111)--(11) (21)--(2) (1)--(0);
                \draw[red,transform canvas={xshift=-1.5pt,yshift=6pt}] (1111)--(111);
                \draw[red,transform canvas={xshift=-0.5pt,yshift=2pt}] (211)--(21) (11)--(1) (1111)--(111);
                \draw[red,transform canvas={xshift=0.5pt,yshift=-2pt}] (211)--(21) (11)--(1) (1111)--(111);
                \draw[red,transform canvas={xshift=1.5pt,yshift=-6pt}] (1111)--(111);
                \draw[red,transform canvas={xshift=-0.75pt,yshift=3pt}] (111)--(11);
                \draw[red,transform canvas={xshift=0.75pt,yshift=-3pt}] (111)--(11);
                
                \draw[dashed] \convexpath{0,1}{0.6cm};
                \draw[dashed] \convexpath{0,2,11}{0.6cm};
                \draw[dashed] \convexpath{0,3,111}{0.6cm};
                \draw[dashed] \convexpath{0,4,31,1111}{0.6cm};
                
                \node at (6.6,5.7) {$k=1$};
                \node at (4.6,5.2) {$k=2$};
                \node at (2.6,4.7) {$k=3$};
                \node at (0.6,4.2) {$k=4$};
            \end{tikzpicture}
         \caption{Hasse diagrams of $\mathcal{M}_{k,G}$ for $k\leq4$ embedded in each other. $l$ red lines connecting two dots denote an elementary slices $lg=g\cup...\cup g$, where $g$ is the minimal nilpotent orbit closure for $G$.}
         \label{fig:growth_inst}
        \end{subfigure}
        
        \begin{subfigure}[b]{1\textwidth}
        \centering
            \begin{tikzpicture}
            \node at (0,5) {};
                \node (0) at (0,0) {$0$};
                
                \begin{scope}[shift={(2,0)}]
                \node (1) at (0,1) {$1$};
                \end{scope}
                
                \begin{scope}[shift={(4,0)}]
                \node (2) at (0,1) {$2$};
                \node (11) at (0,2) {$11$};
                \end{scope}
                
                \begin{scope}[shift={(6,0)}]
                \node (3) at (0,1) {$3$};
                \node (21) at (0,2) {$21$};
                \node (111) at (0,3) {$111$};
                \end{scope}
                
                \begin{scope}[shift={(8,0)}]
                \node (4) at (0,1) {$4$};
                \node (31) at (-0.5,2) {$31$};
                \node (22) at (0.5,2) {$22$};
                \node (211) at (0,3) {$211$};
                \node (1111) at (0,4) {$1111$};
                \end{scope}
                
                \draw (1111)--(211) (22)--(4);
                    \draw[transform canvas={xshift=2pt,yshift=1pt}] (22)--(211);
                    \draw[transform canvas={xshift=-2pt,yshift=-1pt}] (22)--(211);
                \draw[dashed] (4)--(31)--(211);
                \draw (111)--(21);
                \draw[dashed] (21)--(3);
                \draw (11)--(2);
                
                \draw[green] (0)--(1) (0)--(2) (0)--(3) (0)--(4) (1)--(11) (1)--(21) (1)--(31) (2)--(21) (2)--(22) (3)--(31) (11)--(111) (11)--(211) (21)--(211) (111)--(1111);
                
                \draw[dashed] \convexpath{0,1}{0.6cm};
                \draw[dashed] \convexpath{0,11,2}{0.6cm};
                \draw[dashed] \convexpath{0,111,3}{0.6cm};
                \draw[dashed] \convexpath{0,1111,22,4}{0.6cm};
                
                \node at (0.3,1.3) {$k=1$};
                \node at (2.3,2.3) {$k=2$};
                \node at (4.3,3.3) {$k=3$};
                \node at (6.3,4.3) {$k=4$};
            \end{tikzpicture}
         \caption{Hasse diagrams of $\mathrm{Sym}^k(\mathbb{C}^2/\Gamma_G)$ for $k\leq4$ embedded in each other. Green lines denote elementary slices $\mathbb{C}^2/\Gamma_G$.}
         \label{fig:growth_sym}
        \end{subfigure}
         \caption{Illustration how the Hasse diagrams of $\mathcal{M}_{k,G}$ and $\mathrm{Sym}^k(\mathbb{C}^2/\Gamma_G)$ grow, as one increases $k$. While the Hasse diagram of $\mathcal{M}_{k,G}$ grows `downwards' the Hasse diagram of $\mathrm{Sym}^k(\mathbb{C}^2/\Gamma_G)$ grows `upwards'. Note that while the number of leaves is the same in both cases, there are many more elementary transitions in $\mathrm{Sym}^k(\mathbb{C}^2/\Gamma_G)$ than in $\mathcal{M}_{k,G}$. We note that the number of red lines is the same as the number of green lines.}
         \label{fig:Comparison}
\end{figure}
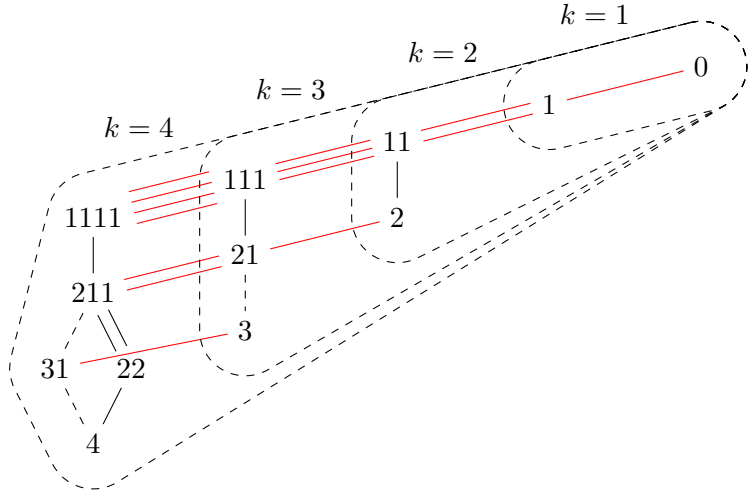
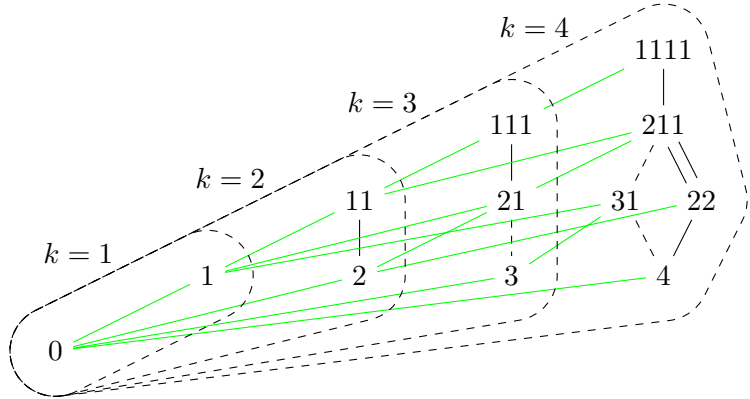

Let $\mathcal{M}_{G,k}$ denote the moduli space of $k$ $G$-instantons on $\mathbb{R}^4$. Following \cite{Bourget:2022ehw} the Hasse diagram of $\mathcal{M}_{G,k}$ is obtained by combining the Hasse diagrams of $\mathrm{Sym}^{k'}(\mathbb{C}^2)$ for $0\leq k'\leq k$, and adding additional transitions (drawn red in Figure \ref{fig:growth_inst}).
It follows from our construction, that the Hasse diagram of $\mathcal{M}_{G,k}$ is contained in the Hasse diagram of $\mathcal{M}_{G,k+1}$, and that the Hasse diagram of $\mathrm{Sym}^k(X)$ is contained in the Hasse diagram of $\mathrm{Sym}^{k+1}(X)$. The way the two Hasse diagrams grow when one increases $k$, however, is quite different. While the Hasse diagram of $\mathrm{Sym}^k(X)$ grows `upwards' when one increases $k$, the Hasse diagram of $\mathcal{M}_{G,k}$ grows `downwards'. This is illustrated in Figure \ref{fig:Comparison}.

\paragraph{An observation about the number of leaves and slices.} The number of symplectic leaves in $\mathcal{M}_{G,k}$ and $\mathrm{Sym}^k(\mathbb{C}^2/\Gamma_G)$ is the same. In $\mathcal{M}_{G,k}$ there are elementary slices which are unions of cones, i.e.\ $g$, $2g$, ..., $kg$. In $\mathrm{Sym}^k(\mathbb{C}^2/\Gamma_G)$ there are no elementary slices which are a union of multiple $\mathbb{C}^2/\Gamma_G$. However, there are many more elementary transitions connecting the various leaves in $\mathrm{Sym}^k(\mathbb{C}^2/\Gamma_G)$ than there are elementary transitions connecting the various leaves in $\mathcal{M}_{G,k}$.

It is interesting to note that since we denote an elementary slice which is a union of $n$ cones by $n$ lines in the Hasse diagram, the number of lines in the Hasse diagrams of $\mathcal{M}_{G,k}$ and $\mathrm{Sym}^k(\mathbb{C}^2/\Gamma_G)$ are the same. More precisely, we show in Appendix \ref{app:proof} that the number of green / red lines, using the color code of Figure \ref{fig:Comparison}, are both equal to 
\begin{equation}
\label{eq:countingLeavesSym}
    \sum\limits_{k'=0}^{k} \sum\limits_{k''=0}^{k'} p(k'') \, . 
\end{equation}

\FloatBarrier

\section{6d SCFTs with Discrete Gauging}
\label{sec:6d}

In this section we consider 6d $\mathcal{N}=(1,0)$ theories living on $n$ M5 branes probing a $\mathbb{C}^2/\mathbb{Z}_k$ singularity, as discussed in the Introduction. When the $n$ M5 branes are separated along the $\mathbb{C}^2/\mathbb{Z}_k$ singularity the low energy theory is the linear quiver \eqref{eq:linearquiver}.
This theory lives on the tensor branch of a 6d $\mathcal{N}=(1,0)$ SCFT. At infinite couplings, i.e.\ at singularities of the tensor branch, these theories exhibit discrete gauging \cite{Hanany:2018vph}. Magnetic quivers for all tensor branch phases of this theory have been computed in \cite{Cabrera:2019izd}.\footnote{Note that parts of the following analysis overlap with certain results of \cite{Hanany:2022itc}, which came out while this work was completed. } 

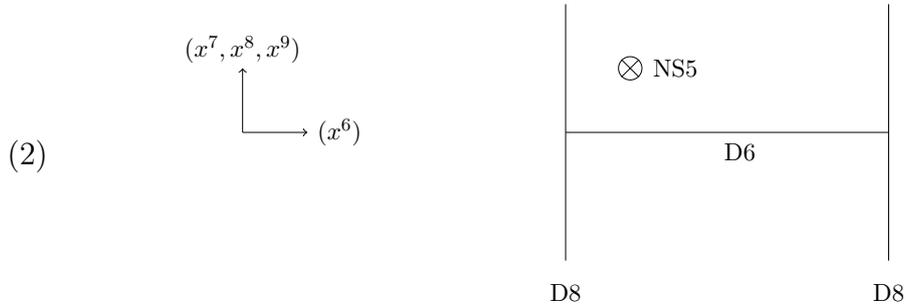
\begin{figure}[t]
\makebox[\textwidth][c]{
\scalebox{.85}{\begin{tikzpicture}
    \node at (0,0) {$\begin{tabular}{c|c|c|c|c|c|c|c|c|c|c}
& $x^0$ & $x^1$ & $x^2$ & $x^3$ & $x^4$ & $x^5$ & $x^6$ & $x^7$ & $x^8$ & $x^9$\\
\hline
NS5 & x & x & x & x & x & x & & & &  \\
\hline
D6 & x & x & x & x & x & x & x & & & \\
\hline
D8 & x & x & x & x & x & x & & x & x & x
\end{tabular}$};
\node at (0,-5) {$\begin{tikzpicture}
        \begin{scope}[shift={(-2,1)}]
        \draw[->] (-3,1)--(-2,1);
        \node at (-1.5,1) {$(x^6)$};
        \draw[->] (-3,1)--(-3,2);
        \node at (-3,2.3) {$(x^7,x^8,x^9)$};
        \end{scope}
        \draw (0,0)--(0,4) (5,0)--(5,4);
        \node at (0,-0.5) {D8};
        \node at (5,-0.5) {D8};
        \draw (0,2)--(5,2);
        \node at (2.7,1.7) {D6};
        \ns{1,3};
        \node at (1.7,3) {NS5};
    \end{tikzpicture}$};
    \node at (-8,0) {\Large(1)};
    \node at (-8,-5) {\Large(2)};
\end{tikzpicture}}}
\caption{(1) The Type IIA set-up: the 'x' mark the spacetime directions spanned by the various branes. (2) Depiction of a brane system.}
\label{fig:setup}
\end{figure}

The goal of this Section is to describe the full moduli space of this theory, using both a the brane system described in the next paragraph, and the techniques developed in the previous sections. Compactifying the M-theory setup, we will work with brane systems in Type IIA. Our conventions for depicting brane systems are summarized in Figure \ref{fig:setup}. Before delving into the details, let us mention a few salient results. 
\begin{itemize}
    \item We fully derive the rank 1 result shown in Figure \ref{fig:coneFibrationZ2}, and its generalization at rank 2, with much greater complexity. 
    \item We examine in detail the points mentioned in the introduction regarding the fibration structure and how it introduces spurious loci in the fibered Hasse diagram, with smooth transitions. 
    \item We characterize the new elementary transverse slices $\mathcal{Y}(n)$, $\mathcal{J}_{23}$ and $\mathcal{J}_{33}$. 
\end{itemize}

Finally, we make an important remark regarding phase transitions in brane diagrams. In several situations, such as linear unitary 3d $\mathcal{N}=4$ quiver theories, or 5d SCFTs on simple brane webs (with only $(p,q)$-fivebranes), the number of leaves on the Higgs branch corresponds to the number of maximal leaves in the full moduli space, and the mixed branches are direct products of a Higgs component and a Coulomb component. When this is the case, singular transitions are always associated to the opening of new branches, and are therefore directly visible in brane systems. This insight was central to the works \cite{Cabrera:2016vvv,Bourget:2019aer} and many others to compute the Hasse diagram of a given branch in a moduli space. Crucially, \emph{the theories studied in the present paper do not satisfy the above property that its mixed branches are direct products}.\footnote{For a simpler example of theories which do not satisfy the property, one can consider $\mathrm{O}(n)$ gauge theories, see Appendix \ref{app:OSo} for a discussion.  } In this case one may be able to identify singular transitions from a brane system by looking at the massless degrees of freedom on a given brane configuration and comparing it to the massless degrees of freedom of a different brane configuration. If the degrees of freedom change, there was a singular transition. This is however only possible if one can characterize the massless degrees of freedom appropriately, which is often not the case at a strongly coupled fixed point. The most versatile and general method we know to compute singular transitions is to characterize a transition in a brane system by computing the magnetic quiver for the moduli needed to move from one phase to another, as this gives access to the singular stratification from a purely geometric point of view. If the magnetic quiver has a singular Coulomb branch, then the brane transition was singular. To summarize, we have at least three methods available to identify singular transitions:
\begin{enumerate}
    \item \textbf{Jump in the number of massless degrees of freedom}. This is the most obvious method, which corresponds to directly reading the phase, but it rarely works for strongly coupled systems. 
    \item \textbf{New Coulomb branch directions}. This is extremely useful and has been used extensively in the literature, but has its limitations, as it relies on the assumption that mixed branches are direct products of Higgs and Coulomb branch directions, and is only sensitive to (the subset of) singularities at which new Coulomb branch directions open up. The present work is dedicated to theories where this assumption is not satisfied. 
    \item \textbf{Magnetic quiver}. This is the most general and versatile method. It requires a careful understanding of the algorithm,  such as rules \eqref{eq:rule1} and \eqref{eq:rule2}. 
\end{enumerate}

\subsection{\texorpdfstring{The case $k=2$ and $n=2$}{The case k=2 and n=2}}
\label{sec:neq2}

We start with the simple case $k=n=2$, which we have already discussed in the Introduction. We will describe the full moduli space as a fibration by the Higgs branch over the tensor branch $\mathcal{M}_T$:  
\begin{equation}
    \pi : \mathcal{M} \longrightarrow \mathcal{M}_T = \mathbb{R}_{\geq 0} \, . 
\end{equation}
There are two cases to consider: finite coupling and SCFT point.

\paragraph{Finite coupling.} The electric quiver is
\begin{equation}
\raisebox{-.5\height}{
    \begin{tikzpicture}
        \node[flavour,label=below:{$2$}] (0) at (0,0) {};
        \node[gaugeBig,label=below:{SU$(2)$}] (1) at (1,0) {};
        \node[flavour,label=below:{$2$}] (2) at (2,0) {};
        \draw (0)--(1)--(2);
    \end{tikzpicture} }\;
\end{equation}
We can depict the brane system at a generic point on the Higgs branch, and the magnetic quiver describing its moduli:
\begin{equation}
\label{eq:2,2top}
        \vcenter{\hbox{\begin{tikzpicture}
        \def\x{2cm};
        \draw (0,-\x)--(0,\x);
        \draw (1,-\x)--(1,\x);
        \draw (4,-\x)--(4,\x);
        \draw (5,-\x)--(5,\x);
        \ns{2,1.5};
        \ns{3,0.5};
        \draw (0,-1)--(1,-1) (1,-1.5)--(4,-1.5) (1,-0.5)--(4,-0.5) (4,-1)--(5,-1);
    \end{tikzpicture}}}
    \hspace{2cm}
    \vcenter{\hbox{\begin{tikzpicture}
        \node[gauge,label=below:{$1$}] (l) at (0,0) {};
        \node[gauge,label=below:{$2$}] (m) at (1,0) {};
        \node[gauge,label=below:{$1$}] (r) at (2,0) {};
        \node[gauge,label=left:{$1$}] (1) at (0.5,1) {};
        \node[gauge,label=right:{$1$}] (2) at (1.5,1) {};
        \draw (l)--(m)--(r) (1)--(m)--(2);
    \end{tikzpicture}}}
\end{equation}
In this phase the gauge group is completely broken. Extra massless states arise, when the gauge group is `unhiggsed' to SU$(2)$, i.e.\ when two D6 branes end on an NS5 brane on the left and right. Keeping in mind the S-rule (at most one D6 brane may be stuck between an NS5 and a D8 brane) we obtain the brane system for this phase and the magnetic quiver describing its moduli:
\begin{equation}
\label{eq:2,2bottom}
        \vcenter{\hbox{\begin{tikzpicture}
        \def\x{1cm};
        \draw (0,-\x)--(0,\x);
        \draw (1,-\x)--(1,\x);
        \draw (4,-\x)--(4,\x);
        \draw (5,-\x)--(5,\x);
        \ns{2,0}l;
        \ns{3,0}r;
        \draw[transform canvas={yshift=-1.5pt}] (0,0)--(l)--(r)--(4,0);
        \draw[transform canvas={yshift=1.5pt}] (1,0)--(l)--(r)--(5,0);
    \end{tikzpicture}}}
    \hspace{2cm}
    \vcenter{\hbox{\begin{tikzpicture}
        \node[gauge,label=below:{$1$}] (l) at (0,0) {};
    \end{tikzpicture}}}
\end{equation}
The elementary slice in the Higgs branch corresponding to the transition from \eqref{eq:2,2top} to \eqref{eq:2,2bottom} is $d_4$. We obtained the Hasse diagram of the Higgs branch of our theory:
\begin{equation}
  \raisebox{-.5\height}{  \begin{tikzpicture}
        \node[hasse] (t) at (0,1) {};
        \node[hasse] (b) at (0,0) {};
        \draw (t)--(b);
        \node at (0.3,0.5) {$d_4$};
    \end{tikzpicture}}
\end{equation}
This Hasse diagram is the diagram of \emph{one} fiber, above a given point on the tensor branch. However when considering $\pi^{-1} (\mathcal{M}_T - \{ 0 \})$, the $\mathbb{Z}_2$ action exchanging the two NS5 branes is manifestly acting. This is the action that ends up being quotiented by at the origin of the branch. We can record this using the notation of \cite{slodowy1980simple,2015arXiv150205770F} as follows: 
\begin{equation}
  \raisebox{-.5\height}{  \begin{tikzpicture}
        \node[hasse] (t) at (0,1) {};
        \node[hasse] (b) at (0,0) {};
        \draw (t)--(b);
        \node at (0.9,0.5) {$(d_4 , \mathbb{Z}_2) $};
    \end{tikzpicture}}
\end{equation}

\paragraph{Infinite coupling.} At infinite coupling the theory has no gauge theoretic description. In the brane system the two NS5 branes are now aligned along the $x^6$-direction.

We can depict the brane system at a generic point on the Higgs branch, and the magnetic quiver describing its moduli:
\begin{equation}
\label{eq:2,2inf,top}
        \vcenter{\hbox{\begin{tikzpicture}
        \def\x{2cm};
        \draw (0,-\x)--(0,\x);
        \draw (1,-\x)--(1,\x);
        \draw (4,-\x)--(4,\x);
        \draw (5,-\x)--(5,\x);
        \ns{2.5,1.5};
        \ns{2.5,0.5};
        \draw (0,-1)--(1,-1) (1,-1.5)--(4,-1.5) (1,-0.5)--(4,-0.5) (4,-1)--(5,-1);
    \end{tikzpicture}}}
    \hspace{2cm}
    \vcenter{\hbox{\begin{tikzpicture}
        \node[gauge,label=below:{$1$}] (l) at (0,0) {};
        \node[gauge,label=below:{$2$}] (m) at (1,0) {};
        \node[gauge,label=below:{$1$}] (r) at (2,0) {};
        \node[gauge,label=left:{$2$}] (1) at (1,1) {};
        \draw (1) to [out=45,in=135,looseness=10] (1);
        \draw (l)--(m)--(r) (1)--(m);
    \end{tikzpicture}}}
\end{equation}
It is less obvious to determine a phase in the brane system where new massless states arise. The Hasse diagram of the Coulomb branch of the magnetic quiver in \eqref{eq:2,2inf,top} is know to be
\begin{equation}
    \label{eq:hasse_b3_next}
  \raisebox{-.5\height}{  \begin{tikzpicture}
        \node[hasse] (t) at (0,1) {};
        \node[hasse] (m) at (0,0) {};
        \node[hasse] (b) at (0,-1) {};
        \draw (t)--(m)--(b);
        \node at (0.3,0.5) {$A_1$};
        \node at (0.3,-0.5) {$b_3$};
    \end{tikzpicture}}\;.
\end{equation}
This can be deduced from the fact that the quiver in \eqref{eq:2,2inf,top} is a magnetic quiver for the closure of the next-to-minimal nilpotent orbit of $\mathfrak{so}(7)$, or from quiver subtraction, as we show below.  
Therefore we know that we can reach a phase in the brane system where extra massless states arise by restricting one modulus in the brane system \eqref{eq:2,2inf,top}. We happen to have one such modulus, the separation between the two NS5 branes. Making these NS5 branes coincide they now share one modulus. Following \cite{Bourget:2022ehw} this modulus appears as a U$(1)$ node of shortness $1/2$ in the magnetic quiver:
\begin{equation}
\label{eq:2,2inf,middle}
        \vcenter{\hbox{\begin{tikzpicture}
        \def\x{2cm};
        \draw (0,-\x)--(0,\x);
        \draw (1,-\x)--(1,\x);
        \draw (4,-\x)--(4,\x);
        \draw (5,-\x)--(5,\x);
        \ns{2.5,0.5};
        \node at (2.7,0.8) {$2$};
        \draw (0,-1)--(1,-1) (1,-1.5)--(4,-1.5) (1,-0.5)--(4,-0.5) (4,-1)--(5,-1);
    \end{tikzpicture}}}
    \hspace{2cm}
    \vcenter{\hbox{\begin{tikzpicture}
        \node[gauge,label=below:{$1$}] (l) at (0,0) {};
        \node[gauge,label=below:{$2$}] (m) at (1,0) {};
        \node[gauge,label=below:{$1$}] (r) at (2,0) {};
        \node[gauge,label=left:{$1$}] (1) at (1,1) {};
        \draw (l)--(m)--(r);
        \draw[transform canvas={xshift=-1.5pt}] (m)--(1);
        \draw[transform canvas={xshift=1.5pt}] (m)--(1);
        \draw (0.8,0.4)--(1,0.6)--(1.2,0.4);
    \end{tikzpicture}}}
\end{equation}
We can now obtain extra massless states much like in the finite coupling case. The two coincident NS5 branes need to team up with 2 D6 branes. Respecting the S-rule we find the brane system and magnetic quiver:
\begin{equation}
\label{eq:2,2inf,bottom}
        \vcenter{\hbox{\begin{tikzpicture}
        \def\x{1cm};
        \draw (0,-\x)--(0,\x);
        \draw (1,-\x)--(1,\x);
        \draw (4,-\x)--(4,\x);
        \draw (5,-\x)--(5,\x);
        \ns{2.5,0}l;
        \node at (2.7,0.3) {2};
        \draw[transform canvas={yshift=-1.5pt}] (0,0)--(l)--(4,0);
        \draw[transform canvas={yshift=1.5pt}] (1,0)--(l)--(5,0);
    \end{tikzpicture}}}
    \hspace{2cm}
    \vcenter{\hbox{\begin{tikzpicture}
        \node[gauge,label=below:{$1$}] (l) at (0,0) {};
    \end{tikzpicture}}}
\end{equation}
By comparing the three phases \eqref{eq:2,2inf,top},\eqref{eq:2,2inf,middle} and \eqref{eq:2,2inf,bottom} we exactly recover the Hasse diagram \eqref{eq:hasse_b3_next}.
Putting everything together, we can draw the Hasse diagram for the full moduli space. This is done in Figure \ref{fig:coneFibrationZ2}. 

The transition between configurations \eqref{eq:2,2top} and \eqref{eq:2,2inf,top}, represented by the red dashed line in Figure \ref{fig:coneFibrationZ2}, is clearly non-singular. The NS5 branes just happen to share one coordinate in $\mathbb{R}^4$, but do not share the other three coordinates before reaching the configuration \eqref{eq:2,2inf,middle}. Therefore, a phase transition takes place between \eqref{eq:2,2top} and \eqref{eq:2,2inf,middle}. In this transition, the D6 and D8 branes are spectators, and the local system involved in the transition preserves 16 supercharges. It is the same transition that would occur in a 6d $\mathcal{N}=(2,0)$ theory. This is the transition denoted by the orange line in Figure \ref{fig:coneFibrationZ2}.

\subsection{\texorpdfstring{The case $k=2$ and $n=3$}{The case k=2 and n=3}}

\label{sec:neq3}

We now consider the case $k=2$, $n=3$. At finite coupling the electric quiver is 
\begin{equation}
\label{eq:theorynequals3}
   \raisebox{-.5\height}{ \begin{tikzpicture}
        \node[flavour,label=below:{$2$}] (0) at (0,0) {};
        \node[gaugeBig,label=below:{SU$(2)$}] (1) at (1,0) {};
        \node[gaugeBig,label=below:{SU$(2)$}] (2) at (2,0) {};
        \node[flavour,label=below:{$2$}] (3) at (3,0) {};
        \draw (0)--(1)--(2)--(3);
    \end{tikzpicture}}\;.
\end{equation}
while the brane system and magnetic quiver are 
\begin{equation}
    \raisebox{-.5\height}{\begin{tikzpicture}
        \def\x{2cm};
        \draw (0,-\x)--(0,\x);
        \draw (1,-\x)--(1,\x);
        \draw (4,-\x)--(4,\x);
        \draw (5,-\x)--(5,\x);
        \ns{1.6,1.5};
        \ns{2.5,1};
        \ns{3.4,0.5};
        \draw (0,-1)--(1,-1) (1,-1.5)--(4,-1.5) (1,-0.5)--(4,-0.5) (4,-1)--(5,-1);
    \end{tikzpicture}    }
   \qquad  
   \qquad 
   \raisebox{-.5\height}{ \begin{tikzpicture}
        \node[gauge,label=below:{$1$}] (l) at (0,0) {};
        \node[gauge,label=below:{$2$}] (m) at (1,0) {};
        \node[gauge,label=below:{$1$}] (r) at (2,0) {};
        \node[gauge,label=above:{$1$}] (1) at (0.2,1) {};
        \node[gauge,label=above:{$1$}] (2) at (1,1) {};
        \node[gauge,label=above:{$1$}] (3) at (1.8,1) {};
        \draw (l)--(m)--(r) (1)--(m)--(2) (m)--(3);
    \end{tikzpicture}}
\end{equation}
From the brane system / magnetic quiver, one identifies 5 distinct ways to reaching a different phase. These correspond to the five brane systems depicted on the left part of Figure \ref{fig:fullMS3}. 

When one of the two tensor moduli (identified with the spacing between the NS5 branes along direction $x^6$) is taken to zero, the story is similar to what happened for $k=n=2$, involving $b_3$ transitions. The configurations and diagrams are shown in the middle part of Figure \ref{fig:fullMS3}. 

Finally, at the origin of the tensor branch, we can depict the brane system at a generic point on the Higgs branch, and the magnetic quiver describing its moduli:
\begin{equation}
\label{eq:2,3inf,top}
        \vcenter{\hbox{\begin{tikzpicture}
        \def\x{2cm};
        \draw (0,-\x)--(0,\x);
        \draw (1,-\x)--(1,\x);
        \draw (4,-\x)--(4,\x);
        \draw (5,-\x)--(5,\x);
        \ns{2.5,1.5};
        \ns{2.5,1};
        \ns{2.5,0.5};
        \draw (0,-1)--(1,-1) (1,-1.5)--(4,-1.5) (1,-0.5)--(4,-0.5) (4,-1)--(5,-1);
    \end{tikzpicture}}}
    \hspace{2cm}
    \vcenter{\hbox{\begin{tikzpicture}
        \node[gauge,label=below:{$1$}] (l) at (0,0) {};
        \node[gauge,label=below:{$2$}] (m) at (1,0) {};
        \node[gauge,label=below:{$1$}] (r) at (2,0) {};
        \node[gauge,label=left:{$3$}] (1) at (1,1) {};
        \draw (1) to [out=45,in=135,looseness=10] (1);
        \draw (l)--(m)--(r) (1)--(m);
    \end{tikzpicture}}}
\end{equation}
Just as in the $n=2$ case we can make two NS5 branes coincide to reach a phase with extra massless states. The brane system is simple, and the magnetic quiver is obtained using the rules of \cite{Bourget:2022ehw}, or directly by quiver subtraction using \eqref{eq:rule1}. The full diagram for the Higgs branch above the origin of the tensor branch is shown in Figure \ref{fig:2,3Full}. Crucially, the closure of the minimal nilpotent orbit of the exceptional Lie algebra $G_2$ is involved in the process, through the subtraction of its magnetic quiver 
\begin{equation}
\label{eq:g2quiver}
\raisebox{-.5\height}{\begin{tikzpicture}
        \node[gauge,label=below:{$1$}] (l) at (0,0) {};
        \node[gauge,label=below:{$2$}] (m) at (1,0) {};
        \node[gauge,label=left:{$1$}] (1) at (1,1) {};
        \draw (l)--(m);
        \draw[transform canvas={xshift=2pt}] (m)--(1);
        \draw[transform canvas={xshift=0pt}] (m)--(1);
        \draw[transform canvas={xshift=-2pt}] (m)--(1);
        \draw (0.8,0.4)--(1,0.5)--(1.2,0.4);
    \end{tikzpicture}}
\end{equation}

The diagram for the full moduli space with respect to the Higgs branch fibration over the tensor branch is summarized on Figure \ref{fig:fullMS3simplified}. As before, it is interesting to notice that several phases are spuriously separated by the fibration, as indicated by the dashed lines which denote smooth transverse spaces. We refer to the caption of this Figure for detailed comments.

\begin{figure}
    \makebox[\textwidth][c]{
    \begin{tikzpicture}
        \node (t) at (0,0) {$\vcenter{\hbox{\scalebox{0.7}{\begin{tikzpicture}
                \def\x{2cm};
                \draw (0,-\x)--(0,\x);
                \draw (1,-\x)--(1,\x);
                \draw (4,-\x)--(4,\x);
                \draw (5,-\x)--(5,\x);
                \ns{2.5,1.5};
                \ns{2.5,1};
                \ns{2.5,0.5};
                \draw (0,-1)--(1,-1) (1,-1.5)--(4,-1.5) (1,-0.5)--(4,-0.5) (4,-1)--(5,-1);
            \end{tikzpicture}}}}
            \hspace{1cm}
            \vcenter{\hbox{\scalebox{0.7}{\begin{tikzpicture}
                \node[gauge,label=below:{$1$}] (l) at (0,0) {};
                \node[gauge,label=below:{$2$}] (m) at (1,0) {};
                \node[gauge,label=below:{$1$}] (r) at (2,0) {};
                \node[gauge,label=left:{$3$}] (1) at (1,1) {};
                \draw (1) to [out=45,in=135,looseness=10] (1);
                \draw (l)--(m)--(r) (1)--(m);
            \end{tikzpicture}}}}$};

        \node (2) at (0,-4) {$\vcenter{\hbox{\scalebox{0.7}{\begin{tikzpicture}
                \def\x{2cm};
                \draw (0,-\x)--(0,\x);
                \draw (1,-\x)--(1,\x);
                \draw (4,-\x)--(4,\x);
                \draw (5,-\x)--(5,\x);
                \ns{2.5,1.5};
                \ns{2.5,0.5};
                \node at (2.7,0.8) {2};
                \draw (0,-1)--(1,-1) (1,-1.5)--(4,-1.5) (1,-0.5)--(4,-0.5) (4,-1)--(5,-1);
            \end{tikzpicture}}}}
            \hspace{1cm}
            \vcenter{\hbox{\scalebox{0.7}{\begin{tikzpicture}
                \node[gauge,label=below:{$1$}] (l) at (0,0) {};
                \node[gauge,label=below:{$2$}] (m) at (1,0) {};
                \node[gauge,label=below:{$1$}] (r) at (2,0) {};
                \node[gauge,label=left:{$1$}] (1) at (0.5,1) {};
                \node[gauge,label=right:{$1$}] (2) at (1.5,1) {};
                \draw (l)--(m)--(r) (1)--(m);
                \draw[transform canvas={xshift={sin(atan(2))},yshift=-{cos(atan(2))}}] (m)--(2);
                \draw[transform canvas={xshift=-{sin(atan(2))},yshift={cos(atan(2))}}] (m)--(2);
                \begin{scope}[shift={(1,0)}]
                    \draw[rotate=60] (0.4,0.2)--(0.6,0)--(0.4,-0.2);
                \end{scope}
                \draw[purple] (1) circle (0.5cm);
                \draw[purple] (2) circle (0.5cm);
            \end{tikzpicture}}}}$};

        \node (3) at (3,-8) {$\vcenter{\hbox{\scalebox{0.7}{\begin{tikzpicture}
                \def\x{2cm};
                \draw (0,-\x)--(0,\x);
                \draw (1,-\x)--(1,\x);
                \draw (4,-\x)--(4,\x);
                \draw (5,-\x)--(5,\x);
                \ns{2.5,0.5};
                \node at (2.7,0.8) {3};
                \draw (0,-1)--(1,-1) (1,-1.5)--(4,-1.5) (1,-0.5)--(4,-0.5) (4,-1)--(5,-1);
            \end{tikzpicture}}}}
            \hspace{1cm}
            \vcenter{\hbox{\scalebox{0.7}{\begin{tikzpicture}
                \node[gauge,label=below:{$1$}] (l) at (0,0) {};
                \node[gauge,label=below:{$2$}] (m) at (1,0) {};
                \node[gauge,label=below:{$1$}] (r) at (2,0) {};
                \node[gauge,label=left:{$1$}] (1) at (1,1) {};
                \draw (l)--(m)--(r);
                \draw[transform canvas={xshift=2pt}] (m)--(1);
                \draw[transform canvas={xshift=0pt}] (m)--(1);
                \draw[transform canvas={xshift=-2pt}] (m)--(1);
                \draw (0.8,0.4)--(1,0.5)--(1.2,0.4);
            \end{tikzpicture}}}}$};
        \node (4) at (-6,-12) {$\vcenter{\hbox{\scalebox{0.7}{\begin{tikzpicture}
                \def\x{1.5cm};
                \draw (0,-\x)--(0,\x);
                \draw (1,-\x)--(1,\x);
                \draw (4,-\x)--(4,\x);
                \draw (5,-\x)--(5,\x);
                \ns{2.5,1};
                \ns{2.5,0}l;
                \node at (2.7,0.3) {2};
                \draw[transform canvas={yshift=-1.5pt}] (0,0)--(l)--(4,0);
                \draw[transform canvas={yshift=1.5pt}] (1,0)--(l)--(5,0);
            \end{tikzpicture}}}}
            \hspace{1cm}
            \vcenter{\hbox{\scalebox{0.7}{\begin{tikzpicture}
                \node[gauge,label=below:{$1$}] (l) at (0,0) {};
                \node[gauge,label=above:{$1$}] (1) at (0,1) {};
                \draw[transform canvas={xshift=-1.5pt}] (l)--(1);
                \draw[transform canvas={xshift=1.5pt}] (l)--(1);
            \end{tikzpicture}}}}$};

        \node (5) at (0,-12) {$\vcenter{\hbox{\scalebox{0.7}{\begin{tikzpicture}
                \def\x{1.5cm};
                \draw (0,-\x)--(0,\x);
                \draw (1,-\x)--(1,\x);
                \draw (4,-\x)--(4,\x);
                \draw (5,-\x)--(5,\x);
                \ns{2.5,0}l;
                \node at (2.7,0.3) {3};
                \draw[transform canvas={yshift=-1.5pt}] (0,0)--(l)--(4,0);
                \draw[transform canvas={yshift=1.5pt}] (1,0)--(l)--(4,0);
                \draw (4,-1)--(5,-1);
            \end{tikzpicture}}}}
            \hspace{1cm}
            \vcenter{\hbox{\scalebox{0.7}{\begin{tikzpicture}
                \node[gauge,label=below:{$1$}] (l) at (0,0) {};
                \node[gauge,label=above:{$1$}] (1) at (0,1) {};
                \draw[transform canvas={xshift=-1.5pt}] (l)--(1);
                \draw[transform canvas={xshift=1.5pt}] (l)--(1);
            \end{tikzpicture}}}}$};
            
        \node (6) at (6,-12) {$\vcenter{\hbox{\scalebox{0.7}{\begin{tikzpicture}
                \def\x{1.5cm};
                \draw (0,-\x)--(0,\x);
                \draw (1,-\x)--(1,\x);
                \draw (4,-\x)--(4,\x);
                \draw (5,-\x)--(5,\x);
                \ns{2.5,0}l;
                \node at (2.7,0.3) {3};
                \draw[transform canvas={yshift=-1.5pt}] (1,0)--(l)--(4,0);
                \draw[transform canvas={yshift=1.5pt}] (1,0)--(l)--(5,0);
                \draw (0,-1)--(1,-1);
            \end{tikzpicture}}}}
            \hspace{1cm}
            \vcenter{\hbox{\scalebox{0.7}{\begin{tikzpicture}
                \node[gauge,label=below:{$1$}] (l) at (0,0) {};
                \node[gauge,label=above:{$1$}] (1) at (0,1) {};
                \draw[transform canvas={xshift=-1.5pt}] (l)--(1);
                \draw[transform canvas={xshift=1.5pt}] (l)--(1);
            \end{tikzpicture}}}}$};

        \node (7) at (0,-16) {$\vcenter{\hbox{\scalebox{0.7}{\begin{tikzpicture}
                \def\x{1.5cm};
                \draw (0,-\x)--(0,\x);
                \draw (1,-\x)--(1,\x);
                \draw (4,-\x)--(4,\x);
                \draw (5,-\x)--(5,\x);
                \ns{2.5,0}l;
                \node at (2.7,0.3) {3};
                \draw[transform canvas={yshift=-1.5pt}] (0,0)--(l)--(4,0);
                \draw[transform canvas={yshift=1.5pt}] (1,0)--(l)--(5,0);
            \end{tikzpicture}}}}
            \hspace{1cm}
            \vcenter{\hbox{\scalebox{0.7}{\begin{tikzpicture}
                \node[gauge,label=below:{$1$}] (l) at (0,0) {};
            \end{tikzpicture}}}}$};
        
        \draw (t)--(2)--(4)--(7) (2)--(3)--(5)--(7) (3)--(6)--(7);
        
        \node at ($(t)!0.5!(2)+(0.5,0)$) {$A_1$};
        \node at ($(2)!0.5!(3)+(0.5,0)$) {$m$};
        \node at ($(2)!0.5!(4)+(0.5,0)$) {$b_3$};
        \node at ($(3)!0.5!(5)+(0.5,0)$) {$g_2$};
        \node at ($(3)!0.5!(6)+(0.5,0)$) {$g_2$};
        \node at ($(4)!0.5!(7)+(0.5,0)$) {$A_1$};
        \node at ($(5)!0.5!(7)+(0.5,0)$) {$A_1$};
        \node at ($(6)!0.5!(7)+(0.5,0)$) {$A_1$};
        
    \end{tikzpicture}}
    \caption{$k=2,n=3$. Different Higgs phases in the brane system along with corresponding magnetic quivers and elementary transitions. The Hasse diagram contains the Hasse diagram for $k=2,n=2$. Note that the decoration on the second quiver from the top imply that there is only one possible $b_3$ subtraction -- there would be three without the decoration -- in other words, the decoration ensures the configuration stays at the same point on the tensor branch.}
    \label{fig:2,3Full}
\end{figure}
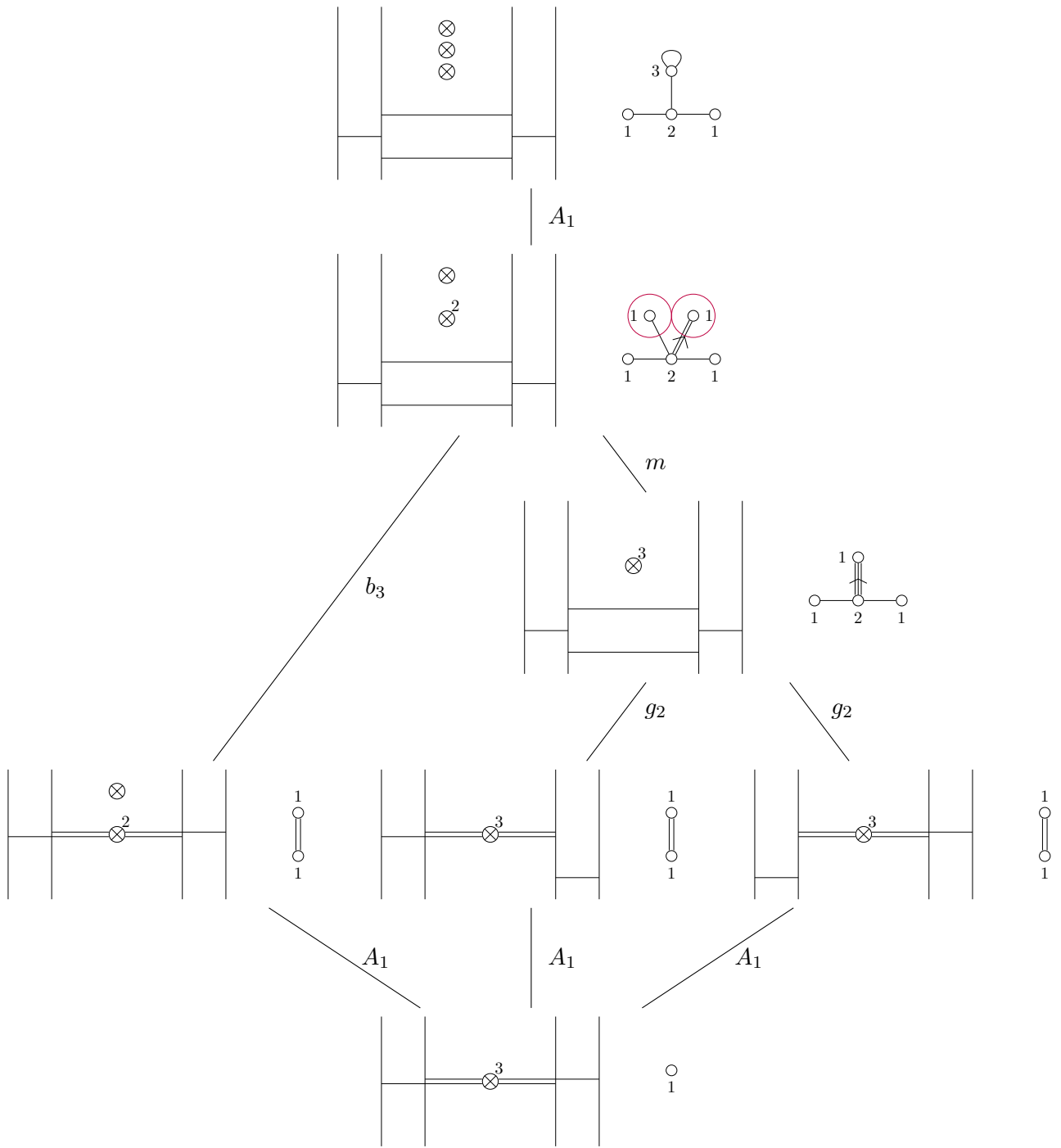

\begin{landscape}
\begin{figure}
\begin{center}
\scalebox{.48}{
\hspace*{-7cm}\begin{tikzpicture}
\def\x{7};
\def\y{-10};
\def\xx{20};
\def\yy{-9};
\def\xxx{25};
\def\yyy{-5};
\draw[draw=white,fill=orange!10,opacity=.5] (\xxx + 16,\yyy + 5)--(\x + 8,\y +17)--(0,15)--(\xx + 8,\yy +17)--(\xxx + 16,\yyy + 5);
\draw[draw=white,fill=red!10,opacity=.5] (0,1)--(\x + 8,\y +7)--(\xxx + 18,\yyy -1)--(\xx + 8,\yy +7)--(0,1);
\draw[draw=white,fill=blue!10,opacity=.5] (\xxx + 18,\yyy + 13)--(\xxx + 23,\yyy + 4)--(\xxx + 18,\yyy -1)--(\xxx + 14,\yyy + 2)--(\xxx + 18,\yyy + 13);
\node[draw,fill=black!5] (za) at (0,15) {\scalebox{.7}{\za}};
\node[draw,fill=black!5] (zba) at (-8,8) {\scalebox{.7}{\zba}};
\node[draw,fill=black!5] (zbb) at (-4,8) {\scalebox{.7}{\zbb}};
\node[draw,fill=black!5] (zbc) at (0,8) {\scalebox{.7}{\zbc}};
\node[draw,fill=black!5] (zbd) at (4,8) {\scalebox{.7}{\zbd}};
\node[draw,fill=black!5] (zbe) at (8,8) {\scalebox{.7}{\zbe}};
\node[draw,fill=black!5] (zc) at (0,1) {\scalebox{.7}{\zc}};
\draw (za)--(zba)--(zc) (za)--(zbb)--(zc) (za)--(zbc)--(zc) (za)--(zbd)--(zc) (za)--(zbe)--(zc) ;
\node[draw,fill=black!5] (zd) at (\x + 8,\y +20) {\scalebox{.7}{\zd}};
\node[draw,fill=black!5] (ze) at (\x + 8,\y +17) {\scalebox{.7}{\ze}};
\node[draw,fill=black!5] (zfa) at (\x + 12,\y +12) {\scalebox{.7}{\zfa}};
\node[draw,fill=black!5] (zfb) at (\x + 8,\y +12) {\scalebox{.7}{\zfb}};
\node[draw,fill=black!5] (zfc) at (\x + 4,\y +12) {\scalebox{.7}{\zfc}};
\node[draw,fill=black!5] (zg) at (\x + 8,\y +7) {\scalebox{.7}{\zg}};
\draw (zd)--(ze)--(zfa)--(zg) (ze)--(zfb)--(zg) (ze)--(zfc)--(zg);
\node[draw,fill=black!5] (zh) at (\xx + 8,\yy +20) {\scalebox{.7}{\zh}};
\node[draw,fill=black!5] (zi) at (\xx + 8,\yy +17) {\scalebox{.7}{\zi}};
\node[draw,fill=black!5] (zja) at (\xx + 12,\yy +12) {\scalebox{.7}{\zja}};
\node[draw,fill=black!5] (zjb) at (\xx + 8,\yy +12) {\scalebox{.7}{\zjb}};
\node[draw,fill=black!5] (zjc) at (\xx + 4,\yy +12) {\scalebox{.7}{\zjc}};
\node[draw,fill=black!5] (zk) at (\xx + 8,\yy +7) {\scalebox{.7}{\zk}};
\draw (zh)--(zi)--(zja)--(zk) (zi)--(zjb)--(zk) (zi)--(zjc)--(zk);
\node[draw,fill=black!5] (zl) at (\xxx + 18,\yyy + 11) {\scalebox{.7}{\zl}};
\node[draw,fill=black!5] (zm) at (\xxx + 18,\yyy + 8) {\scalebox{.7}{\zm}};
\node[draw,fill=black!5] (zn) at (\xxx + 23,\yyy + 4) {\scalebox{.7}{\zn}};
\node[draw,fill=black!5] (zo) at (\xxx + 16,\yyy + 5) {\scalebox{.7}{\zo}};
\node[draw,fill=black!5] (zp) at (\xxx + 18,\yyy + 2) {\scalebox{.7}{\zp}};
\node[draw,fill=black!5] (zq) at (\xxx + 14,\yyy + 2) {\scalebox{.7}{\zq}};
\node[draw,fill=black!5] (zr) at (\xxx + 18,\yyy -1) {\scalebox{.7}{\zr}};
\draw (zl)--(zm)--(zn)--(zr)--(zp)--(zo)--(zq)--(zr) (zo)--(zm);
\draw[red,dashed] (za)--(zd)--(zl)--(zh)--(za);
\draw[red] (zc)--(zg)--(zr)--(zk)--(zc);
\draw[red] (zba)--(zfc);
\draw[red] (zbd)--(zfb);
\draw[red] (zbe)--(zfa);
\draw[red] (zbe)--(zja);
\draw[red] (zbb)--(zjb);
\draw[red] (zba)--(zjc);
\draw[red] (zfa)--(zp);
\draw[red] (zfc)--(zq);
\draw[red] (zja)--(zp);
\draw[red] (zjc)--(zq);
\draw[red,dashed] (ze)--(zm);
\draw[red,dashed] (zi)--(zm);
\draw[red,dashed] (zfb)--(zn);
\draw[red,dashed] (zjb)--(zn);
\draw[orange,very thick] (za)--(zi)--(zo);
\draw[orange,very thick] (za)--(ze)--(zo);
\end{tikzpicture}}
\end{center}
\caption{Hasse diagram with respect to the Higgs branch fibration over the tensor branch for the full moduli space of theory \eqref{eq:theorynequals3}. Black lines represent Higgs branch transitions, red lines represent $\mathcal{N}=(1,0)$ tensor branch transitions, and orange lines represent $\mathcal{N}=(2,0)$ tensor branch transitions. The red shaded area depicts the pure tensor branch, and the blue shaded area depicts the pure Higgs branch. Each phase of the theory is represented by a brane diagram. A more schematic version of this figure is shown in Figure \ref{fig:fullMS3simplified}.   }
\label{fig:fullMS3}
\end{figure}
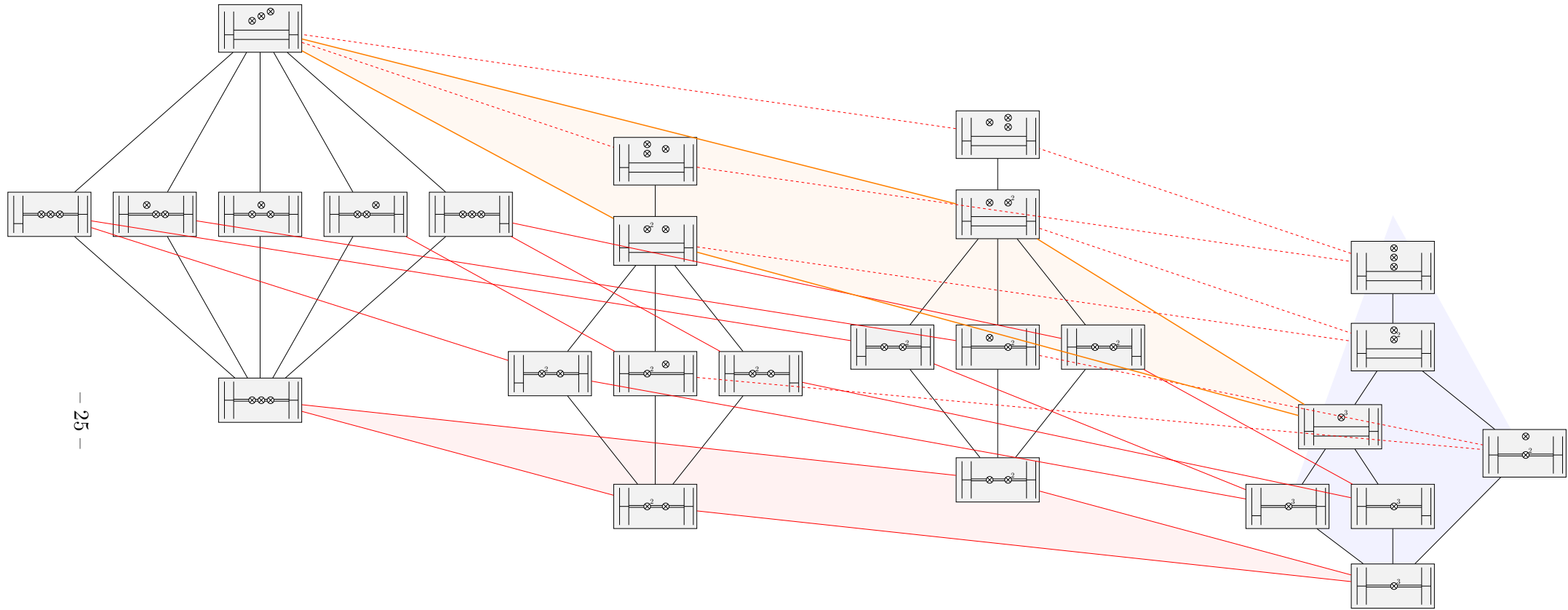
\end{landscape}   
     
     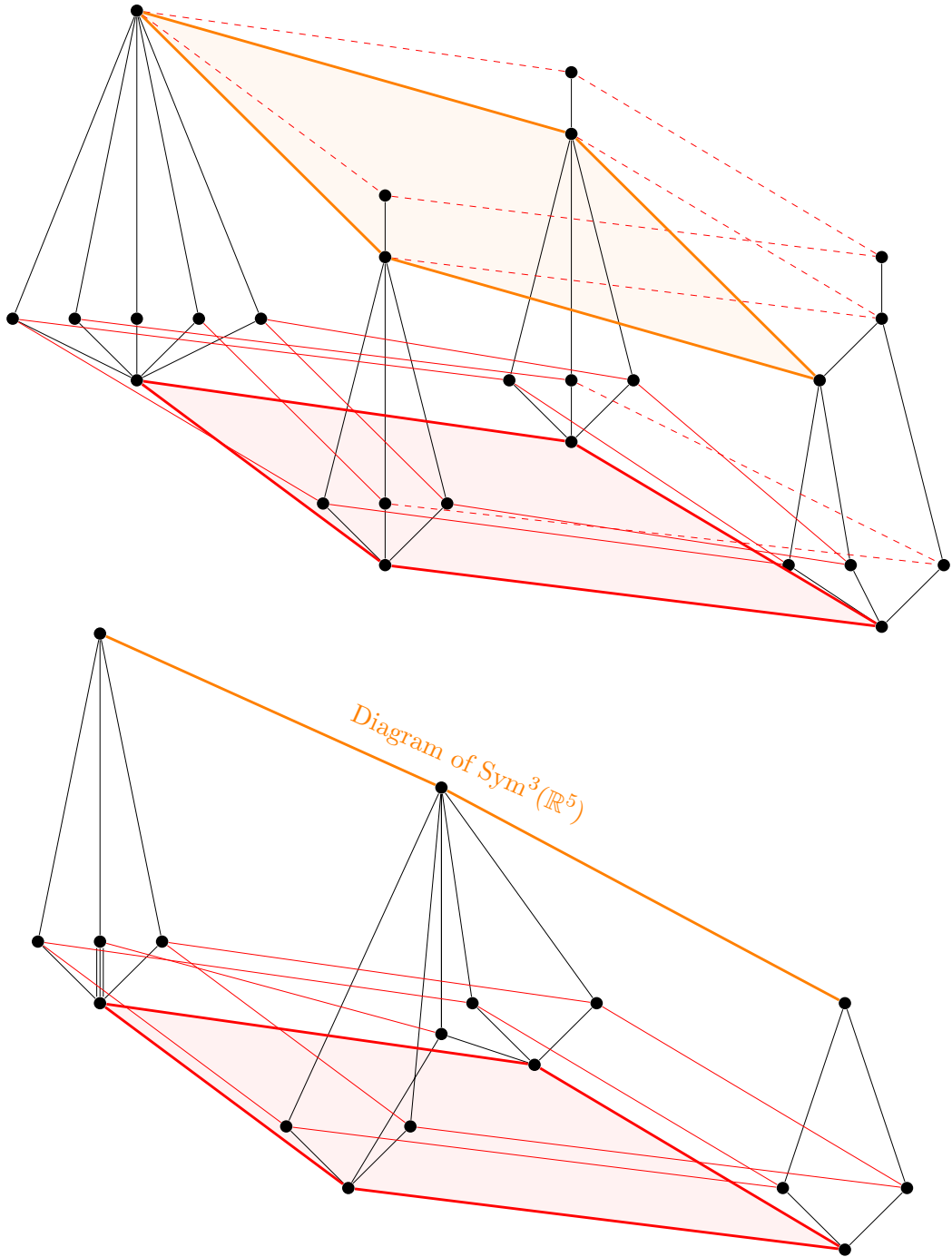
\begin{figure}
 \begin{center} 
\vspace*{-4cm}\begin{tikzpicture}
\node at (0,10) {\scalebox{.9}{\begin{tikzpicture}
\def\x{4};
\def\y{-3};
\def\xx{7};
\def\yy{-1};
\def\xxx{12};
\def\yyy{-4};
\draw[draw=white,fill=orange!10,opacity=.5] (0,6)--(\x,\y+5)--(\xxx -1 ,\yyy+4)--(\xx ,\yy +5)--(0,6);
\draw[draw=white,fill=red!10,opacity=.5] (0,0)--(\x,\y)--(\xxx ,\yyy)--(\xx ,\yy )--(0,0);
\draw[draw=white,fill=blue!10,opacity=.5] (\xxx + 18,\yyy + 13)--(\xxx + 23,\yyy + 4)--(\xxx + 18,\yyy -1)--(\xxx + 14,\yyy + 2)--(\xxx + 18,\yyy + 13);
\node[hasse] (za) at (0,6) {};
\node[hasse] (zba) at (-2,1) {};
\node[hasse] (zbb) at (-1,1) {};
\node[hasse] (zbc) at (0,1) {};
\node[hasse] (zbd) at (1,1) {};
\node[hasse] (zbe) at (2,1) {};
\node[hasse] (zc) at (0,0) {};
\draw (za)--(zba)--(zc) (za)--(zbb)--(zc) (za)--(zbc)--(zc) (za)--(zbd)--(zc) (za)--(zbe)--(zc) ;
\node[hasse] (zd) at (\x + 0,\y +6) {};
\node[hasse] (ze) at (\x + 0,\y +5) {};
\node[hasse] (zfa) at (\x +1,\y +1) {};
\node[hasse] (zfb) at (\x + 0,\y +1) {};
\node[hasse] (zfc) at (\x - 1,\y +1) {};
\node[hasse] (zg) at (\x + 0,\y +0) {};
\draw (zd)--(ze)--(zfa)--(zg) (ze)--(zfb)--(zg) (ze)--(zfc)--(zg);
\node[hasse] (zh) at (\xx + 0,\yy +6) {};
\node[hasse] (zi) at (\xx + 0,\yy +5) {};
\node[hasse] (zja) at (\xx +1,\yy +1) {};
\node[hasse] (zjb) at (\xx + 0,\yy +1) {};
\node[hasse] (zjc) at (\xx - 1,\yy +1) {};
\node[hasse] (zk) at (\xx + 0,\yy +0) {};
\draw (zh)--(zi)--(zja)--(zk) (zi)--(zjb)--(zk) (zi)--(zjc)--(zk);
\node[hasse] (zl) at (\xxx + 0,\yyy + 6) {};
\node[hasse] (zm) at (\xxx + 0,\yyy + 5) {};
\node[hasse] (zn) at (\xxx + 1,\yyy + 1) {};
\node[hasse] (zo) at (\xxx -1,\yyy + 4) {};
\node[hasse] (zp) at (\xxx -.5,\yyy + 1) {};
\node[hasse] (zq) at (\xxx -1.5,\yyy + 1) {};
\node[hasse] (zr) at (\xxx + 0,\yyy +0) {};
\draw (zl)--(zm)--(zn)--(zr)--(zp)--(zo)--(zq)--(zr) (zo)--(zm);
\draw[red, dashed] (za)--(zd)--(zl)--(zh)--(za);
\draw[red, very thick] (zc)--(zg)--(zr)--(zk)--(zc);
\draw[red] (zba)--(zfc);
\draw[red] (zbd)--(zfb);
\draw[red] (zbe)--(zfa);
\draw[red] (zbe)--(zja);
\draw[red] (zbb)--(zjb);
\draw[red] (zba)--(zjc);
\draw[red] (zfa)--(zp);
\draw[red] (zfc)--(zq);
\draw[red] (zja)--(zp);
\draw[red] (zjc)--(zq);
\draw[red,dashed] (ze)--(zm);
\draw[red,dashed] (zi)--(zm);
\draw[red,dashed] (zfb)--(zn);
\draw[red,dashed] (zjb)--(zn);
\draw[orange,very thick] (za)--(zi)--(zo);
\draw[orange,very thick] (za)--(ze)--(zo);
\end{tikzpicture}}};
\node at (-10,0) {\scalebox{.9}{ \begin{tikzpicture}
\def\x{4};
\def\y{-3};
\def\xx{7};
\def\yy{-1};
\def\xxx{12};
\def\yyy{-4};
\draw[draw=white,fill=red!10,opacity=.5] (0,0)--(\x,\y)--(\xxx ,\yyy)--(\xx ,\yy )--(0,0);
\node[hasse] (za) at (0,6) {};
\node[hasse] (zba) at (-1,1) {};
\node[hasse] (zbc) at (0,1) {};
\node[hasse] (zbe) at (1,1) {};
\node[hasse] (zc) at (0,0) {};
\draw (za)--(zba)--(zc) (za)--(zbc)--(zc) (za)--(zbe)--(zc);
\draw[transform canvas={xshift=-1.5pt}] (zbc)--(zc);
\draw[transform canvas={xshift=1.5pt}] (zbc)--(zc);
\node[hasse] (zei) at (\x + 1.5,\y +6.5) {};
\node[hasse] (zfa) at (\x +1,\y +1) {};
\node[hasse] (zfjb) at (\x + 0+1.5,\y +1+1.5) {};
\node[hasse] (zfc) at (\x - 1,\y +1) {};
\node[hasse] (zg) at (\x + 0,\y +0) {};
\draw (zei)--(zfa)--(zg) (zei)--(zfjb)--(zg) (zei)--(zfc)--(zg);
\node[hasse] (zja) at (\xx +1,\yy +1) {};
\node[hasse] (zjc) at (\xx - 1,\yy +1) {};
\node[hasse] (zk) at (\xx + 0,\yy +0) {};
\draw (zei)--(zja)--(zk) (zei)--(zfjb)--(zk) (zei)--(zjc)--(zk);
\node[hasse] (zo) at (\xxx +0,\yyy + 4) {};
\node[hasse] (zp) at (\xxx +1,\yyy + 1) {};
\node[hasse] (zq) at (\xxx -1,\yyy + 1) {};
\node[hasse] (zr) at (\xxx + 0,\yyy +0) {};
\draw (zr)--(zp)--(zo)--(zq)--(zr);
\draw[red,very thick] (zc)--(zg)--(zr)--(zk)--(zc);
\draw[red] (zba)--(zfc);
\draw[red] (zbe)--(zfa);
\draw[red] (zbe)--(zja);
\draw[red] (zba)--(zjc);
\draw[red] (zfa)--(zp);
\draw[red] (zfc)--(zq);
\draw[red] (zja)--(zp);
\draw[red] (zjc)--(zq);
\draw[red] (zbc)--(zfjb);
\draw[orange,very thick] (za)--(zei)--(zo);
\end{tikzpicture}}};
\node at (-10,2.6) {\rotatebox{-24}{\textcolor{orange}{Diagram of $\mathrm{Sym}^3(\mathbb{R}^5)$}}};
\end{tikzpicture}
 \end{center}
 \caption{\textit{(Top)} Simplified version of Figure \ref{fig:fullMS3} representing the Hasse diagram of the full moduli space of the theory $k=2$, $n=3$ with respect to the fibration of the Higgs branch over the tensor branch. Red lines are $\mathcal{N}=(1,0)$ tensor transitions (dashed lines for smooth transverse slices), orange lines are $\mathcal{N}=(1,0)$ tensor transitions, black lines are Higgs branch transitions (the type is not specified as it can be read from the brane systems in Figure \ref{fig:fullMS3}). The red shaded area represents the tensor branch, with the origin at the bottom right, and the most generic point at the top left. The orange shaded area is the tensor branch for the 6d $\mathcal{N}=(2,0)$ theory of type $A_{n-1} = A_2$. \textit{(Bottom)} Phase diagram for the same theory. The points in the top diagram representing the same phase have been identified. The 6d $\mathcal{N}=(2,0)$ tensor branch is seen to have the structure of the third symmetric product of $\mathbb{R}^5$, as expected. }
 \label{fig:fullMS3simplified}
 \end{figure}   
   
\FloatBarrier

\subsection{\texorpdfstring{The case $k=2$ and $n \geq 4$}{The case k=2 and n>=4}}

Let us now turn to \eqref{eq:linearquiver} with $n\geq 4$, still with $k=2$. 
We skip the finite coupling discussion and go to infinite coupling at the origin of the $n-1$ dimensional tensor branch straight away, where a new phenomenon occurs. 
We can depict the brane system at a generic point on the Higgs branch, and the magnetic quiver describing its moduli:
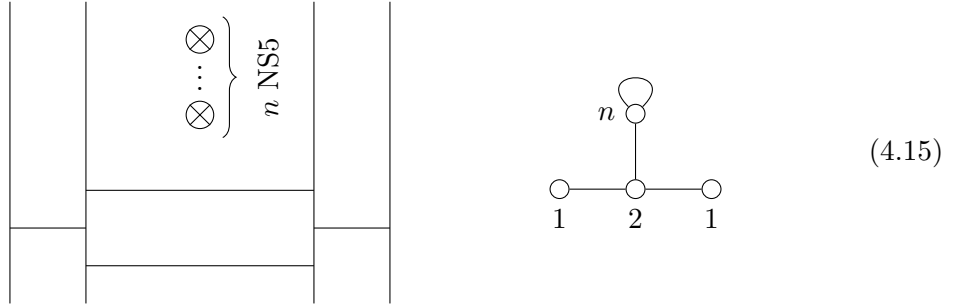
\begin{equation}
        \vcenter{\hbox{\begin{tikzpicture}
        \def\x{2cm};
        \draw (0,-\x)--(0,\x);
        \draw (1,-\x)--(1,\x);
        \draw (4,-\x)--(4,\x);
        \draw (5,-\x)--(5,\x);
        \ns{2.5,1.5};
        \node at (2.5,1.1) {$\vdots$};
        \ns{2.5,0.5};
		\draw [decorate,decoration={brace,amplitude=5pt}] (2.8,1.8)--(2.8,0.2);
		\node[rotate=90] at (3.4,1) {$n$ NS5};
        \draw (0,-1)--(1,-1) (1,-1.5)--(4,-1.5) (1,-0.5)--(4,-0.5) (4,-1)--(5,-1);
    \end{tikzpicture}}}
    \hspace{2cm}
    \vcenter{\hbox{\begin{tikzpicture}
        \node[gauge,label=below:{$1$}] (l) at (0,0) {};
        \node[gauge,label=below:{$2$}] (m) at (1,0) {};
        \node[gauge,label=below:{$1$}] (r) at (2,0) {};
        \node[gauge,label=left:{$n$}] (1) at (1,1) {};
        \draw (1) to [out=45,in=135,looseness=10] (1);
        \draw (l)--(m)--(r) (1)--(m);
    \end{tikzpicture}}}
\end{equation}
There are many possible transitions, and we will focus only on a new type of transition not previously discussed in the physics literature. After making $4 \leq d \leq n$ NS5 branes coincide, we can align them with D6 branes. Suppressing all branes not involved in this elementary transition, we have the following:
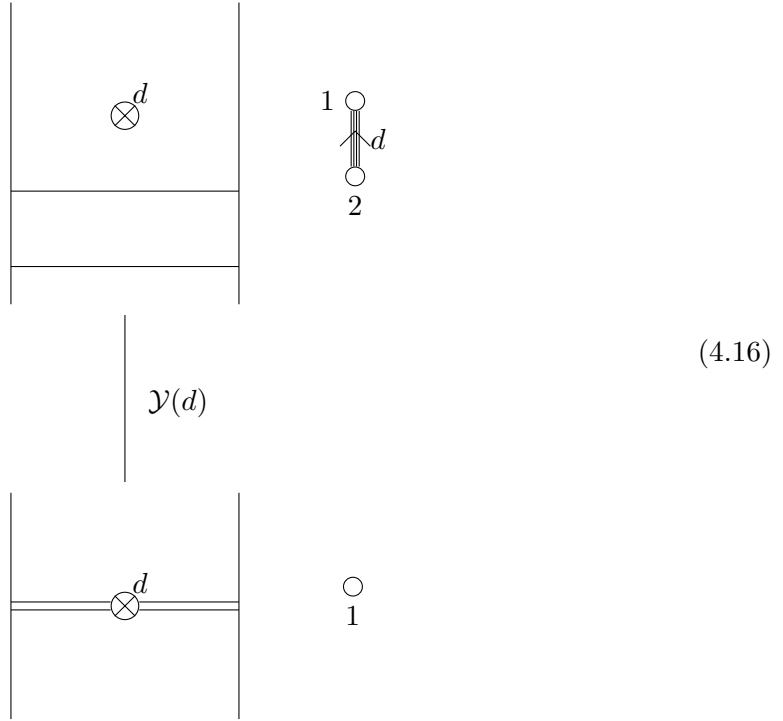
\begin{equation}
\label{eq:Ytransition}
\raisebox{-.5\height}{
\begin{tikzpicture}
    \node (t) at (0,0) {\begin{tikzpicture}
        \def\x{2cm};
        \draw (1,-\x)--(1,\x);
        \draw (4,-\x)--(4,\x);
        \ns{2.5,0.5};
        \node at (2.7,0.8) {$d$};
        \draw (1,-1.5)--(4,-1.5) (1,-0.5)--(4,-0.5);
    \end{tikzpicture}};
        \node (tr) at (3,0) {\begin{tikzpicture}
        \node[gauge,label=below:{$2$}] (0) at (1,0) {};
        \node[gauge,label=left:{$1$}] (1) at (1,1) {};
            \draw[transform canvas={xshift=-1.5pt}] (0)--(1);
            \draw[transform canvas={xshift=-0.5pt}] (0)--(1);
            \draw[transform canvas={xshift=0.5pt}] (0)--(1);
            \draw[transform canvas={xshift=1.5pt}] (0)--(1);
        \draw (0.8,0.4)--(1,0.6)--(1.2,0.4);
        \node at (1.3,0.5) {$d$};
    \end{tikzpicture}};
    \node (b) at (0,-6) {\begin{tikzpicture}
                \def\x{1.5cm};
                \draw (1,-\x)--(1,\x);
                \draw (4,-\x)--(4,\x);
                \ns{2.5,0}l;
                \node at (2.7,0.3) {$d$};
                \draw[transform canvas={yshift=-1.5pt}] (1,0)--(l)--(4,0);
                \draw[transform canvas={yshift=1.5pt}] (1,0)--(l)--(4,0);
            \end{tikzpicture}};
            \node (br) at (3,-6) {\begin{tikzpicture}
                \node[gauge,label=below:{$1$}] (l) at (0,0) {};
            \end{tikzpicture}};
            \draw (t)--(b);
            \node at ($(t)!0.55!(b)+(0.7,0)$) {$\mathcal{Y}(d)$};
\end{tikzpicture}}
\end{equation}
The elementary slice $\mathcal{Y}(d)$ was only recently discovered by mathematicians \cite{2021arXiv211215494B} and is described in Appendix \ref{app:slice}. Note that for $d<4$, the quiver on top of \eqref{eq:Ytransition} has an underbalanced node, and accordingly the transitions that are performed are of type $b_3$ (see \eqref{eq:2,2inf,middle}) or $g_2$ (see \eqref{eq:g2quiver}). 
The Hasse diagram for $k=2,n=4$ is given in Figure \ref{fig:2,4Hasse}. 

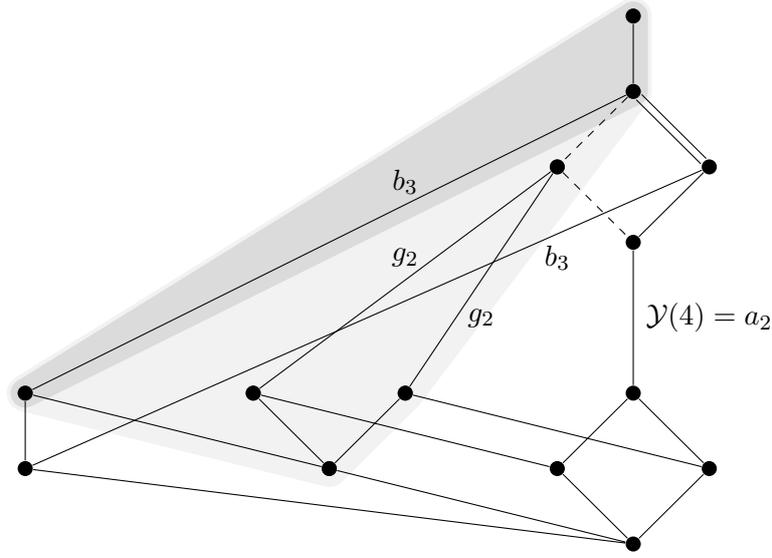
\begin{figure}
    \centering
         \begin{tikzpicture}
        \node[hasse] (1) at (0,0) {};
        \node[hasse] (2) at (0,-1) {};
        \node[hasse] (3) at (-1,-2) {};
        \node[hasse] (4) at (1,-2) {};
        \node[hasse] (5) at (0,-3) {};
        \node[hasse] (6) at (0,-5) {};
        \node[hasse] (7) at (-1,-6) {};
        \node[hasse] (8) at (1,-6) {};
        \node[hasse] (9) at (0,-7) {};
        \node[hasse] (10) at (-3,-5) {};
        \node[hasse] (11) at (-5,-5) {};
        \node[hasse] (12) at (-8,-5) {};
        \node[hasse] (13) at (-4,-6) {};
        \node[hasse] (14) at (-8,-6) {};
        \draw[fill=black!35,opacity=.5,draw=white] \convexpath{1,2,12}{0.2cm};
        \draw[fill=black!10,opacity=.5,draw=white] \convexpath{1,2,10,13,12}{0.25cm};
        \draw (1)--(2) (6)--(7)--(9) (4)--(5) (6)--(8)--(9) (12)--(14)--(9) (12)--(13)--(9) (10)--(13) (11)--(13) (11)--(7) (10)--(8);
        \draw[transform canvas={xshift=-1pt,yshift=-1pt}] (2)--(4);
        \draw[transform canvas={xshift=1pt,yshift=1pt}] (2)--(4);
        \draw[dashed] (2)--(3)--(5);
        \draw (5)--(6);
        \draw (2)--(12) (4)--(14);
        \draw (3)--(11) (3)--(10);
        \node[hasse] at (1) {};
        \node[hasse] at (2) {};
        \node[hasse] at (12) {};
        \node[hasse] at (10) {};
        \node[hasse] at (13) {};
        \node[hasse] at (3) {};
        \node[hasse] at (11) {};
        \node at (1,-4) {$\mathcal{Y}(4)=a_2$};
        \node at (-2,-4) {$g_2$};
        \node at (-3,-3.2) {$g_2$};
        \node at (-1,-3.2) {$b_3$};
        \node at (-3,-2.2) {$b_3$};
    \end{tikzpicture}
    \caption{Higgs branch Hasse diagram for $k=2,n=4$ at the origin of the tensor branch. The height of a dot is proportional to the dimension of the leaf it stands for. As before black lines with no label stand for $A_1$ transitions, and dashed lines for $m$ transitions. In the light gray area we can recognize the Hasse diagram for $k=2,n=3$ and in the dark gray area the Hasse diagram for $k=2,n=2$.   }
    \label{fig:2,4Hasse}
\end{figure}

\subsection{General Case}

As made clear by the previous case studies, the phase diagram for the theory of $n$ M5 branes probing a $\mathbb{C}^2 / \mathbb{Z}_k$ singularity shows an extraordinary complexity as $k$ and $n$ grow. However, this complexity is only combinatorial and the transverse slices -- i.e. the phase transitions -- are fully understood in principle, and can be worked out in any particular example.

Let us begin by studying the theory on the most generic point of the tensor branch, i.e.\ at finite coupling. The Type IIA brane system at the origin of the Higgs branch is
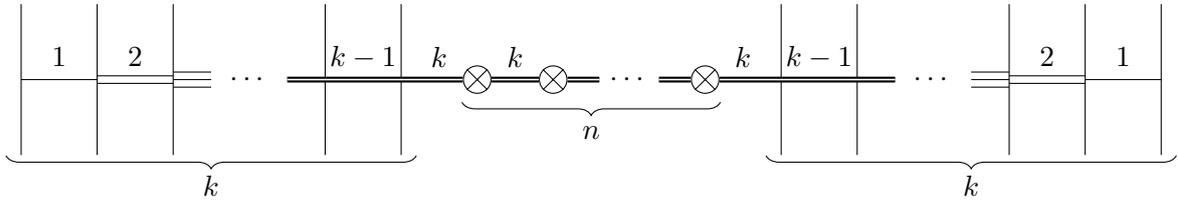
\begin{equation*}
    \vcenter{\hbox{\begin{tikzpicture}
        \def\x{1cm};
        \draw (0,-\x)--(0,\x);
        \draw (1,-\x)--(1,\x);
        \draw (2,-\x)--(2,\x);
        \node at (3,0) {$\cdots$};
        \draw (4,-\x)--(4,\x);
        \draw (5,-\x)--(5,\x);
        \draw (10,-\x)--(10,\x);
        \draw (11,-\x)--(11,\x);
        \node at (12,0) {$\cdots$};
        \draw (13,-\x)--(13,\x);
        \draw (14,-\x)--(14,\x);
        \draw (15,-\x)--(15,\x);
        \ns{6,0}1;
        \ns{7,0}2;
        \node (3) at (8,0) {$\cdots$};
        \ns{9,0}4;
        \draw (0,0)--(1,0) (1,0.05)--(2,0.05) (1,-0.05)--(2,-0.05) (2,0.1)--(2.5,0.1) (2,0)--(2.5,0) (2,-0.1)--(2.5,-0.1);
        \draw[thick,double] (3.5,0)--(4,0)--(5,0)--(1)--(2)--(3)--(4)--(10,0)--(11,0)--(11.5,0);
        \draw (12.5,0.1)--(13,0.1) (12.5,0)--(13,0) (12.5,-0.1)--(13,-0.1) (13,0.05)--(14,0.05) (13,-0.05)--(14,-0.05) (14,0)--(15,0);
        \node at (0.5,0.3) {$1$};
        \node at (1.5,0.3) {$2$};
        \node at (4.5,0.3) {$k-1$};
        \node at (5.5,0.3) {$k$};
        \node at (6.5,0.3) {$k$};
        \node at (9.5,0.3) {$k$};
        \node at (10.5,0.3) {$k-1$};
        \node at (13.5,0.3) {$2$};
        \node at (14.5,0.3) {$1$};
        \draw [decorate,decoration={brace,amplitude=5pt},xshift=0pt,yshift=0pt]
    (9.2,-0.3)--(5.8,-0.3) node [black,midway,yshift=-0.4cm] {$n$};
        \draw [decorate,decoration={brace,amplitude=5pt},xshift=0pt,yshift=0pt]
    (5.2,-\x-0.2)--(-0.2,-\x-0.2) node [black,midway,yshift=-0.4cm] {$k$};
        \draw [decorate,decoration={brace,amplitude=5pt},xshift=0pt,yshift=0pt]
    (15.2,-\x-0.2)--(9.8,-\x-0.2) node [black,midway,yshift=-0.4cm] {$k$};
    \end{tikzpicture}}}
\end{equation*}
The brane system on a generic point of the Higgs branch is
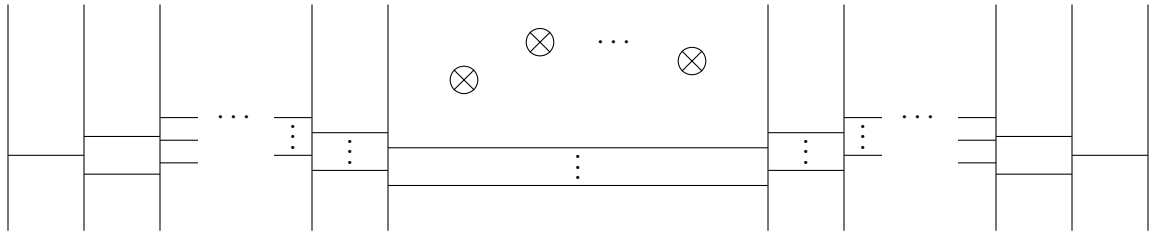
\begin{equation*}
    \vcenter{\hbox{\begin{tikzpicture}
        \def\x{1.5cm};
        \draw (0,-\x)--(0,\x);
        \draw (1,-\x)--(1,\x);
        \draw (2,-\x)--(2,\x);
        \node at (3,0) {$\cdots$};
        \draw (4,-\x)--(4,\x);
        \draw (5,-\x)--(5,\x);
        \draw (10,-\x)--(10,\x);
        \draw (11,-\x)--(11,\x);
        \node at (12,0) {$\cdots$};
        \draw (13,-\x)--(13,\x);
        \draw (14,-\x)--(14,\x);
        \draw (15,-\x)--(15,\x);
        \ns{6,0.5}1;
        \ns{7,1}2;
        \node (3) at (8,1) {$\cdots$};
        \ns{9,0.75}4;
        \draw (0,-0.5)--(1,-0.5) (1,-0.25)--(2,-0.25) (1,-0.75)--(2,-0.75) (2,0)--(2.5,0) (2,-0.3)--(2.5,-0.3) (2,-0.6)--(2.5,-0.6);
        \draw (3.5,0)--(4,0) (3.5,-0.5)--(4,-0.5);
        \node at (3.75,-0.15) {$\vdots$};
        \draw (4,-0.2)--(5,-0.2) (4,-0.7)--(5,-0.7);
        \node at (4.5,-0.35) {$\vdots$};
        \draw (5,-0.4)--(10,-0.4) (5,-0.9)--(10,-0.9);
        \node at (7.5,-0.55) {$\vdots$};
        \draw (10,-0.2)--(11,-0.2) (10,-0.7)--(11,-0.7);
        \node at (10.5,-0.35) {$\vdots$};
        \draw (11,0)--(11.5,0) (11,-0.5)--(11.5,-0.5);
        \node at (11.25,-0.15) {$\vdots$};
        \draw (12.5,0)--(13,0) (12.5,-0.3)--(13,-0.3) (12.5,-0.6)--(13,-0.6) (13,-0.25)--(14,-0.25) (13,-0.75)--(14,-0.75) (14,-0.5)--(15,-0.5);
    \end{tikzpicture}}}
\end{equation*}
The geometry of the Higgs branch is encapsulated in its magnetic quiver, which can be read off from the brane system \cite{Cabrera:2019izd}
\begin{equation}
\raisebox{-.5\height}{\begin{tikzpicture}
        \node[gauge,label=below:{$1$}] (1) at (0,0) {};
        \node[gauge,label=below:{$2$}] (2) at (1,0) {};
        \node (3) at (2,0) {$\cdots$};
        \node[gauge,label=below:{$k-1$}] (4) at (3,0) {};
        \node[gauge,label=below:{$k$}] (5) at (4,0) {};
        \node[gauge,label=below:{$k-1$}] (6) at (5,0) {};
        \node (7) at (6,0) {$\cdots$};
        \node[gauge,label=below:{$2$}] (8) at (7,0) {};
        \node[gauge,label=below:{$1$}] (9) at (8,0) {};
        \draw (1)--(2)--(3)--(4)--(5)--(6)--(7)--(8)--(9);
        \node[gauge,label=left:{$1$}] (5l) at (4-0.5,1) {};
        \node at (4,1) {$\cdots$};
        \node[gauge,label=right:{$1$}] (5r) at (4+0.5,1) {};
        \draw (5)--(5l) (5)--(5r);
        \draw [decorate,
    decoration = {brace}] (4-0.7,1.3) --  (4+0.7,1.3);
    \node at (4,1.6) {$n$};
    \end{tikzpicture}}
\end{equation}

\paragraph{Structure at the origin of the tensor branch. }
At the origin of the tensor branch, one gets the following magnetic quiver, which involves discrete gauging: 
\begin{equation}
 \raisebox{-.5\height}{ \begin{tikzpicture}
        \node[gauge,label=below:{$1$}] (1) at (0,0) {};
        \node[gauge,label=below:{$2$}] (2) at (1,0) {};
        \node (3) at (2,0) {$\cdots$};
        \node[gauge,label=below:{$k-1$}] (4) at (3,0) {};
        \node[gauge,label=below:{$k$}] (5) at (4,0) {};
        \node[gauge,label=below:{$k-1$}] (6) at (5,0) {};
        \node (7) at (6,0) {$\cdots$};
        \node[gauge,label=below:{$2$}] (8) at (7,0) {};
        \node[gauge,label=below:{$1$}] (9) at (8,0) {};
        \draw (1)--(2)--(3)--(4)--(5)--(6)--(7)--(8)--(9);
        \node[gauge,label=right:{$n$}] (5l) at (4,1) {};
        \draw (5l) to [out=45,in=135,looseness=8] (5l);
        \draw (5)--(5l);
    \end{tikzpicture} }
\end{equation}
The Hasse diagram is very complicated, and cannot be depicted in general. A part of it is shown in Figure \ref{fig:HasseDiscreteGauging}. In particular for $k=2$ the magnetic quiver simplifies, and the Hasse diagram contains as a subdiagram 
\begin{equation}
    \raisebox{-.5\height}{ \begin{tikzpicture}[xscale=.03cm,yscale=.03cm]
    \draw[draw=white,fill=black!10] (0,11) ellipse (.3 and 1);
\node[hasse] (2) at (0,4) {};
\node[hasse] (4) at (-2,6) {};
\node[hasse] (3) at (2,6) {};
\node[hasse] (5) at (0,8) {};
\node[hasse] (6) at (0,10) {};
\node[hasse] (7) at (0,12) {};
\draw (5)--(6);
\draw (5)--(4)--(2)--(3)--(5);
\node at (-.6,9) {$\mathcal{Y}(n)$};
\node at (-1.6,7) {$A_1$};
\node at (1.6,7) {$A_1$};
\node at (-1.6,5) {$A_1$};
\node at (1.6,5) {$A_1$};
\node at (-1.9,11) {$\mathrm{Sym}^{n}(\mathbb{C}^2)/\mathbb{C}^2$};
 \end{tikzpicture}}
\end{equation}
where the gray area stands for the diagram of $\mathrm{Sym}^{n}(\mathbb{C}^2)/\mathbb{C}^2$. 
We have not drawn the $b_3$, $g_2$, and $\mathcal{Y}(k)$ with $4\leq k< n$ transitions which are also present in the Hasse diagram. The bottom part of this diagram, drawn in black lines, can be reproduced using the tools from the atomic classification of 6d SCFTs \cite{DelZotto:2014hpa,Heckman:2016ssk,Heckman:2018pqx,Hassler:2019eso,Baume:2021qho}.\footnote{We thank Craig Lawrie for drawing our attention on this agreement. } 

\paragraph{New transitions. }
At a point on an arbitrary locus on the tensor branch, one is confronted with the situation 
\begin{equation}
    \raisebox{-.5\height}{\begin{tikzpicture}
        \def\x{2cm};
        \draw (1,-\x)--(1,\x);
        \draw (4,-\x)--(4,\x);
        \ns{1.5,0.5};
        \node at (1.7,0.8) {$d_1$};
        \ns{2.5,0.5};
        \node at (2.7,0.8) {$d_2$};
        \node at (3.5,0.5) {$\cdots$};
        \draw (1,-1.5)--(4,-1.5) (1,-0.5)--(4,-0.5);
    \end{tikzpicture}}
\end{equation}
characterized by a partition $\mathbf{d} = [d_1 , d_2 ,\dots]$ of an integer $d_1 + d_2 + \dots \geq 4$ (if $d_1 + d_2 + \dots < 4$, the singular transition involves additional D6 branes as shown in Sections \ref{sec:neq2} and \ref{sec:neq3}). If $\mathbf{d}$ is a partition of 4, the transitions already have been worked out in the previous subsection. If $\mathbf{d}$ has a sub-partition that is a partition of an integer $\geq 4$, then a transition will be associated to that sub-partition, and we can focus on that one. This means that \emph{elementary transitions are associated to partitions of integers $\geq 4$ which do not contain a strict sub-partition of a number $\geq 4$}. The full list of such partitions is given in Table \ref{tab:partitions4Irreductible}. Inspecting this table we notice that two entries have not been covered yet. The associated isolated symplectic singularities are dubbed $\mathcal{J}_{3,2}$ and $\mathcal{J}_{3,3}$, and are studied in detail in Appendix \ref{app:Jslices}. 

\begin{table}
\centering 
\begin{tabular}{cc} \hline 
Partition & Transition \\ \hline 
$[1^4]$ & $d_4$ \\ 
$[2,1^2]$ & $b_3$ \\ 
$[2^2]$ & $d_3 = a_3$ \\ 
$[3,1]$ & $g_2$ \\ 
$[4]$ & $\mathcal{Y}(4) = a_2$  \\ \hline 
$[3,2]$ & $\mathcal{J}_{3,2}$  \\ 
$[3^2]$ & $\mathcal{J}_{3,3}$   \\ \hline 
$[d]$ (with $d \geq 4$) & $\mathcal{Y}(d)$  \\ \hline  
\end{tabular}
\caption{List of partitions of integers $\geq 4$ which do not contain a strict sub-partition of a number $\geq 4$. For each such partition we indicate the isolated symplectic singularity. }
\label{tab:partitions4Irreductible}
\end{table}

\begin{figure}
\centering
\begin{tikzpicture}[xscale=.04cm,yscale=.04cm]
\draw[draw=white,fill=black!10] (0,11) ellipse (.3 and 1);
\node (4) at (0,7) {};
\node[hasse] (5) at (0,8) {};
\node[hasse] (6) at (0,10) {};
\node[hasse] (7) at (0,12) {};
\draw (5)--(6);
\draw[dotted] (4)--(5);
\node at (-.6,9) {$\mathcal{Y}(n)$};
\node at (-1.5,11) {$\mathrm{Sym}^{n}(\mathbb{C}^2)/\mathbb{C}^2$};
\node at (5,8) {\raisebox{-.5\height}{   \begin{tikzpicture}
        \node[gauge,label=below:{$1$}] (1) at (0,0) {};
        \node[gauge,label=below:{$2$}] (2) at (1,0) {};
        \node (3) at (2,0) {$\cdots$};
        \node[gauge,label=below:{$k-1$}] (4) at (3,0) {};
        \node[gauge,label=below:{$k-2$}] (5) at (4,0) {};
        \node[gauge,label=below:{$k-1$}] (6) at (5,0) {};
        \node (7) at (6,0) {$\cdots$};
        \node[gauge,label=below:{$2$}] (8) at (7,0) {};
        \node[gauge,label=below:{$1$}] (9) at (8,0) {};
        \draw (1)--(2)--(3)--(4)--(5)--(6)--(7)--(8)--(9);
        \node[gauge,label=right:{$1$}] (5l) at (4,1) {};
        \draw[transform canvas={xshift=-1.5pt}] (5)--(5l);
        \draw[transform canvas={xshift=+1.5pt}] (5)--(5l);
        \draw (5)--(5l);
        \draw[transform canvas={xshift=-1pt,yshift=+1pt}] (4)--(5l);
        \draw[transform canvas={xshift=+1pt,yshift=-1pt}] (4)--(5l);
        \draw[transform canvas={xshift=-1pt,yshift=-1pt}] (6)--(5l);
        \draw[transform canvas={xshift=+1pt,yshift=+1pt}] (6)--(5l);
        \node (10) at (6,.5) {$n-4$};
        \draw[->] (10)--(4.2,.5);
    \end{tikzpicture} }};
\node at (5,10) {\raisebox{-.5\height}{  \begin{tikzpicture}
        \node[gauge,label=below:{$1$}] (1) at (0,0) {};
        \node[gauge,label=below:{$2$}] (2) at (1,0) {};
        \node (3) at (2,0) {$\cdots$};
        \node[gauge,label=below:{$k-1$}] (4) at (3,0) {};
        \node[gauge,label=below:{$k$}] (5) at (4,0) {};
        \node[gauge,label=below:{$k-1$}] (6) at (5,0) {};
        \node (7) at (6,0) {$\cdots$};
        \node[gauge,label=below:{$2$}] (8) at (7,0) {};
        \node[gauge,label=below:{$1$}] (9) at (8,0) {};
        \draw (1)--(2)--(3)--(4)--(5)--(6)--(7)--(8)--(9);
        \node[gauge,label=right:{$1$}] (5l) at (4,1) {};
        \draw[transform canvas={xshift=-1.5pt}] (5)--(5l);
        \draw[transform canvas={xshift=+1.5pt}] (5)--(5l);
        \draw (5)--(5l);
        \draw (4-.2,.5-.2)--(4,.5+.2)--(4+.2,.5-.2);
        \node at (4.5,.5) {$n$};
    \end{tikzpicture}}};
    \node at (5,12) {\raisebox{-.5\height}{ \begin{tikzpicture}
        \node[gauge,label=below:{$1$}] (1) at (0,0) {};
        \node[gauge,label=below:{$2$}] (2) at (1,0) {};
        \node (3) at (2,0) {$\cdots$};
        \node[gauge,label=below:{$k-1$}] (4) at (3,0) {};
        \node[gauge,label=below:{$k$}] (5) at (4,0) {};
        \node[gauge,label=below:{$k-1$}] (6) at (5,0) {};
        \node (7) at (6,0) {$\cdots$};
        \node[gauge,label=below:{$2$}] (8) at (7,0) {};
        \node[gauge,label=below:{$1$}] (9) at (8,0) {};
        \draw (1)--(2)--(3)--(4)--(5)--(6)--(7)--(8)--(9);
        \node[gauge,label=right:{$n$}] (5l) at (4,1) {};
        \draw (5l) to [out=45,in=135,looseness=8] (5l);
        \draw (5)--(5l);
    \end{tikzpicture} }};
\end{tikzpicture}   
 \caption{Part of the beginning of the Hasse diagram for the theory \eqref{eq:linearquiver}, for $n \geq 4$. The gray area represents the Hasse diagram of the symmetric product. In the full diagram, several lines would emerge from that area (see Figure \ref{fig:2,4Hasse}).  }
 \label{fig:HasseDiscreteGauging}
\end{figure}

\section*{Acknowledgements}

We would like to thank Philip Argyres, Amihay Hanany, Daniel Juteau, Craig Lawrie, Paul Levy, Carlo Meneghelli, Travis Schedler, Marcus Sperling, Alexander Thomas and Zhenghao Zhong for stimulating discussions.
We also thank the Aspen Center for Physics, which is supported by National Science Foundation grant PHY-1607611, and especially the organizers and participants of the `Geometrization of S(QFT) in $D\leq6$' workshop for a stimulating environment which triggered some of the work in this paper.

AB is supported by the ERC Consolidator Grant 772408-Stringlandscape, and by the LabEx ENS-ICFP: ANR-10-LABX-0010/ANR-10-IDEX-0001-02 PSL*.
JFG is supported by STFC grants ST/P000762/1 and ST/T000791/1.

\appendix

\section{New Isolated Symplectic Singularities}

\subsection{\texorpdfstring{The slice $\mathcal{Y}(d)$ from \cite{2021arXiv211215494B}}{The slice Y(d)}}
\label{app:slice}
The classification of isolated symplectic singularities is an open problem in mathematics. Recently a new family of normal isolated symplectic singularities of quaternionic dimension 2 was found by a team of mathematicians \cite{2021arXiv211215494B}, which they call $\mathcal{Y}(d)$, $d\geq4$, where $\mathcal{Y}(4)=a_2$. We conjecture
\begin{equation}
\label{eq:Yd}
    \mathcal{C}\left(
        \vcenter{\hbox{
        \begin{tikzpicture}
            \node[gauge,label=below:{$2$}] (0) at (0,0) {};
            \node[gauge,label=below:{$1$}] (1) at (1,0) {};
            \draw[transform canvas={yshift=-1.5pt}] (0)--(1);
            \draw[transform canvas={yshift=-0.5pt}] (0)--(1);
            \draw[transform canvas={yshift=0.5pt}] (0)--(1);
            \draw[transform canvas={yshift=1.5pt}] (0)--(1);
            \draw (0.4,0.2)--(0.6,0)--(0.4,-0.2);
            \node at (0.5,0.4) {$d$};
        \end{tikzpicture}
        }}\right)=\mathcal{Y}(d)\qquad\textnormal{for }d\geq 4
\end{equation}
based on a Hilbert series computation. The (unrefined) Hilbert series of $\mathcal{Y}(d)$ was computed in \cite[eq. (6.4)]{2021arXiv211215494B} and reads
\begin{equation}
    \mathrm{HS}(\mathcal{Y}(d))=\frac{\sum_{i=0}^{d-2}t^{2i}+(d-1)t^{d-2}}{(1-t^2)^2(1-t^{d-2})^2}\;,
\end{equation}
which matches the Hilbert series of the Coulomb branch in \eqref{eq:Yd}.

The global symmetry is SU$(2)$ and the highest weight generating function is\footnote{We thank Amihay Hanany, Daniel Juteau and Zhenghao Zhong for discussions on this Hilbert series and HWG.}
\begin{equation}
    \mathrm{HWG}(\ref{eq:Yd})=\mathrm{PE}[\mu^2t^2+t^4+\mu^d(t^{d-2}+t^d)-\mu^{2d}t^{2d}]\;,
\end{equation}
where $\mu$ denotes the highest weight fugacity of SU$(2)$. $\mathcal{Y}(d)$ is realized as the singularity associated to the brane transition \eqref{eq:Yd}.

\paragraph{$\mathbf{\mathbb{Z}_d}$ quotient of $\mathbf{\mathcal{Y}(d)}$.} As pointed out in \cite{2021arXiv211215494B}, $\mathcal{Y}(d)$ is a $\mathbb{Z}_d$ cover of the slice between the sub-sub-regular $[d-2,2]$ orbit and the regular $[d]$ orbit in the nilpotent cone of $\mathfrak{sl}(d)$, which is also realized as the affine Grassmannian slice $\overline{\left[\mathcal{W}_{\mathrm{SL}_2/\mathbb{Z}_2}\right]}_{[0,d-4]}^{[0,d]}$, which is nothing but the Coulomb branch of U$(2)$ with $d$ fundamental hypers.

\begin{figure}
    \centering
\begin{tikzpicture}
        \node (3) at (7,0) {$
            \begin{tikzpicture}
                \node at (-0.5,-1.5) {$\mathcal{Y}(d)$};
                \node[hasse] (t) at (0,0) {};
                \node[hasse] (b) at (0,-3) {};
                \draw (t)--(b);
            \end{tikzpicture}
        $};
        \node (4) at (12,0) {$
            \begin{tikzpicture}
                \node at (-0.5,-0.75) {$A_{d-1}$};
                \node at (-0.5,-2.25) {$A_{d-3}$};
                \node[hasse] (t) at (0,0) {};
                \node[hasse] (m) at (0,-1.5) {};
                \node[hasse] (b) at (0,-3) {};
                \draw (t)--(m)--(b);
                \draw (0.5,0.2)--(0.7,0.2)--(0.7,-3.2)--(0.5,-3.2);
                \node at (2.2,-1.5) {$\overline{\left[\mathcal{W}_{\mathrm{SL}_2/\mathbb{Z}_2}\right]}_{[0,d-4]}^{[0,d]}$};
            \end{tikzpicture}
        $};
        \draw[->] (3)--(4);
        \node at (9,0.5) {$\mathbb{Z}_d$};
    \end{tikzpicture}
    \caption{$\mathbb{Z}_d$ quotient of $\mathcal{Y}(d)$ visible from the quiver.}
    \label{fig:YdQuot}
\end{figure}
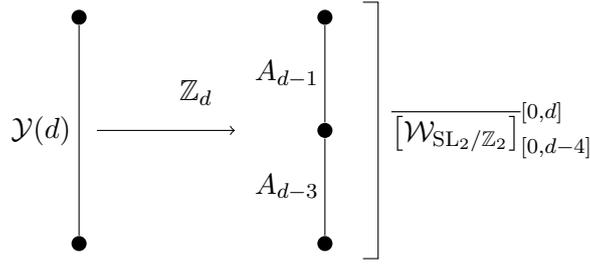

This is immediately visible from the magnetic quiver: promoting the U$(1)$ node of `shortness' $d$ to a long node -- or in other words, changing the ungauging scheme \cite{Hanany:2020jzl} -- the Coulomb branch is quotiented by $\mathbb{Z}_d$. We get
\begin{equation}
    \mathcal{C}\left(
        \vcenter{\hbox{
        \begin{tikzpicture}
            \node[gauge,label=below:{$2$}] (0) at (0,0) {};
            \node[gauge,label=below:{$1$}] (1) at (1,0) {};
            \draw[transform canvas={yshift=-1.5pt}] (0)--(1);
            \draw[transform canvas={yshift=-0.5pt}] (0)--(1);
            \draw[transform canvas={yshift=0.5pt}] (0)--(1);
            \draw[transform canvas={yshift=1.5pt}] (0)--(1);
            \draw (0.4,0.2)--(0.6,0)--(0.4,-0.2);
            \node at (0.5,0.4) {$d$};
        \end{tikzpicture}
        }}\right)\quad \overset{\mathbb{Z}_d}{\longrightarrow} \quad
    \mathcal{C}\left(
    \vcenter{\hbox{
        \begin{tikzpicture}
            \node[gauge,label=below:{$2$}] (0) at (0,0) {};
            \node[gauge,label=below:{$1$}] (1) at (1,0) {};
            \draw[transform canvas={yshift=-1.5pt}] (0)--(1);
            \draw[transform canvas={yshift=-0.5pt}] (0)--(1);
            \draw[transform canvas={yshift=0.5pt}] (0)--(1);
            \draw[transform canvas={yshift=1.5pt}] (0)--(1);
            \node at (0.5,0.4) {$d$};
        \end{tikzpicture}
    }}\right)
    =
    \mathcal{C}\left(
    \vcenter{\hbox{
        \begin{tikzpicture}
            \node[gauge,label=below:{$2$}] (0) at (0,0) {};
            \node[flavour,label=right:{$d$}] (f) at (0,1) {};
            \draw (f)--(0);
        \end{tikzpicture}
        }}\right)\;.
\end{equation}
We depict this quotient in Figure \ref{fig:YdQuot}.

\subsection{\texorpdfstring{The slices $\mathcal{J}_{k_1 , k_2}$}{The slices J}}
\label{app:Jslices}

Consider the singularity  
\begin{equation}
\label{eq:new_maybe}
        \mathcal{J}_{k_1,k_2}=\mathcal{C}\left(
        \vcenter{\hbox{
        \begin{tikzpicture}
            \node[gauge,label=below:{$2$}] (0) at (0,0) {};
            \node[gauge,label=below:{$1$}] (1) at (1,0) {};
            \node[gauge,label=below:{$1$}] (-1) at (-1,0) {};
            \draw[transform canvas={yshift=-1.5pt}] (0)--(1);
            \draw[transform canvas={yshift=-0.5pt}] (0)--(1);
            \draw[transform canvas={yshift=0.5pt}] (0)--(1);
            \draw[transform canvas={yshift=1.5pt}] (0)--(1);
            \draw (0.4,0.2)--(0.6,0)--(0.4,-0.2);
            \node at (0.5,0.4) {$k_2$};
            \draw[transform canvas={yshift=-1.5pt}] (0)--(-1);
            \draw[transform canvas={yshift=-0.5pt}] (0)--(-1);
            \draw[transform canvas={yshift=0.5pt}] (0)--(-1);
            \draw[transform canvas={yshift=1.5pt}] (0)--(-1);
            \draw (-0.4,0.2)--(-0.6,0)--(-0.4,-0.2);
            \node at (-0.5,0.4) {$k_1$};
        \end{tikzpicture}
        }}\right)
\end{equation}
Which is realized for example as the singularity associated to the transition
\begin{equation}
\raisebox{-.5\height}{
    \begin{tikzpicture}
        \node at (0,0) {$\begin{tikzpicture}
                \def\x{1.5cm};
                \draw (0,-\x)--(0,\x);
                \draw (5,-\x)--(5,\x);
                \ns{1.5,0.5}l;
                \node at (1.7,0.8) {$k_1$};
                \ns{3.5,0}r;
                \node at (3.7,0.3) {$k_2$};
                \draw (0,-0.8)--(5,-0.8) (0,-1.2)--(5,-1.2);
            \end{tikzpicture}$};
        \node at (0,-5) {$\begin{tikzpicture}
                \def\x{1.5cm};
                \draw (0,-\x)--(0,\x);
                \draw (5,-\x)--(5,\x);
                \ns{1.5,0}l;
                \node at (1.7,0.3) {$k_1$};
                \ns{3.5,0}r;
                \node at (3.7,0.3) {$k_2$};
                \draw[transform canvas={yshift=-1.5pt}] (0,0)--(l)--(r)--(5,0);
                \draw[transform canvas={yshift=1.5pt}] (0,0)--(l)--(r)--(5,0);
            \end{tikzpicture}$};
        \draw (0,-2)--(0,-4);
        \node at (0.7,-3) {$\mathcal{J}_{k_1,k_2}$};
    \end{tikzpicture}}
\end{equation}
in the brane system.

The Hilbert series for $\mathcal{J}_{k_1 , k_2}$ (with $k_1 , k_2
\geq 1$ and $(k_1 , k_2) \neq (1,1)$) is 
\begin{equation}
    \frac{\frac{(2 k-3) t^{k-2}+(2 k-3) t^{2 k}+(1-2 k) t^{2 k-2}+t^{3 k-4}+t^{3 k-2}+(1-2 k)
   t^k+t^2+1}{\left(t^2-1\right)^2}+(k_1-1) (k_2-1) t^{k-4} \left(t^k+t^2\right)}{\left(1-t^2\right)^3
   \left(1-t^{k-2}\right)^3}
\end{equation}
where $k = k_1  + k_2$.

We have quiver subtraction and brane transitions at our disposal to compute the Hasse diagrams for various $k_1$ and $k_2$, summarized in Table \ref{tab:JHasses}. As argued in Table \ref{tab:partitions4Irreductible}, for two specific values, $(k_1,k_2)=(3,3)$ and $(k_1,k_2)=(2,3)\sim(3,2)$, there is no known quiver to subtract. Furthermore the S-rule implies that there is a single transition in the brane system. This indicates, that $\mathcal{J}_{3,3}$ and $\mathcal{J}_{2,3}=\mathcal{J}_{3,2}$ are elementary slices.

From the magnetic quiver is it clear that these Coulomb branches have a $\mathbb{Z}_{k_1}$ and $\mathbb{Z}_{k_2}$ quotients by promoting the U$(1)$ node of `shortness' $k_i$ to a long node -- or in other words, changing the ungauging scheme \cite{Hanany:2020jzl}.

\begin{table}
    \centering
    \begin{tabular}{c}
        $\begin{tikzpicture}
        \node at (-0.25,0.75) {$k_2$};
        \node at (-0.75,0.25) {$k_1$};
        \draw (-1,1)--(0,0);
        \draw (-1,1)--(-1,-9) (0,1)--(0,-9) (3,1)--(3,-9) (5,1)--(5,-9) (7,1)--(7,-9) (9,1)--(9,-9);
        \draw (-1,1)--(9,1) (-1,0)--(9,0) (-1,-3)--(9,-3) (-1,-5)--(9,-5) (-1,-7)--(9,-7) (-1,-9)--(9,-9);
        \node at (-0.5,-1.5) {$\geq4$};
        \node at (-0.5,-4) {$3$};
        \node at (-0.5,-6) {$2$};
        \node at (-0.5,-8) {$1$};
        \node at (1.5,0.5) {$\geq4$};
        \node at (4,0.5) {$3$};
        \node at (6,0.5) {$2$};
        \node at (8,0.5) {$1$};
        \node at (1.5,-1.5) {$
            \begin{tikzpicture}
                \node[hasse] (0) at (0,0) {};
                \node[hasse] (-1) at (-0.5,-1) {};
                \node[hasse] (1) at (0.5,-1) {};
                \node[hasse] (2) at (0,-2) {};
                \draw (0)--(-1)--(2)--(1)--(0);
                \node at (-0.75,-0.25) {$\mathcal{Y}(k_1)$};
                \node at (0.75,-0.25) {$\mathcal{Y}(k_2)$};
                \node at (-0.75,-1.75) {$A_1$};
                \node at (0.75,-1.75) {$A_1$};
                \node at (0,-2.5) {su$(2)^2$};
            \end{tikzpicture}
        $};
        \node at (4,-1.5) {$
            \begin{tikzpicture}
                \node[hasse] (0) at (0,0) {};
                \node[hasse] (1) at (0,-1) {};
                \node[hasse] (2) at (0,-2) {};
                \draw (2)--(1)--(0);
                \node at (-0.5,-0.5) {$\mathcal{Y}(k_1)$};
                \node at (-0.5,-1.5) {$A_1$};
                \node at (0,-2.5) {su$(2)^2$};
            \end{tikzpicture}
        $};
        \node at (6,-1.5) {$
            \begin{tikzpicture}
                \node[hasse] (0) at (0,0) {};
                \node[hasse] (1) at (0,-1) {};
                \node[hasse] (2) at (0,-2) {};
                \draw (2)--(1)--(0);
                \node at (-0.5,-0.5) {$\mathcal{Y}(k_1)$};
                \node at (-0.5,-1.5) {$A_1$};
                \node at (0,-2.5) {su$(2)^2$};
            \end{tikzpicture}
        $};
        \node at (8,-1.5) {$
            \begin{tikzpicture}
                \node[hasse] (0) at (0,0) {};
                \node[hasse] (1) at (0,-1) {};
                \node[hasse] (2) at (0,-2) {};
                \draw (2)--(1)--(0);
                \node at (-0.5,-0.5) {$\mathcal{Y}(k_1)$};
                \node at (-0.5,-1.5) {$A_1$};
                \node at (0,-2.5) {su$(2)^2$};
            \end{tikzpicture}
        $};
        \node at (4,-4) {$
            \begin{tikzpicture}
                \node[hasse] (0) at (0,0) {};
                \node[hasse] (1) at (0,-1) {};
                \draw (0)--(1);
                \node at (-0.5,-0.5) {$\mathcal{J}_{33}$};
                \node at (0,-1.5) {su$(2)^2$};
            \end{tikzpicture}
        $};
        \node at (6,-4) {$
            \begin{tikzpicture}
                \node[hasse] (0) at (0,0) {};
                \node[hasse] (1) at (0,-1) {};
                \draw (0)--(1);
                \node at (-0.5,-0.5) {$\mathcal{J}_{32}$};
                \node at (0,-1.5) {su$(2)^2$};
            \end{tikzpicture}
        $};
        \node at (8,-4) {$
            \begin{tikzpicture}
                \node[hasse] (0) at (0,0) {};
                \node[hasse] (1) at (0,-1) {};
                \draw (0)--(1);
                \node at (-0.5,-0.5) {$g_2$};
                \node at (0,-1.5) {G${}_2$};
            \end{tikzpicture}
        $};
        \node at (6,-6) {$
            \begin{tikzpicture}
                \node[hasse] (0) at (0,0) {};
                \node[hasse] (1) at (0,-1) {};
                \draw (0)--(1);
                \node at (-0.5,-0.5) {$d_3$};
                \node at (0,-1.5) {so$(6)$};
            \end{tikzpicture}
        $};
        \node at (8,-6) {$
            \begin{tikzpicture}
                \node[hasse] (0) at (0,0) {};
                \node at (-0.5,0) {$\mathbb{H}^3$};
                \node at (0,-0.5) {sp$(3)$};
            \end{tikzpicture}
        $};
        \node at (8,-8) {$
            \begin{tikzpicture}
                \node at (0,-0.5) {bad};
            \end{tikzpicture}
        $};
    \end{tikzpicture}$ 
    \end{tabular}
    \caption{We depict the Hasse diagram and the global symmetry of the slices $\mathcal{J}_{k_1,k_2}$ for various values of $k_1$ and $k_2$. It is symmetric in $k_1$ and $k_2$.}
    \label{tab:JHasses}
\end{table}

\subsubsection{\texorpdfstring{The slice $\mathcal{J}_{33}$}{The slice J33}}
Let
\begin{equation}
    \mathcal{J}_{33}=\mathcal{C}\left(
        \vcenter{\hbox{
        \begin{tikzpicture}
            \node[gauge,label=below:{$2$}] (0) at (0,0) {};
            \node[gauge,label=below:{$1$}] (1) at (1,0) {};
            \node[gauge,label=below:{$1$}] (-1) at (-1,0) {};
            \draw[transform canvas={yshift=-2pt}] (0)--(1);
            \draw[transform canvas={yshift=0pt}] (0)--(1);
            \draw[transform canvas={yshift=2pt}] (0)--(1);
            \draw (0.4,0.2)--(0.6,0)--(0.4,-0.2);
            \node at (0.5,0.4) {$3$};
            \draw[transform canvas={yshift=-2pt}] (0)--(-1);
            \draw[transform canvas={yshift=0pt}] (0)--(-1);
            \draw[transform canvas={yshift=2pt}] (0)--(-1);
            \draw (-0.4,0.2)--(-0.6,0)--(-0.4,-0.2);
            \node at (-0.5,0.4) {$3$};
        \end{tikzpicture}
        }}
    \right)
\end{equation}
with Hilbert series
\begin{equation}
    \mathrm{HS}\left(\mathcal{J}_{33}\right)=\frac{1+3t^2+18t^4+14t^6+18t^8+3t^{10}+t^{12}}{(1-t^2)^3(1-t^4)^3}\;.
\end{equation}
The character expansion of the refined Hilbert series reads 
\begin{equation}
    \begin{split}
        \mathrm{HS}\left(\mathcal{J}_{33}\right)=&1+([2,0]+[0,2])t^2+([3,3]+[2,2]+[4,0]+[0,4]+[0,0])t^4\\
        &+([5,3]+[3,5]+[6,0]+[0,6]+[4,2]+[2,4]+[3,3]+[2,0]+[0,2])t^6\\
        &+([6,6]+[7,3]+[3,7]+[5,5]+[8,0]+[0,8]+[6,2]+[2,6]+[5,3]+[3,5]\\
        &+[4,4]+[3,3]+[4,0]+[0,4]+[2,2]+[0,0])t^8+\dots\;,
    \end{split}
\end{equation}
and the plethystic logarithm of the refined Hilbert series reads
\begin{equation}
    \begin{split}
        \mathrm{PL}\left(\mathrm{HS}\left(\mathcal{J}_{33}\right)\right)&=([2,0]+[0,2])t^2+([3,3]-[0,0])t^4+(-[3,3]-[3,1]-[1,3])t^6\\
        &+([3,3]+[3,1]+[1,3]+[1,1]-[6,2]-[2,6]-[4,4]-[4,0]-[0,4]-[2,2]-[0,0])t^8\;.
    \end{split}
\end{equation}
From this we gather that the global symmetry is $\mathrm{SO}(4)=\mathrm{Spin}(4)/\mathbb{Z}_2$, where the $\mathbb{Z}_2$ is the diagonal subgroup of the $\mathbb{Z}_2\times\mathbb{Z}_2$ center of $\mathrm{Spin}(4)$.

\paragraph{$\mathbf{\mathbb{Z}_3}$ quotient of $\mathbf{\mathcal{J}_{33}}$.}
Note that $\mathcal{J}_{33}$ is a $\mathbb{Z}_3$ cover of the affine Grassmannian slice $\overline{\left[\mathcal{W}_{G_2}\right]}_{[0,2]}^{[0,3]}$ with magnetic quiver \cite{Bourget:2021siw}:
\begin{equation}
    \label{eq:Z3quot}
    \vcenter{\hbox{
        \begin{tikzpicture}
            \node[gauge,label=below:{$2$}] (0) at (0,0) {};
            \node[gauge,label=below:{$1$}] (1) at (1,0) {};
            \node[gauge,label=below:{$1$}] (-1) at (-1,0) {};
            \draw[transform canvas={yshift=-2pt}] (0)--(1);
            \draw[transform canvas={yshift=0pt}] (0)--(1);
            \draw[transform canvas={yshift=2pt}] (0)--(1);
            \node at (0.5,0.4) {$3$};
            \draw[transform canvas={yshift=-2pt}] (0)--(-1);
            \draw[transform canvas={yshift=0pt}] (0)--(-1);
            \draw[transform canvas={yshift=2pt}] (0)--(-1);
            \draw (-0.4,0.2)--(-0.6,0)--(-0.4,-0.2);
            \node at (-0.5,0.4) {$3$};
        \end{tikzpicture}
    }}
    =
    \vcenter{\hbox{
        \begin{tikzpicture}
            \node[gauge,label=below:{$2$}] (0) at (0,0) {};
            \node[gauge,label=below:{$1$}] (1) at (1,0) {};
            \node[gauge,label=below:{$1$}] (-1) at (-1,0) {};
            \draw[transform canvas={yshift=-2pt}] (0)--(1);
            \draw[transform canvas={yshift=0pt}] (0)--(1);
            \draw[transform canvas={yshift=2pt}] (0)--(1);
            \draw (0.4,0.2)--(0.6,0)--(0.4,-0.2);
            \node at (0.5,0.4) {$3$};
            \draw[transform canvas={yshift=-2pt}] (0)--(-1);
            \draw[transform canvas={yshift=0pt}] (0)--(-1);
            \draw[transform canvas={yshift=2pt}] (0)--(-1);
            \node at (-0.5,0.4) {$3$};
        \end{tikzpicture}
    }}
    =
    \vcenter{\hbox{
        \begin{tikzpicture}
            \node[gauge,label=below:{$2$}] (0) at (0,0) {};
            \node[gauge,label=below:{$1$}] (-1) at (-1,0) {};
            \node[flavour,label=right:{$3$}] (f) at (0,1) {};
            \draw[transform canvas={yshift=-2pt}] (0)--(-1);
            \draw[transform canvas={yshift=0pt}] (0)--(-1);
            \draw[transform canvas={yshift=2pt}] (0)--(-1);
            \draw (-0.4,0.2)--(-0.6,0)--(-0.4,-0.2);
            \node at (-0.5,0.4) {$3$};
            \draw (f)--(0);
        \end{tikzpicture}
        }}
\end{equation}
and Hilbert series
\begin{equation}
    \mathrm{HS}\left(\overline{\left[\mathcal{W}_{G_2}\right]}_{[0,2]}^{[0,3]}\right)=\frac{1 + 2 t^2 + 8 t^4 + 11 t^6 + 14 t^8 + 11 t^{10} + 8 t^{12} + 2 t^{14} + t^{16}}{(1 - t^2)^2 (1 - t^4)^3 (1 - t^6)}\;.
\end{equation}
Indeed we have $\frac{\mathrm{Vol}\left(\mathcal{J}_{33}\right)}{\mathrm{Vol}\left(\overline{\left[\mathcal{W}_{G_2}\right]}_{[0,2]}^{[0,3]}\right)}=3$ indicating the $\mathbb{Z}_3$ quotient. From the magnetic quiver of $\mathcal{J}_{33}$ we can see, that this $\mathbb{Z}_3$ quotient can be taken in two different (but isomorphic) ways, indicated by the two unframed quivers in \eqref{eq:Z3quot}.

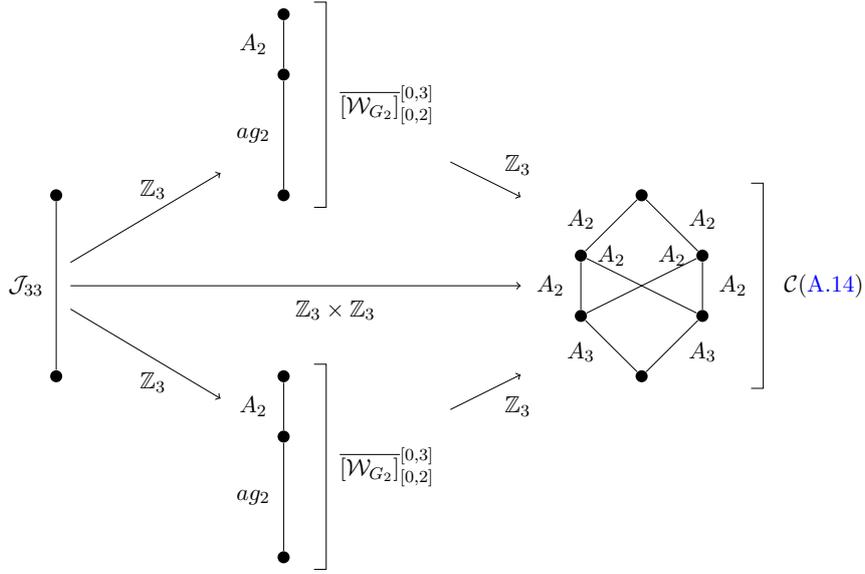
\begin{figure}
    \centering
        \scalebox{0.8}{\begin{tikzpicture}
        \node (3) at (7,0) {$
            \begin{tikzpicture}
                \node at (-0.5,-1.5) {$\mathcal{J}_{33}$};
                \node[hasse] (t) at (0,0) {};
                \node[hasse] (b) at (0,-3) {};
                \draw (t)--(b);
            \end{tikzpicture}
        $};
        \node (4) at (12,3) {$
            \begin{tikzpicture}
                \node at (-0.5,-0.5) {$A_2$};
                \node at (-0.5,-2) {$ag_2$};
                \node[hasse] (t) at (0,0) {};
                \node[hasse] (m) at (0,-1) {};
                \node[hasse] (b) at (0,-3) {};
                \draw (t)--(m)--(b);
                \draw (0.5,0.2)--(0.7,0.2)--(0.7,-3.2)--(0.5,-3.2);
                \node at (1.7,-1.5) {$\overline{\left[\mathcal{W}_{G_2}\right]}_{[0,2]}^{[0,3]}$};
            \end{tikzpicture}
        $};
        \node (5) at (12,-3) {$
            \begin{tikzpicture}
                \node at (-0.5,-0.5) {$A_2$};
                \node at (-0.5,-2) {$ag_2$};
                \node[hasse] (t) at (0,0) {};
                \node[hasse] (m) at (0,-1) {};
                \node[hasse] (b) at (0,-3) {};
                \draw (t)--(m)--(b);
                \draw (0.5,0.2)--(0.7,0.2)--(0.7,-3.2)--(0.5,-3.2);
                \node at (1.7,-1.5) {$\overline{\left[\mathcal{W}_{G_2}\right]}_{[0,2]}^{[0,3]}$};
            \end{tikzpicture}
        $};
        \node (6) at (18,0) {$
            \begin{tikzpicture}
                \node at (-1,-0.4) {$A_2$};
                \node at (-1.5,-1.5) {$A_2$};
                \node at (-1,-2.6) {$A_3$};
                \node at (1,-0.4) {$A_2$};
                \node at (1.5,-1.5) {$A_2$};
                \node at (1,-2.6) {$A_3$};
                \node at (-.5,-1) {$A_2$};
                \node at (.5,-1) {$A_2$};
                \node[hasse] (10) at (0,0) {};
                \node[hasse] (21) at (-1,-1) {};
                \node[hasse] (22) at (1,-1) {};
                \node[hasse] (31) at (-1,-2) {};
                \node[hasse] (32) at (1,-2) {};
                \node[hasse] (40) at (0,-3) {};
                \draw (10)--(21)--(31)--(40)--(32)--(22)--(10) (21)--(32) (31)--(22);
                \draw (1.8,0.2)--(2,0.2)--(2,-3.2)--(1.8,-3.2);
                \node at (3,-1.5) {$\mathcal{C}\eqref{eq:Z33quot}$};
            \end{tikzpicture}
        $};
        \draw[->] (3)--(4);
        \draw[->] (3)--(5);
        \draw[->] (4)--(6);
        \draw[->] (5)--(6);
        \draw[->] (3)--(6);
        \node at (9,1.6) {$\mathbb{Z}_3$};
        \node at (9,-1.6) {$\mathbb{Z}_3$};
        \node at (15,2) {$\mathbb{Z}_3$};
        \node at (15,-2) {$\mathbb{Z}_3$};
        \node at (12,-.4) {$\mathbb{Z}_3\times\mathbb{Z}_3$};
    \end{tikzpicture}}
    \caption{Quotients of $\mathcal{J}_{33}$ which are visible from the quiver.}
    \label{fig:newSliceQuotients2}
\end{figure}

\paragraph{$\mathbf{\mathbb{Z}_3\times\mathbb{Z}_3}$ quotient of $\mathbf{\mathcal{J}_{33}}$.}
After the first $\mathbb{Z}_3$ quotient one can do a further $\mathbb{Z}_3$ quotient with magnetic quiver
\begin{equation}
\label{eq:Z33quot}
    \vcenter{\hbox{
        \begin{tikzpicture}
            \node[gauge,label=below:{$2$}] (0) at (0,0) {};
            \node[gauge,label=below:{$1$}] (-1) at (-1,0) {};
            \node[gauge,label=below:{$1$}] (1) at (1,0) {};
            \draw[transform canvas={yshift=-2pt}] (0)--(-1);
            \draw[transform canvas={yshift=0pt}] (0)--(-1);
            \draw[transform canvas={yshift=2pt}] (0)--(-1);
            \node at (-0.5,0.4) {$3$};
            \draw[transform canvas={yshift=-2pt}] (0)--(1);
            \draw[transform canvas={yshift=0pt}] (0)--(1);
            \draw[transform canvas={yshift=2pt}] (0)--(1);
            \node at (0.5,0.4) {$3$};
        \end{tikzpicture}
        }}
\end{equation}
and Hilbert series
\begin{equation}
    \mathrm{HS}\left(\mathcal{C}\eqref{eq:Z33quot}\right)=\frac{1 + t^2 + 3 t^4 + 9 t^6 + 11 t^8 + 8 t^{10} + 11 t^{12} + 9 t^{14} + 3 t^{16} + t^{18} + t^{20}}{(1 - t^2) (1 - t^4)^3 (1 - t^6)^2}\;.
\end{equation}
Indeed we have $\frac{\mathrm{Vol}\left(\overline{\left[\mathcal{W}_{G_2}\right]}_{[0,2]}^{[0,3]}\right)}{\mathrm{Vol}\left(\mathcal{C}\eqref{eq:Z33quot}\right)}=3$ indicating the second $\mathbb{Z}_3$ quotient, and $\frac{\mathrm{Vol}\left(\mathcal{J}_{33}\right)}{\mathrm{Vol}\left(\mathcal{C}\eqref{eq:Z33quot}\right)}=9$ indicating a quotient of order $9$ which we deduce to be $\mathbb{Z}_3\times\mathbb{Z}_3$ based on the commutative diagram in Figure \ref{fig:newSliceQuotients2}.

\subsubsection{\texorpdfstring{The slice $\mathcal{J}_{32}=\mathcal{J}_{23}$}{The slice J32}}
\begin{equation}
    \mathcal{J}_{32}=\mathcal{C}\left(
        \vcenter{\hbox{
        \begin{tikzpicture}
            \node[gauge,label=below:{$2$}] (0) at (0,0) {};
            \node[gauge,label=below:{$1$}] (1) at (1,0) {};
            \node[gauge,label=below:{$1$}] (-1) at (-1,0) {};
            \draw[transform canvas={yshift=-1.5pt}] (0)--(1);
            \draw[transform canvas={yshift=1.5pt}] (0)--(1);
            \draw (0.4,0.2)--(0.6,0)--(0.4,-0.2);
            \node at (0.5,0.4) {$2$};
            \draw[transform canvas={yshift=-2pt}] (0)--(-1);
            \draw[transform canvas={yshift=0pt}] (0)--(-1);
            \draw[transform canvas={yshift=2pt}] (0)--(-1);
            \draw (-0.4,0.2)--(-0.6,0)--(-0.4,-0.2);
            \node at (-0.5,0.4) {$3$};
        \end{tikzpicture}
        }}
    \right)
\end{equation}
with Hilbert series
\begin{equation}
    \mathrm{HS}\left(\mathcal{J}_{32}\right)=\frac{1+3t^2+9t^3+5t^4+5t^5+9t^6+3t^7+t^9}{(1-t^2)^3(1-t^3)^3}\;.
\end{equation}
The character expansion of the refined Hilbert series reads
\begin{equation}
    \begin{split}
        \mathrm{HS}\left(\mathcal{J}_{23}\right)=&1+([2,0]+[0,2])t^2+([2,3])t^3+([2,2]+[4,0]+[0,4]+[0,0])t^4\\
        &+([2,5]+[4,3]+[2,4])t^5\\
        &+([4,6]+[4,2]+[2,4]+[6,0]+[0,6]+[2,0]+[0,2])t^6\\
        &+([2,7]+[6,3]+[4,5]+[2,5]+[4,3]+[2,3])t^7+\dots\;,
    \end{split}
\end{equation}
and the plethystic logarithm of the refined Hilbert series reads
\begin{equation}
    \begin{split}
        \mathrm{PL}\left(\mathrm{HS}\left(\mathcal{J}_{23}\right)\right)&=([2,0]+[0,2])t^2+([2,3])t^3-([0,0])t^4-([2,3]+[2,1]+[0,3])t^5\\
        &-([2,4]+[4,2]+[2,0]+[0,6]+[0,2])t^6\\
        &+([2,3]+[2,1]+[0,3]+[0,1])t^7+\dots\;.
    \end{split}
\end{equation}
From this we gather that the global symmetry is $\mathrm{SemiSpin}(4)=\mathrm{Spin}(4)/\mathbb{Z}_2$, where the $\mathbb{Z}_2$ is one of the factors in the $\mathbb{Z}_2\times\mathbb{Z}_2$ center of $\mathrm{Spin}(4)$.

\paragraph{$\mathbf{\mathbb{Z}_2}$ quotient of $\mathbf{\mathcal{J}_{32}}$.}
$\mathcal{J}_{32}$ is a $\mathbb{Z}_2$ cover of the affine Grassmannian slice $\overline{\left[\mathcal{W}_{G_2}\right]}_{[0,1]}^{[0,2]}$ with magnetic quiver\cite{Bourget:2021siw}:
\begin{equation}
        \vcenter{\hbox{
        \begin{tikzpicture}
            \node[gauge,label=below:{$2$}] (0) at (0,0) {};
            \node[gauge,label=below:{$1$}] (1) at (1,0) {};
            \node[gauge,label=below:{$1$}] (-1) at (-1,0) {};
            \draw[transform canvas={yshift=-1.5pt}] (0)--(1);
            \draw[transform canvas={yshift=1.5pt}] (0)--(1);
            \node at (0.5,0.4) {$2$};
            \draw[transform canvas={yshift=-2pt}] (0)--(-1);
            \draw[transform canvas={yshift=0pt}] (0)--(-1);
            \draw[transform canvas={yshift=2pt}] (0)--(-1);
            \draw (-0.4,0.2)--(-0.6,0)--(-0.4,-0.2);
            \node at (-0.5,0.4) {$3$};
        \end{tikzpicture}
        }}
        =
    \vcenter{\hbox{
        \begin{tikzpicture}
            \node[gauge,label=below:{$2$}] (0) at (0,0) {};
            \node[gauge,label=below:{$1$}] (-1) at (-1,0) {};
            \node[flavour,label=right:{$2$}] (f) at (0,1) {};
            \draw[transform canvas={yshift=-2pt}] (0)--(-1);
            \draw[transform canvas={yshift=0pt}] (0)--(-1);
            \draw[transform canvas={yshift=2pt}] (0)--(-1);
            \draw (-0.4,0.2)--(-0.6,0)--(-0.4,-0.2);
            \node at (-0.5,0.4) {$3$};
            \draw (f)--(0);
        \end{tikzpicture}
        }}
\end{equation}
and Hilbert series
\begin{equation}
    \mathrm{HS}\left(\overline{\left[\mathcal{W}_{G_2}\right]}_{[0,1]}^{[0,2]}\right)=\frac{1 + 2 t^2 + 5 t^3 + 4 t^4 + 6 t^5 + 6 t^6 + 4 t^7 + 5 t^8 + 
 2 t^9 + t^{11}}{(1 - t^2)^2 (1 - t^3)^3 (1 - t^4)}\;.
\end{equation}
Indeed we have 
\begin{equation}
    \frac{\mathrm{Vol}\left(\mathcal{J}_{23}\right)}{\mathrm{Vol}\left(\overline{\left[\mathcal{W}_{G_2}\right]}_{[0,1]}^{[0,2]}\right)}=2 \, , 
\end{equation}
indicating the $\mathbb{Z}_2$ quotient.

\paragraph{$\mathbf{\mathbb{Z}_3}$ quotient of $\mathbf{\mathcal{J}_{32}}$.}
$\mathcal{J}_{32}$ is also a $\mathbb{Z}_3$ cover of the affine Grassmannian slice $\overline{\left[\mathcal{W}_{\mathrm{Sp}_2/\mathbb{Z}_2}\right]}_{[0,1]}^{[0,3]}$ with magnetic quiver\cite{Bourget:2021siw}:
\begin{equation}
        \vcenter{\hbox{
        \begin{tikzpicture}
            \node[gauge,label=below:{$2$}] (0) at (0,0) {};
            \node[gauge,label=below:{$1$}] (1) at (1,0) {};
            \node[gauge,label=below:{$1$}] (-1) at (-1,0) {};
            \draw[transform canvas={yshift=-1.5pt}] (0)--(1);
            \draw[transform canvas={yshift=1.5pt}] (0)--(1);
            \draw (0.4,0.2)--(0.6,0)--(0.4,-0.2);
            \node at (0.5,0.4) {$2$};
            \draw[transform canvas={yshift=-2pt}] (0)--(-1);
            \draw[transform canvas={yshift=0pt}] (0)--(-1);
            \draw[transform canvas={yshift=2pt}] (0)--(-1);
            \node at (-0.5,0.4) {$3$};
        \end{tikzpicture}
        }}
        =
    \vcenter{\hbox{
        \begin{tikzpicture}
            \node[gauge,label=below:{$2$}] (0) at (0,0) {};
            \node[gauge,label=below:{$1$}] (-1) at (-1,0) {};
            \node[flavour,label=right:{$3$}] (f) at (0,1) {};
            \draw[transform canvas={yshift=-1.5pt}] (0)--(-1);
            \draw[transform canvas={yshift=1.5pt}] (0)--(-1);
            \draw (-0.4,0.2)--(-0.6,0)--(-0.4,-0.2);
            \node at (-0.5,0.4) {$2$};
            \draw (f)--(0);
        \end{tikzpicture}
        }}
\end{equation}
and Hilbert series
\begin{equation}
    \mathrm{HS}\left(\overline{\left[\mathcal{W}_{\mathrm{Sp}_2/\mathbb{Z}_2}\right]}_{[0,1]}^{[0,3]}\right)=\frac{1 + t^2 + 3 t^3 + t^4 + t^5 + 3 t^6 + t^7 + t^9}{(1 - t^2)^3 (1 - t^3)^3}\;.
\end{equation}
Indeed we have 
\begin{equation}
    \frac{\mathrm{Vol}\left(\mathcal{J}_{23}\right)}{\mathrm{Vol}\left(\overline{\left[\mathcal{W}_{\mathrm{Sp}_2/\mathbb{Z}_2}\right]}_{[0,1]}^{[0,3]}\right)}=3 \, , 
\end{equation}
indicating the $\mathbb{Z}_3$ quotient.

\begin{figure}
    \centering
        \scalebox{0.8}{\begin{tikzpicture}
        \node (3) at (7,0) {$
            \begin{tikzpicture}
                \node at (-0.5,-1.5) {$\mathcal{J}_{23}$};
                \node[hasse] (t) at (0,0) {};
                \node[hasse] (b) at (0,-3) {};
                \draw (t)--(b);
            \end{tikzpicture}
        $};
        \node (4) at (12,3) {$
            \begin{tikzpicture}
                \node at (-0.5,-0.5) {$A_1$};
                \node at (-0.5,-2) {$cg_2$};
                \node[hasse] (t) at (0,0) {};
                \node[hasse] (m) at (0,-1) {};
                \node[hasse] (b) at (0,-3) {};
                \draw  (t)--(m)--(b);
                \draw (0.5,0.2)--(0.7,0.2)--(0.7,-3.2)--(0.5,-3.2);
                \node at (1.7,-1.5) {$\overline{\left[\mathcal{W}_{G_2}\right]}_{[0,1]}^{[0,2]}$};
            \end{tikzpicture}
        $};
        \node (5) at (12,-3) {$
            \begin{tikzpicture}
                \node at (-0.5,-0.5) {$A_2$};
                \node at (-0.5,-2) {$ac_2$};
                \node[hasse] (t) at (0,0) {};
                \node[hasse] (m) at (0,-1) {};
                \node[hasse] (b) at (0,-3) {};
                \draw  (t)--(m)--(b);
                \draw (0.5,0.2)--(0.7,0.2)--(0.7,-3.2)--(0.5,-3.2);
                \node at (2,-1.5) {$\overline{\left[\mathcal{W}_{\mathrm{Sp}_2/\mathbb{Z}_2}\right]}_{[0,1]}^{[0,3]}$};
            \end{tikzpicture}
        $};
        \node (6) at (18,0) {$
            \begin{tikzpicture}
                \node at (-1,-0.4) {$A_1$};
                \node at (-1.5,-1.5) {$A_2$};
                \node at (-1,-2.6) {$A_2$};
                \node at (1,-0.4) {$A_2$};
                \node at (1.5,-1.5) {$A_1$};
                \node at (1,-2.6) {$A_2$};
                \node at (-.5,-1) {$A_2$};
                \node at (.5,-1) {$A_1$};
                \node[hasse] (10) at (0,0) {};
                \node[hasse] (21) at (-1,-1) {};
                \node[hasse] (22) at (1,-1) {};
                \node[hasse] (31) at (-1,-2) {};
                \node[hasse] (32) at (1,-2) {};
                \node[hasse] (40) at (0,-3) {};
                \draw (10)--(21)--(31)--(40)--(32)--(22)--(10) (21)--(32) (31)--(22);
                \draw (1.8,0.2)--(2,0.2)--(2,-3.2)--(1.8,-3.2);
                \node at (3,-1.5) {$\mathcal{C}\eqref{eq:Z6quot}$};
            \end{tikzpicture}
        $};
        \draw[->] (3)--(4);
        \draw[->] (3)--(5);
        \draw[->] (4)--(6);
        \draw[->] (5)--(6);
        \draw[->] (3)--(6);
        \node at (9,1.6) {$\mathbb{Z}_2$};
        \node at (9,-1.6) {$\mathbb{Z}_3$};
        \node at (15,2) {$\mathbb{Z}_3$};
        \node at (15,-2) {$\mathbb{Z}_2$};
        \node at (12,-.4) {$\mathbb{Z}_6$};
    \end{tikzpicture}}
    \caption{Quotients of $\mathcal{J}_{23}$ which are visible from the quiver.}
    \label{fig:newSliceQuotients3}
\end{figure}
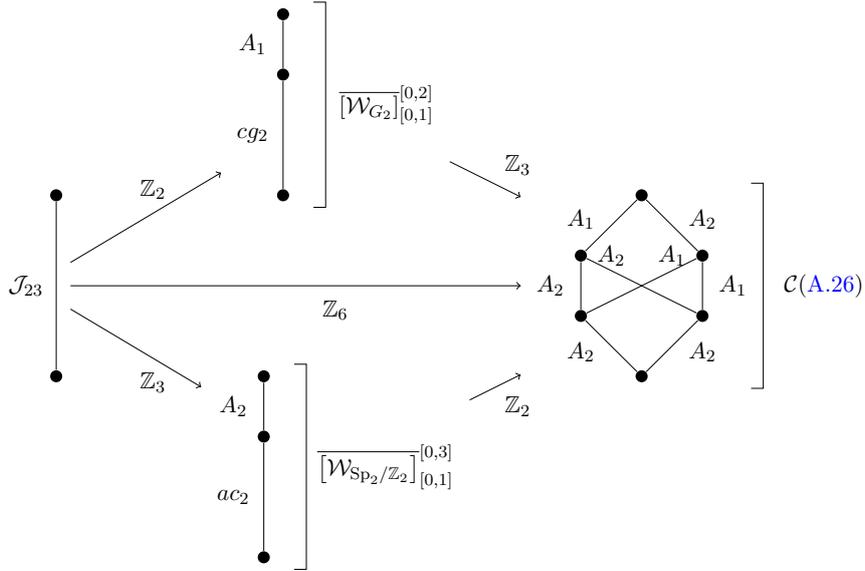

\paragraph{$\mathbf{\mathbb{Z}_2\times\mathbb{Z}_3=\mathbb{Z}_6}$ quotient of $\mathbf{\mathcal{J}_{32}}$.}
$\mathcal{J}_{32}$ is also a $\mathbb{Z}_2\times\mathbb{Z}_3=\mathbb{Z}_6$ cover of the Coulomb branch of
\begin{equation}
    \label{eq:Z6quot}
        \vcenter{\hbox{
        \begin{tikzpicture}
            \node[gauge,label=below:{$2$}] (0) at (0,0) {};
            \node[gauge,label=below:{$1$}] (1) at (1,0) {};
            \node[gauge,label=below:{$1$}] (-1) at (-1,0) {};
            \draw[transform canvas={yshift=-1.5pt}] (0)--(1);
            \draw[transform canvas={yshift=1.5pt}] (0)--(1);
            \node at (0.5,0.4) {$2$};
            \draw[transform canvas={yshift=-2pt}] (0)--(-1);
            \draw[transform canvas={yshift=0pt}] (0)--(-1);
            \draw[transform canvas={yshift=2pt}] (0)--(-1);
            \node at (-0.5,0.4) {$3$};
        \end{tikzpicture}
        }}
\end{equation}
with Hilbert series
\begin{equation}
    \mathrm{HS}\left(\mathcal{C}\eqref{eq:Z6quot}\right)=\frac{1 + 2 t^3 + 2 t^4 + 4 t^5 + 3 t^6 + 3 t^8 + 4 t^9 + 2 t^{10} + 2 t^{11} + t^{14}}{(1 - t^2)^2 (1 - t^3)^2 (1 - t^4) (1 - t^6)}\;.
\end{equation}
From the structure of the quiver (\ref{eq:Z6quot}) we expect that its Coulomb branch is a slice in the double affine Grassmannian of $\hat{A}_1$. 
One can check that
\begin{equation}
    \frac{\mathrm{Vol}\left(\overline{\left[\mathcal{W}_{G_2}\right]}_{[0,1]}^{[0,2]}\right)}{\mathrm{Vol}\left(\mathcal{C}\eqref{eq:Z6quot}\right)}=3 \, , \frac{\mathrm{Vol}\left(\overline{\left[\mathcal{W}_{\mathrm{Sp}_2/\mathbb{Z}_2}\right]}_{[0,1]}^{[0,3]}\right)}{\mathrm{Vol}\left(\mathcal{C}\eqref{eq:Z6quot}\right)}=2 \, , \frac{\mathrm{Vol}\left(\mathcal{J}_{23}\right)}{\mathrm{Vol}\left(\mathcal{C}\eqref{eq:Z6quot}\right)}=6 \, . 
\end{equation}
This can be represented by the commutative diagram in Figure \ref{fig:newSliceQuotients3}.

\subsubsection{\texorpdfstring{Why $\mathcal{J}_{33}$ and $\mathcal{J}_{23}$ are isolated symplectic singularities}{Why J33 and J32 are isolated symplectic singularities}}

Without an explicit description in terms of generators and relations it is hard to tell whether a singularity is an isolated symplectic singularity. We do not have such a description of $\mathcal{J}_{33}$ and $\mathcal{J}_{23}$, and the Hilbert Series indicates that such a description would be very complicated and hence difficult to obtain. We will only give a collection of arguments why we conjecture both singularities to be isolated symplectic singularities.

First an observation about global symmetry is in order. There are isolated symplectic singularities with a global symmetry algebra which is a product of an abelian and a simple non-abelian part. For example the slices called $h_{n,k}$ and $\bar{h}_{n,k}$ in \cite{Bourget:2021siw}, for $k>2$, have a $\mathfrak{u}(n) = \mathfrak{u}(1) \oplus \mathfrak{su}(n)$ global symmetry algebra. However no isolated symplectic singularity with a global symmetry algebra which is a product of multiple simple non-abelian parts is known to the authors. Both slices $\mathcal{J}_{33}$ and $\mathcal{J}_{23}$, have an $\mathfrak{su}(2)\oplus \mathfrak{su}(2)$ global symmetry, which is unusual but not impossible to the best of our knowledge. 

The minimal nilpotent orbit of $\mathrm{O}(4)$ has an $\mathfrak{su}(2)\oplus \mathfrak{su}(2)$ global symmetry, but it is a union of two $A_1$ singularities, and hence it is not a symplectic singularity in the sense of Beauville \cite{beauville2000symplectic} which requires normality. Nevertheless `non-normal symplectic singularities' appear as elementary slices in symplectic singularities. E.g.\ the singularities $A_{2k-1}\cup A_{2k-1}$ and other unions appear as elementary slices in the nilcone of classical and exceptional algebras \cite{Kraft1982,Cabrera:2017njm,2015arXiv150205770F}. Furthermore the singularities $m$, $m'$ and $\mu$ of \cite{2015arXiv150205770F} -- which are not unions of symplectic singularities (they have only two leaves) but are still non-normal -- appear as elementary slices in the nilcone of exceptional algebras.

Therefore there are several questions to address for $\mathcal{J}_{33}$ and $\mathcal{J}_{23}$.
\begin{itemize}
    \item Are they elementary slices?
    \item If they are elementary, are they not a union, i.e.\ are there only two leaves?
    \item If they are not a union, are they normal?
\end{itemize}
If they are normal elementary slices, then they are isolated symplectic singularities.

There are several arguments for the slices $\mathcal{J}_{33}$ and $\mathcal{J}_{23}$ to be elementary and also normal.
\begin{enumerate}
    \item From 6d physics and the associated brane system, following the arguments given at the end of Section \ref{sec:6d}. 
    \item Quotients: When we have a symplectic singularity $S$ with a $\mathbb{Z}_k$ quotient $S'$ then the number of symplectic leaves of $S'$ is often greater than the number of symplectic leaves of $S$. Examples are $S=\mathbb{C}^{2n},b_n,d_n,...$ . Known counterexamples are $ADE$ singularities whose $Z_k$ quotients are again an $ADE$ singularity.\\
    Since the $\mathbb{Z}_{k_i}$ quotients of $\mathcal{J}_{33}$ and $\mathcal{J}_{23}$ have known Hasse diagrams which are linear with three leaves, we can expect their coverings to have only two leaves, which would mean that the slices are elementary and also not a union.
    \item Hilbert series: All known Hilbert series of non-normal singularities have non palindromic numerators. The numerators in the Hilbert series for $\mathcal{J}_{33}$ and $\mathcal{J}_{23}$ are palindromic, which suggests the slices are normal, and hence also not a union. 
\end{enumerate}
We therefore conclude conjecturally that \emph{ $\mathcal{J}_{33}$ and $\mathcal{J}_{23}$ are new isolated symplectic singularities.}

\section{The Case for Decorated Quivers}
\label{app:decorated}

Decorated quivers were introduced in \cite{Bourget:2022ehw}. In this Appendix we briefly review the salient facts about these quivers and explain why they are needed. 

Before doing this, we remind the properties of magnetic quivers. Given a theory with 8 supercharges, one can produce a Hasse diagram which can be interpreted in several equivalent ways: 
\begin{enumerate}
    \item[$(i)$] As the \textbf{phase diagram} of the theory on its Higgs branch. The elementary transitions correspond to phase transitions where a minimal set of fields becomes massless / massive, generalizing the usual Higgs mechanism. 
    \item[$(ii)$] As the \textbf{brane configuration diagram}, if the theory can be realized on a brane system. The elementary transitions correspond to minimal brane recombinations preserving supersymmetry;
    \item[$(iii)$] As the \textbf{singularity diagram} of the Higgs branch, viewed as a symplectic singularity. This is purely geometric, and the elementary transitions correspond to minimal degenerations. 
    \item[$(iv)$] As the \textbf{quiver subtraction diagram}, produced by the quiver subtraction algorithm \cite{Bourget:2021siw} when a magnetic quiver is available for the Higgs branch. 
\end{enumerate}
In the simplest cases (e.g. for a pair of 3d $\mathcal{N}=4$ unitary and linear mirror quivers), one can easily relate all four interpretations above: quiver subtraction $(iv)$ corresponds to the Higgs mechanism $(i)$ via 3d mirror symmetry, to brane recombination $(ii)$ through Kraft-Procesi transitions, and to the geometry $(iii)$ using the monopole formula. 

However it was pointed out in \cite{Bourget:2022ehw} that for moduli spaces of instantons, the correspondence with $(iv)$ naively breaks down.  This is an important problem, as method $(iv)$ is the most straightforward to effectively compute the phase diagram.  It can be restored by introducing \emph{decorated quivers}. More generally, these appear whenever repeated identical quiver subtraction is needed, see the rule on page \pageref{rule2}.

A strong case can be made for this addition, as it perfectly fits with all other approaches $(i)$, $(ii)$ and $(iii)$. The relation between brane systems and decorated quiver is made precise, for brane webs, in \cite[Appendix C]{Bourget:2022ehw}. The agreement with field theory methods has been explicitly checked using the ADHM construction. The consistency with independently computed singular stratifications has been demonstrated for the $G_2$ nilcone in \cite[Appendix A]{Bourget:2022ehw}, and e.g. for symmetric products in the present work -- see Figure \ref{fig:Higgsing0}, which agrees with the mathematical literature. 

The previous paragraph argues that introducing decorated quivers is \emph{sufficient} to restore the agreement, and provides an efficient way to compute phase diagrams whenever a magnetic quiver is available. One could however ask whether this is \emph{necessary}: does there exist a simpler fix that wouldn't need such a vast and exotic extension of the notion of quivers? We can safely argue that indeed these quivers are necessary. The main argument for decorations is the very natural string theory interpretation which we will comment on in the next subsection. But another reason, of a more mathematical nature, is as compelling. It was shown in \cite{Nakajima:2019olw} that Coulomb branches of quivers, including non simply laced edges, are always normal. However the geometry includes non-normal transverse slices, like $m$, discussed at length in the present work, showing that a major generalization is indeed needed if one wishes to use quivers to describe these singularities.
Moreover, decorated quivers turn out to be needed to describe moduli spaces, as opposed to mere transverse slices. This is illustrated in the next paragraph. 
Before that, let us mention that one aspect of decorated quivers remains open at the moment, namely the direct computation of the Hilbert series of the corresponding symplectic singularity. We are currently investigating this question, and the results will be presented in a future work \cite{MonopoleDecorated}.

\subsection*{Instantons in Product Groups}

\begin{figure}
    \makebox[\textwidth][c]{
    \begin{tikzpicture}[scale=0.9]
        \node (20) at (-6,10) {$\begin{tikzpicture}[scale=1.5]
                \node[seven] (o) at (0,0) {};
                \node[seven] (l) at (1,0) {};
                \node[seven] (lu) at (1,1) {};
                \node[seven] (u) at (2,1) {};
                \node[seven] (d) at (2,-1) {};
                \node[seven] (rd) at (3,-1) {};
                \node[seven] (r) at (3,0) {};
                \draw[magenta] (o)--(l);
                \draw[orange,transform canvas={yshift=-2pt}] (l)--(r);
                \draw[orange,transform canvas={yshift=2pt}] (l)--(r);
                \draw[red,transform canvas={xshift=-2pt}] (u)--(d);
                \draw[red,transform canvas={xshift=2pt}] (u)--(d);
                \draw[blue,transform canvas={xshift=-2pt,yshift=-2pt}] (lu)--(rd);
                \draw[blue,transform canvas={xshift=2pt,yshift=2pt}] (lu)--(rd);
            \end{tikzpicture}$};
        \node (11) at (0,10) {$\begin{tikzpicture}[scale=1.5]
                \node[seven] (o) at (0,0) {};
                \node[seven] (l) at (1,0) {};
                \node[seven] (lu) at (1,1) {};
                \node[seven] (u) at (2,1) {};
                \node[seven] (d) at (2,-1) {};
                \node[seven] (rd) at (3,-1) {};
                \node[seven] (r) at (3,0) {};
                \draw[magenta] (o)--(l);
                \draw[orange,transform canvas={yshift=-2pt}] (l)--(r);
                \draw[red,transform canvas={xshift=-2pt}] (u)--(d);
                \draw[blue,transform canvas={xshift=2pt,yshift=2pt}] (lu)--(rd);
                \draw[cyan,transform canvas={xshift=-5pt,yshift=1pt}] (lu)--(2,0)--(d) (2,0)--(r);
                \draw[olive,transform canvas={xshift=1pt,yshift=-5pt}] (rd)--(2,0)--(u) (2,0)--(l);
            \end{tikzpicture}$};
        \node (02) at (6,10) {$\begin{tikzpicture}[scale=1.5]
                \node[seven] (o) at (0,0) {};
                \node[seven] (l) at (1,0) {};
                \node[seven] (lu) at (1,1) {};
                \node[seven] (u) at (2,1) {};
                \node[seven] (d) at (2,-1) {};
                \node[seven] (rd) at (3,-1) {};
                \node[seven] (r) at (3,0) {};
                \draw[magenta] (o)--(l);
                \draw[cyan,transform canvas={xshift=-5pt,yshift=1pt}] (lu)--(2,0)--(d) (2,0)--(r);
                \draw[cyan,transform canvas={xshift=-1pt,yshift=5pt}] (lu)--(2,0)--(d) (2,0)--(r);
                \draw[olive,transform canvas={xshift=1pt,yshift=-5pt}] (rd)--(2,0)--(u) (2,0)--(l);
                \draw[olive,transform canvas={xshift=5pt,yshift=-1pt}] (rd)--(2,0)--(u) (2,0)--(l);
            \end{tikzpicture}$};
        \node (10) at (-3,5) {$\begin{tikzpicture}[scale=1.5]
                \node[seven] (o) at (0,0) {};
                \node[seven] (l) at (1,0) {};
                \node[seven] (lu) at (1,1) {};
                \node[seven] (u) at (2,1) {};
                \node[seven] (d) at (2,-1) {};
                \node[seven] (rd) at (3,-1) {};
                \node[seven] (r) at (3,0) {};
                \draw[magenta] (o)--(l);
                \draw[orange,transform canvas={yshift=-2pt}] (l)--(r);
                \draw[red,transform canvas={xshift=-2pt}] (u)--(d);
                \draw[blue,transform canvas={xshift=-2pt,yshift=-2pt}] (lu)--(rd);
                \draw[green,transform canvas={yshift=2pt}] (l)--(r);
                \draw[green,transform canvas={xshift=2pt}] (u)--(d);
                \draw[green,transform canvas={xshift=2pt,yshift=2pt}] (lu)--(rd);
            \end{tikzpicture}$};
        \node (01) at (3,5) {$\begin{tikzpicture}[scale=1.5]
                \node[seven] (o) at (0,0) {};
                \node[seven] (l) at (1,0) {};
                \node[seven] (lu) at (1,1) {};
                \node[seven] (u) at (2,1) {};
                \node[seven] (d) at (2,-1) {};
                \node[seven] (rd) at (3,-1) {};
                \node[seven] (r) at (3,0) {};
                \draw[magenta] (o)--(l);
                \draw[cyan,transform canvas={xshift=-5pt,yshift=1pt}] (lu)--(2,0)--(d) (2,0)--(r);
                \draw[olive,transform canvas={xshift=1pt,yshift=-5pt}] (rd)--(2,0)--(u) (2,0)--(l);
                \draw[green] (l)--(r) (lu)--(rd) (u)--(d);
            \end{tikzpicture}$};
        \node (00) at (0,0) {$\begin{tikzpicture}[scale=1.5]
                \node[seven] (o) at (0,0) {};
                \node[seven] (l) at (1,0) {};
                \node[seven] (lu) at (1,1) {};
                \node[seven] (u) at (2,1) {};
                \node[seven] (d) at (2,-1) {};
                \node[seven] (rd) at (3,-1) {};
                \node[seven] (r) at (3,0) {};
                \draw[magenta] (o)--(l);
                \draw[green,transform canvas={yshift=-3pt}] (l)--(r);
                \draw[green,transform canvas={xshift=-3pt}] (u)--(d);
                \draw[green,transform canvas={xshift=-3pt,yshift=-3pt}] (lu)--(rd);
                \draw[green] (l)--(r) (lu)--(rd) (u)--(d);
            \end{tikzpicture}$};
        \node (0) at (0,-5) {$\begin{tikzpicture}[scale=1.5]
                \node[seven] (o) at (0,0) {};
                \node[seven] (l) at (1,0) {};
                \node[seven] (lu) at (1,1) {};
                \node[seven] (u) at (2,1) {};
                \node[seven] (d) at (2,-1) {};
                \node[seven] (rd) at (3,-1) {};
                \node[seven] (r) at (3,0) {};
                \draw[magenta] (o)--(l);
                \draw[green,thick] (l)--(r) (lu)--(rd) (u)--(d);
                \node at (1.6,-0.4) {\color{green}$2$};
            \end{tikzpicture}$};
        \draw (20)--(10)--(11)--(01)--(02) (00)--(0);
        \draw[double] (10)--(00)--(01);
        \node at (-5,7.5) {$a_2$};
        \node at (-1,7.5) {$a_1$};
        \node at (1,7.5) {$a_2$};
        \node at (5,7.5) {$a_1$};
        \node at (-2,2.5) {$2a_2$};
        \node at (2,2.5) {$2a_1$};
        \node at (-0.5,-2.5) {$A_1$};
    \end{tikzpicture}}
    \caption{The seven distinct brane phases arranged in a Hasse diagram. The Hasse diagrams of two $\mathrm{SU}(3)$ (two $\mathrm{SU}(2)$) instantons is the subdiagram from the bottom to the top left (top right).}
    \label{fig:twoInstBW}
\end{figure}
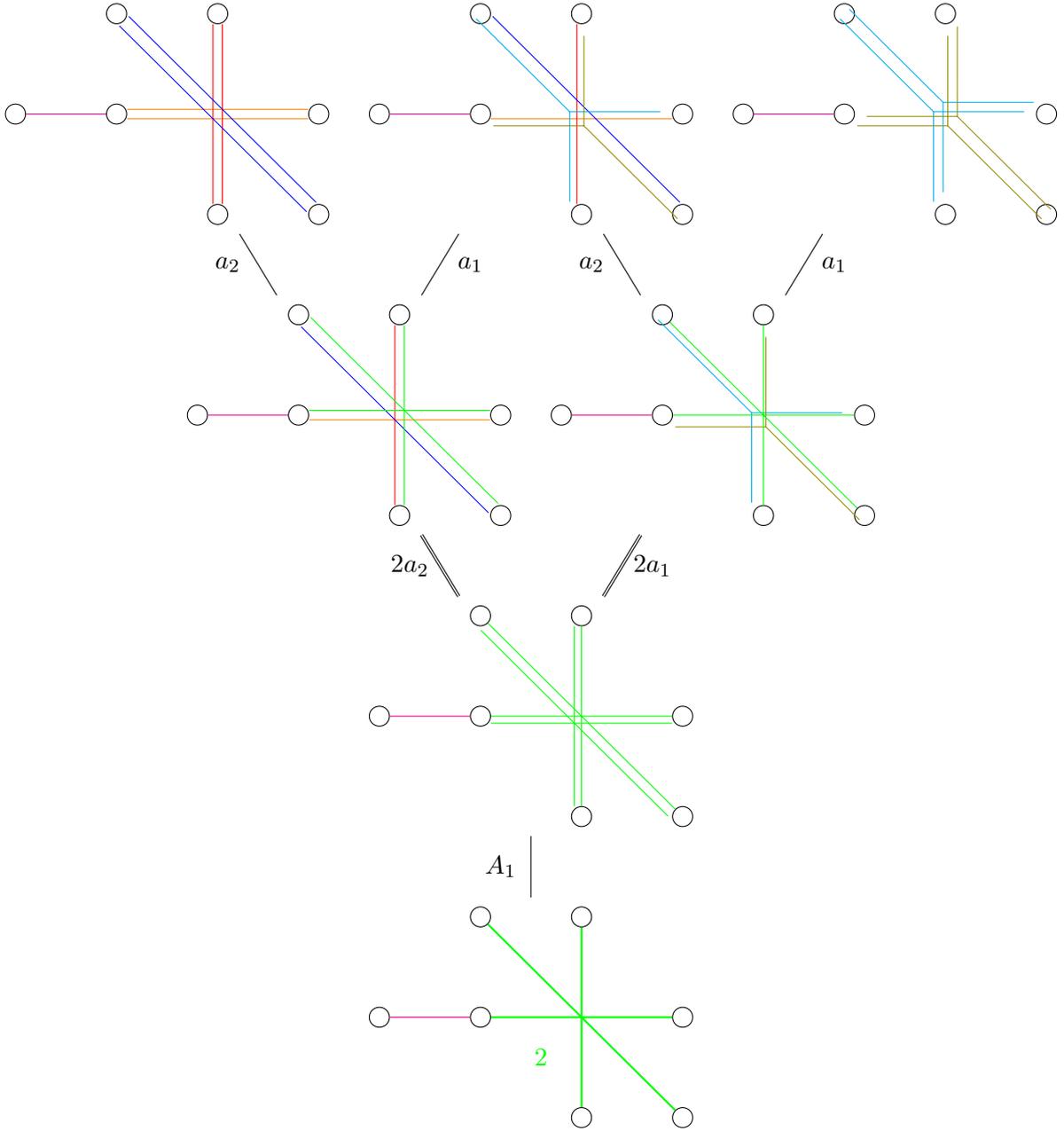

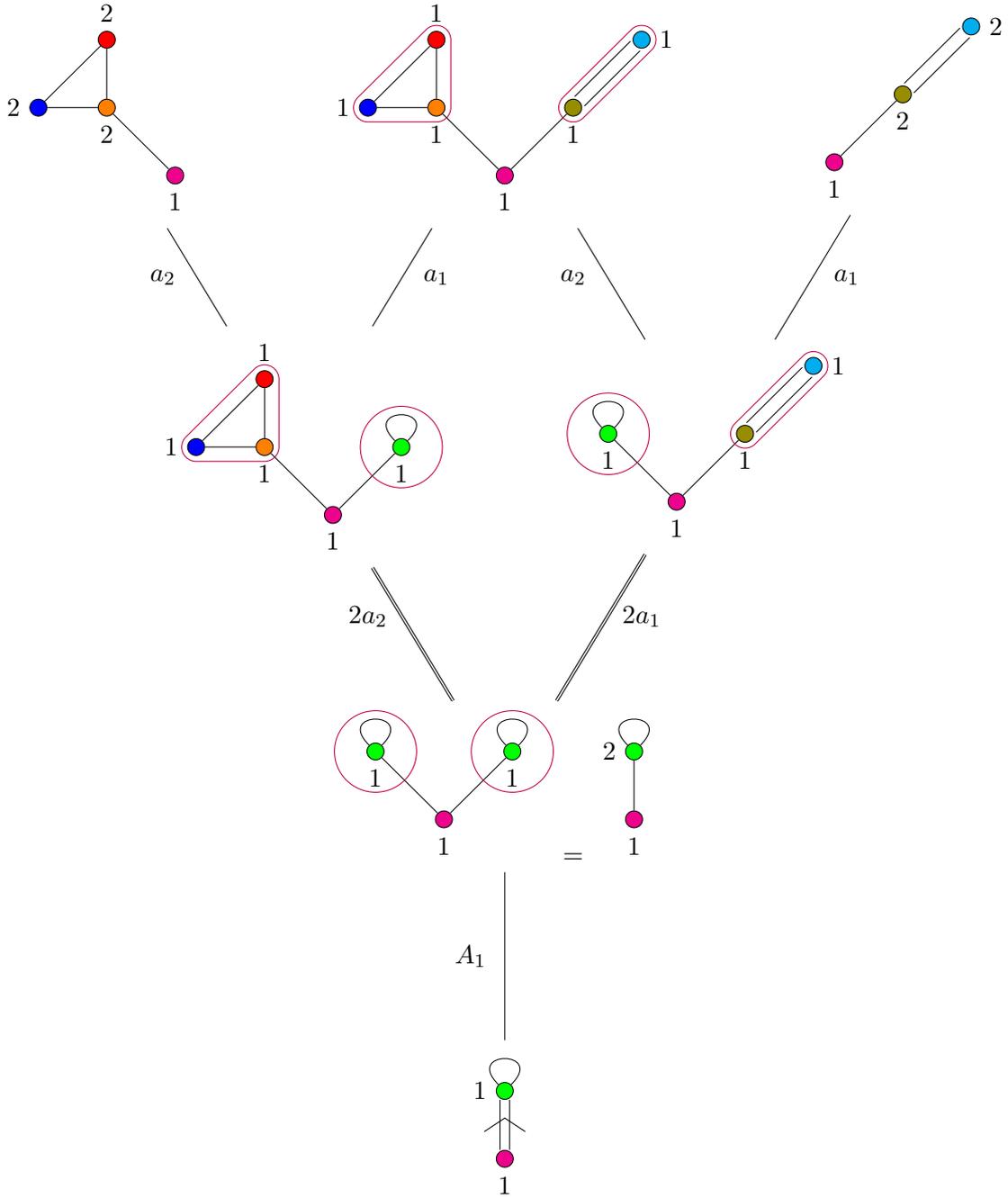
\begin{figure}
    \makebox[\textwidth][c]{
    \begin{tikzpicture}
        \node (20) at (-6,10) {$\begin{tikzpicture}
                \node[gaugem,label=below:{1}] (m) at (0,0) {};
                \node[gaugeo,label=below:{2}] (l1) at (-1,1) {};
                \node[gaugeb,label=left:{2}] (l2) at (-2,1) {};
                \node[gauger,label=above:{2}] (l3) at (-1,2) {};
                \draw (m)--(l1)--(l2)--(l3)--(l1);
            \end{tikzpicture}$};
        \node (11) at (0,10) {$\begin{tikzpicture}
                \node[gaugem,label=below:{1}] (m) at (0,0) {};
                \node[gaugeo,label=below:{1}] (l1) at (-1,1) {};
                \node[gaugeb,label=left:{1}] (l2) at (-2,1) {};
                \node[gauger,label=above:{1}] (l3) at (-1,2) {};
                \node[gaugeol,label=below:{1}] (r1) at (1,1) {};
                \node[gaugec,label=right:{1}] (r2) at (2,2) {};
                \draw (r1)--(m)--(l1)--(l2)--(l3)--(l1);
                \draw[transform canvas={xshift=-2pt,yshift=2pt}] (r1)--(r2);
                \draw[transform canvas={xshift=2pt,yshift=-2pt}] (r1)--(r2);
                \draw[purple] \convexpath{l1,l2,l3}{6pt};
                \draw[purple] \convexpath{r1,r2}{6pt};
            \end{tikzpicture}$};
        \node (02) at (6,10) {$\begin{tikzpicture}
                \node[gaugem,label=below:{1}] (m) at (0,0) {};
                \node[gaugeol,label=below:{2}] (r1) at (1,1) {};
                \node[gaugec,label=right:{2}] (r2) at (2,2) {};
                \draw (r1)--(m);
                \draw[transform canvas={xshift=-2pt,yshift=2pt}] (r1)--(r2);
                \draw[transform canvas={xshift=2pt,yshift=-2pt}] (r1)--(r2);
            \end{tikzpicture}$};
        \node (10) at (-3,5) {$\begin{tikzpicture}
                \node[gaugem,label=below:{1}] (m) at (0,0) {};
                \node[gaugeo,label=below:{1}] (l1) at (-1,1) {};
                \node[gaugeb,label=left:{1}] (l2) at (-2,1) {};
                \node[gauger,label=above:{1}] (l3) at (-1,2) {};
                \node[gaugeg,label=below:{1}] (r) at (1,1) {};
                \draw (r1)--(m)--(l1)--(l2)--(l3)--(l1);
                \draw (r) to [out=45,in=135,looseness=10] (r);
                \draw[purple] \convexpath{l1,l2,l3}{6pt};
                \draw[purple] (r) circle (0.6cm);
            \end{tikzpicture}$};
        \node (01) at (3,5) {$\begin{tikzpicture}
                \node[gaugem,label=below:{1}] (m) at (0,0) {};
                \node[gaugeg,label=below:{1}] (l) at (-1,1) {};
                \node[gaugeol,label=below:{1}] (r1) at (1,1) {};
                \node[gaugec,label=right:{1}] (r2) at (2,2) {};
                \draw (r1)--(m)--(l);
                \draw (l) to [out=45,in=135,looseness=10] (l);
                \draw[transform canvas={xshift=-2pt,yshift=2pt}] (r1)--(r2);
                \draw[transform canvas={xshift=2pt,yshift=-2pt}] (r1)--(r2);
                \draw[purple] (l) circle (0.6cm);
                \draw[purple] \convexpath{r1,r2}{6pt};
            \end{tikzpicture}$};
        \node (00) at (0,0) {$\begin{tikzpicture}
                \node[gaugem,label=below:{1}] (m) at (0,0) {};
                \node[gaugeg,label=below:{1}] (l) at (-1,1) {};
                \node[gaugeg,label=below:{1}] (r) at (1,1) {};
                \draw (r)--(m)--(l);
                \draw (l) to [out=45,in=135,looseness=10] (l);
                \draw (r) to [out=45,in=135,looseness=10] (r);
                \draw[purple] (r) circle (0.6cm);
                \draw[purple] (l) circle (0.6cm);
            \end{tikzpicture}$ = $\begin{tikzpicture}
                \node[gaugem,label=below:{1}] (m) at (0,0) {};
                \node[gaugeg,label=left:{2}] (n) at (0,1) {};
                \draw (m)--(n);
                \draw (n) to [out=45,in=135,looseness=10] (n);
            \end{tikzpicture}$};
        \node (0) at (0,-5) {$\begin{tikzpicture}
                \node[gaugem,label=below:{1}] (m) at (0,0) {};
                \node[gaugeg,label=left:{1}] (n) at (0,1) {};
                \draw[transform canvas={xshift=-2pt}] (m)--(n);
                \draw[transform canvas={xshift=2pt}] (m)--(n);
                \draw (-0.3,0.4)--(0,0.6)--(0.3,0.4);
                \draw (n) to [out=45,in=135,looseness=10] (n);
            \end{tikzpicture}$};
        \draw (20)--(10)--(11)--(01)--(02) (00)--(0);
        \draw[double] (10)--(00)--(01);
        \node at (-5,7.5) {$a_2$};
        \node at (-1,7.5) {$a_1$};
        \node at (1,7.5) {$a_2$};
        \node at (5,7.5) {$a_1$};
        \node at (-2,2.5) {$2a_2$};
        \node at (2,2.5) {$2a_1$};
        \node at (-0.5,-2.5) {$A_1$};
    \end{tikzpicture}}
    \caption{Magnetic quivers read from the brane web decompositions in Figure \ref{fig:twoInstBW}. Note that the top middle magnetic quiver, corresponding to a maximal brane web decomposition, is decorated. The Hasse diagram is obtained using the quiver subtraction rules of Section \ref{sec:unionCones}.}
    \label{fig:twoInstMQ}
\end{figure}

Decorated quivers were first conceived to address a problem in quiver subtraction of a conventional (non-decorated) quiver. In this paragraph we show that decorated quivers arise in their own right as magnetic quivers for certain moduli spaces. The simplest example is in the moduli space of instantons of a product group. Let us consider the moduli space of two instantons in $\mathrm{SU}(3)\times\mathrm{SU}(2)$. This moduli space is realised as the Higgs branch of the 5d $\mathcal{N}=1$ theory living on the brane web
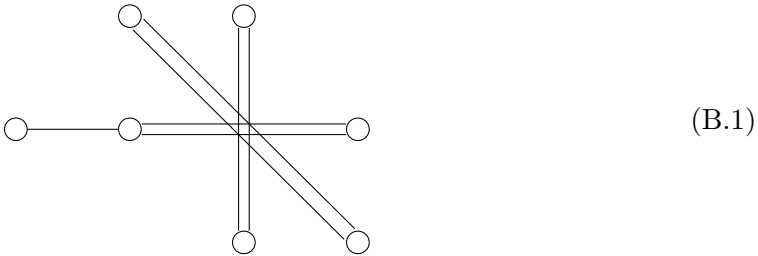
\begin{equation}
\raisebox{-.5\height}{
    \begin{tikzpicture}[scale=1.5]
        \node[seven] (o) at (0,0) {};
        \node[seven] (l) at (1,0) {};
        \node[seven] (lu) at (1,1) {};
        \node[seven] (u) at (2,1) {};
        \node[seven] (d) at (2,-1) {};
        \node[seven] (rd) at (3,-1) {};
        \node[seven] (r) at (3,0) {};
        \draw (o)--(l);
        \draw[transform canvas={yshift=-2pt}] (l)--(r);
        \draw[transform canvas={yshift=2pt}] (l)--(r);
        \draw[transform canvas={xshift=-2pt}] (u)--(d);
        \draw[transform canvas={xshift=2pt}] (u)--(d);
        \draw[transform canvas={xshift=-2pt,yshift=-2pt}] (lu)--(rd);
        \draw[transform canvas={xshift=2pt,yshift=2pt}] (lu)--(rd);
    \end{tikzpicture}}
\end{equation}
This brane web has three maximal decompositions. Corresponding to the three choices: both instantons in $\mathrm{SU}(3)$, one instanton in $\mathrm{SU}(3)$ and one instanton in $\mathrm{SU}(2)$, both instantons in $\mathrm{SU}(2)$. We depict all three maximal decompositions, as well as all other phases of the brane web in Figure \ref{fig:twoInstBW}. From these brane web decompositions we can read off (decorated) magnetic quivers following the rules of \cite[Appendix C]{Bourget:2022ehw}. We summarize the magnetic quivers for each phase in Figure \ref{fig:twoInstMQ}. Importantly the cone in the moduli space for one instanton in $\mathrm{SU}(3)$ and one instanton in $\mathrm{SU}(2)$ has a decorated magnetic quiver. Without the decoration one would not obtain the Hasse diagram predicted from the branes and other physical considerations.

\newpage

\section{Non-singular Transitions in the Hasse Diagram of the full Moduli Space}
\label{app:16susy}

\subsection{Theories with 16 Supercharges}

\paragraph{4d $\mathcal{N}=4$ SYM. }
We start with the description of the vacuum moduli space of 4d $\mathcal{N}=4$ SYM with simple gauge group $G$. This is $\mathcal{M} = \mathbb{C}^{3r} / W$ where $r$ is the rank of $G$ and $W$ is its Weyl group. The Weyl group acts simultaneously on the three $\mathbb{C}^r$ factors. A choice of an $\mathcal{N}=2$ subalgebra inside the $\mathcal{N}=4$ algebra corresponds to a choice inside $\mathcal{M}$ of a Higgs branch part $\mathcal{M}_H$ and a Coulomb branch part $\mathcal{M}_C$, respectively of dimension $2r$ and $r$. We are interested here in the interplay between the Hasse diagrams of $\mathcal{M}$, $\mathcal{M}_H$ and $\mathcal{M}_C$. 

Consider for simplicity $G = \mathrm{SU}(2)$, with Weyl group $W = \mathbb{Z}_2$. The moduli space $\mathcal{M} = \mathbb{C}^{3} / \mathbb{Z}_2$, viewed as an algebraic variety, clearly has the following singular stratification: 
\begin{equation}
    \begin{tikzpicture}
        \node[hasse,label=left:{0}] (0) at (0,0) {};
        \node[hasse,label=left:{3}] (1) at (0,1) {};
        \draw (0)--(1);
        \node[label=right:{$\mathbb{C}^{3} / \mathbb{Z}_2$}] at ($(0)!0.5!(1)$) {};
    \end{tikzpicture} 
\end{equation}
as there is a unique singular point, located at the origin. 

We introduce coordinates $z_1 , z_2 , z_3$ on $\mathbb{C}^3$, so that the $\mathbb{Z}_2$ action is $(z_1 , z_2 , z_3) \sim (-z_1 , -z_2 ,- z_3)$. The variety $\mathcal{M}$ can then be described as the spectrum of $\mathbb{C}[Z_{ij}] / I$ where we have introduced variables $Z_{ij} = z_i z_j$ for $1 \leq i , j \leq 3$ and the ideal $I$ is defined by 
\begin{equation}
    I = (Z_{ij} - Z_{ji} , Z_{ij} Z_{kl} - Z_{ik} Z_{jl}) \, . 
\end{equation}
Explicitly, there are six quadratic equations: 
\begin{equation}
\label{eq:6eq}
    \begin{array}{ccc}
        Z_{11} Z_{22} = Z_{12}^2 & \qquad & Z_{11} Z_{23} = Z_{12} Z_{13} \\        
        Z_{22} Z_{33} = Z_{23}^2 &\qquad  & Z_{22} Z_{13} = Z_{12} Z_{23} \\        
        Z_{33} Z_{11} = Z_{13}^2 & \qquad & Z_{33} Z_{12} = Z_{13} Z_{23} 
    \end{array}
\end{equation}

The Higgs branch can be defined by $z_3 = 0$, which in terms of the $Z_{ij}$ variables translate into $Z_{i,3}=0$ for all $i$. The above ring reduces to $\mathbb{C}[Z_{11} , Z_{12} , Z_{22}] / (Z_{11} Z_{22} - Z_{12}^2)$, which corresponds to $\mathbb{C}^2 / \mathbb{Z}_2$. On the other hand the Coulomb branch can be defined by $z_1 = z_2 = 0$, whereby the ring reduces to $\mathbb{C}[Z_{33}]$, so the Coulomb branch is $\mathbb{C} \simeq \mathbb{C}/ \mathbb{Z}_2$. Finally, the mixed branch can be described as a fibration over either the Higgs or Coulomb branch. The fibers can be called respectively Coulomb and Higgs branches, keeping in mind that their geometry can change as the base point is changed. 

Consider for instance $\mathcal{M}_C$ to be the base of the fibration. We have already seen that the fiber above the origin of the Coulomb branch is $\mathcal{M}_H = \mathbb{C}^2 / \mathbb{Z}_2$. Let us now look at the fiber above a generic point $Z_{33} \neq 0$. We can now solve three of the equations of \eqref{eq:6eq} as 
\begin{equation}
    Z_{11} = \frac{Z_{13}^2}{Z_{33}} \, , \qquad 
    Z_{22} = \frac{Z_{23}^2}{Z_{33}} \, , \qquad 
    Z_{12} = \frac{Z_{13}Z_{23}}{Z_{33}} \, ,  
\end{equation}
and the three other equations become trivial, so the fiber is described by its coordinate ring $\mathbb{C}[Z_{13} , Z_{23}]$, and it is simply $\mathbb{C}^2$. 

Similarly, going on a generic point on the Higgs branch, we can assume without loss of generality that $Z_{11} \neq 0$, and a similar procedure leads to the fiber having coordinate ring $\mathbb{C}[Z_{13}]$. We can then draw the following Hasse diagram: 
\begin{equation}
        \raisebox{-.5\height}{ \begin{tikzpicture}
        \node[hasse,label=below:{0}] (0) at (0,0) {};
        \node[hasse,label=left:{1}] (1) at (-1,1) {};
        \node[hasse,label=right:{2}] (2) at (1,1) {};
        \node[hasse,label=above:{3}] (3) at (0,2) {};
        \draw[red] (0)--(1) ;
        \draw[blue] (0)--(2);
        \draw[red,dashed] (3)--(2);
        \draw[blue,dashed] (1)--(3);
        \node at (-1,.3) {$\mathbb{C}$};
        \node at (-1,1.7) {$\mathbb{C}^2$};
        \node at (1.2,.3) {$\mathbb{C}^2 / \mathbb{Z}_2$};
        \node at (1,1.7) {$\mathbb{C} $};
    \end{tikzpicture} }
\end{equation}
A few comments are in order. First, the fact that the Coulomb branch has a free coordinate ring indicates it does not have complex structure singularity, but it nevertheless has a metric singularity at the origin (see \cite{Argyres:2018wxu} for a detailed discussion of this point). It corresponds to the planar special K\"ahler singularity with Kodaira type $I_0^*$.  Second, the choice of $\mathcal{N}=2$ subalgebra inside the $\mathcal{N}=4$ algebra defines the Higgs and Coulomb branches as sub-varieties, and as such, the fibration of one above the other should be seen as a fibration with section. The fibers come equipped with a $\mathbb{Z}_2$ action, whose fixed point is this section. We use the notation $(X , \mathbb{Z}_2)$ to denote a fiber $X$ with such an action, following \cite{slodowy1980simple,2015arXiv150205770F}. With these amendments, the full Hasse diagram for SU(2) $\mathcal{N}=4$ SYM with a choice of $\mathcal{N}=2$ subalgebra is 
\begin{equation}
     \raisebox{-.5\height}{   \begin{tikzpicture}
        \node[hasse,label=below:{0}] (0) at (0,0) {};
        \node[hasse,label=left:{1}] (1) at (-1,1) {};
        \node[hasse,label=right:{2}] (2) at (1,1) {};
        \node[hasse,label=above:{3}] (3) at (0,2) {};
        \draw[red] (0)--(1);
        \draw[blue] (0)--(2);
        \node at (-1.2,.3) {$I_0^* \equiv \mathbb{C}$};
        \node at (-1.2,1.7) {$(\mathbb{C}^2 , \mathbb{Z}_2)$};
        \node at (1.2,.3) {$\mathbb{C}^2 / \mathbb{Z}_2$};
        \node at (1.2,1.7) {$(\mathbb{C} , \mathbb{Z}_2)$};
        \draw[red,dashed] (3)--(2);
        \draw[blue,dashed] (1)--(3);
    \end{tikzpicture} }
\end{equation}

\paragraph{3d $\mathcal{N}=8$ SYM. } A very similar situation describes the three dimensional theory, though more symmetric as the Coulomb and Higgs branch are now isomorphic. The Hasse diagram for the SU(2) 3d $\mathcal{N}=8$ theory with a choice of $\mathcal{N}=4$ subalgebra is 
\begin{equation}
    \raisebox{-.5\height}{    \begin{tikzpicture}
            \node[hasse,label=below:{0}] (0) at (0,0) {};
            \node[hasse,label=above:{4}] (4) at (0,2) {};
            \draw (0)--(4);
            \node at (-0.6,1) {$\mathbb{C}^4/\mathbb{Z}_2$};
        \end{tikzpicture}} \qquad\rightarrow\qquad
       \raisebox{-.5\height}{ \begin{tikzpicture}
        \node[hasse,label=below:{0}] (0) at (0,0) {};
        \node[hasse,label=left:{2}] (1) at (-1,1) {};
        \node[hasse,label=right:{2}] (2) at (1,1) {};
        \node[hasse,label=above:{4}] (3) at (0,2) {};
        \draw[red] (0)--(1);
        \draw[red,dashed] (3)--(2);
        \draw[blue] (0)--(2);
        \draw[blue,dashed] (1)--(3);
        \node at (-1,.3) {$\mathbb{C}^2/\mathbb{Z}_2$};
        \node at (-1.2,1.7) {$(\mathbb{C}^2,\mathbb{Z}_2)$};
        \node at (1,.3) {$\mathbb{C}^2 / \mathbb{Z}_2$};
        \node at (1.2,1.7) {$(\mathbb{C}^2,\mathbb{Z}_2)$};
    \end{tikzpicture} }
\end{equation}
We can sketch the full moduli space with a choice of Higgs and Coulomb as follows:
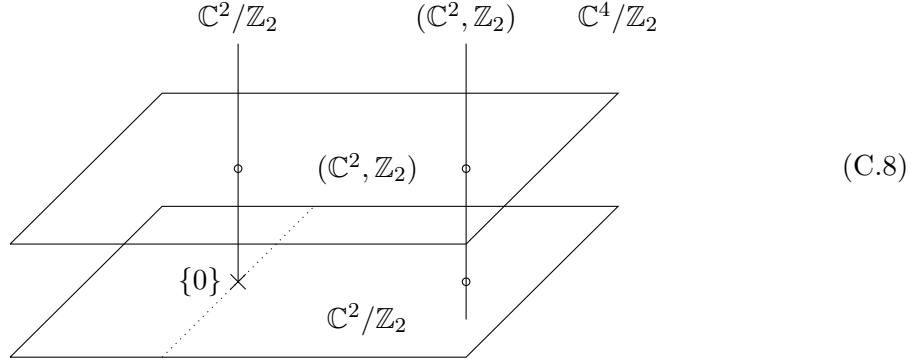
\begin{equation}
   \raisebox{-.5\height}{ \begin{tikzpicture}
        \node (o) at (0,3.5) {$\mathbb{C}^2/\mathbb{Z}_2$};
        \node (g) at (3,3.5) {$(\mathbb{C}^2,\mathbb{Z}_2)$};
        \draw (-3,-1)--(3,-1)--(5,1)--(-1,1)--(-3,-1);
        \draw (0-0.1,0-0.1)--(0+0.1,0+0.1) (0-0.1,0+0.1)--(0+0.1,0-0.1);
        \draw[dotted] (-1,-1)--(1,1);
        \node at (1.7,-0.5) {$\mathbb{C}^2/\mathbb{Z}_2$};
        \draw (0,0)--(o) (3,-0.5)--(g);
        \draw (3,0) circle (0.05cm);
        \draw (0,1.5) circle (0.05cm);
        \draw (3,1.5) circle (0.05cm);
        \draw (-3,-1+1.5)--(3,-1+1.5)--(5,1+1.5)--(-1,1+1.5)--(-3,-1+1.5);
        \node at (1.7,1.5) {$(\mathbb{C}^2,\mathbb{Z}_2)$};
        \node at (5,3.5) {$\mathbb{C}^4/\mathbb{Z}_2$};
        \node at (-0.5,0) {$\{0\}$};
    \end{tikzpicture}}
\end{equation}

\subsection{SO(3) with fundamental flavours}

\label{app:OSo}

In the last section on theories with 16 supercharges we have seen, that the unnatural split of the moduli space in a Higgs and Coulomb part leads to an overcounting of leaves when in the Hasse diagram of the full moduli space, accompanied by the presence of non-singular slices
\begin{equation}
    (\mathbb{C}^{2n},\mathbb{Z}_k)\;,
\end{equation}
which are quotiented to
\begin{equation}
    \mathbb{C}^{2n}/\mathbb{Z}_k\;.
\end{equation}

We observe a similar phenomenon in the full moduli space of theories with less supersymmetry as well. This was already pointed out in the realm of the enhanced Coulomb branch in \cite{Argyres:2016xmc}. It is instructive to look at the 3d $\mathcal{N}=4$ gauge theories with O$(3)$ or SO$(3)$ gauge group and $N$ fundamental flavours.

\paragraph{Orthogonal gauge group.} Consider the 3d $\mathcal{N}=4$ theory
\begin{equation}
    \raisebox{-.5\height}{\begin{tikzpicture}
        \node[gauge,label=below:{O$(3$)}] (1) at (0,0) {};
        \node[flavour,label=below:{$C_N$}] (2) at (1,0) {};
        \draw (1)--(2);
    \end{tikzpicture}}
\end{equation}
The full moduli space Hasse diagram of this theory was already discussed in detail in \cite{Grimminger:2020dmg}. We will nevertheless explain it here. We use blue lines for Higgs branch directions and red lines for Coulomb branch directions.

The Higgs branch Hasse diagram is \cite{Bourget:2019aer}:
\begin{equation}
   \raisebox{-.5\height}{ \begin{tikzpicture}
        \node[hasse] (0) at (0,0) {};
        \node[hasse] (1) at (1,1) {};
        \node[hasse] (2) at (2,2) {};
        \node[hasse] (3) at (3,3) {};
        \draw[blue] (0)--(1)--(2)--(3);
        \node at (1,0.5) {$c_N$};
        \node at (2.2,1.5) {$c_{N-1}$};
        \node at (3.2,2.5) {$c_{N-2}$};
        \node at (1.1,0) {\color{purple}\scriptsize$\mathrm{O}(3)-[C_N]$};
        \node at (2.2,1) {\color{purple}\scriptsize$\mathrm{O}(2)-[C_{N-1}]$};
        \node at (3.3,2) {\color{purple}\scriptsize$\mathrm{O}(1)-[C_{N-2}]$};
        \node at (3.3,3) {\color{purple}\scriptsize$\emptyset$};
    \end{tikzpicture}}\;.
\end{equation}
Where we have added in purple what the theory is higgsed to on a given stratum. On the most generic stratum the theory is completely higgsed. The Coulomb branch Hasse diagram of the O$(1)$ theory is trivial. The Coulomb branch Hasse diagram of the O$(2)$ is
\begin{equation}
   \raisebox{-.5\height}{ \begin{tikzpicture}
        \node[hasse] (0) at (0,0) {};
        \node[hasse] (c1) at (-1,1) {};
        \draw[red] (0)--(c1);
        \node at (0.2,0.5) {$D_{N+1}$};
        \node at (-1.2,0) {\color{purple}\scriptsize$\mathrm{O}(2)-[C_{N-1}]$}; 
        \node at (-1.3,1) {\color{purple}\scriptsize$\emptyset$}; 
    \end{tikzpicture}}\;.
\end{equation}

The Coulomb branch Hasse diagram of the full O$(3)$ theory is
\begin{equation}
   \raisebox{-.5\height}{ \begin{tikzpicture}
        \node[hasse] (0) at (0,0) {}; 
        \node[hasse] (c1) at (-1,1) {};
        \draw[red] (0)--(c1);
        \node at (-1,0.5) {$D_{N+1}$};
        \node at (-1.1,0) {\color{purple}\scriptsize$\mathrm{O}(3)-[C_N]$}; 
        \node at (-2.1,1) {\color{purple}\scriptsize$\mathrm{O}(1)-[C_N]$};
    \end{tikzpicture}}\;.
\end{equation}
On the most general stratum, there is an O$(1)$ gauge theory which is left over. Its Higgs branch Hasse diagram is
\begin{equation}
   \raisebox{-.5\height}{ \begin{tikzpicture}
        \node[hasse] (0) at (0,0) {};
        \node[hasse] (1) at (1,1) {};
        \draw[blue] (0)--(1);
        \node at (1,0.5) {$c_N$};
        \node at (1.1,0) {\color{purple}\scriptsize$\mathrm{O}(1)-[C_N]$};
        \node at (1.3,1) {\color{purple}\scriptsize$\emptyset$};
    \end{tikzpicture}}\;.
\end{equation}
Putting this all together, we obtain the Hasse diagram of the full moduli space \cite{Grimminger:2020dmg}:
\begin{equation}
   \raisebox{-.5\height}{ \begin{tikzpicture}
        \node[hasse] (0) at (0,0) {};
        \node[hasse] (1) at (1,1) {};
        \node[hasse] (2) at (2,2) {};
        \node[hasse] (3) at (3,3) {};
        \node[hasse] (c0) at (-1,1) {};
        \node[hasse] (c1) at (0,2) {};
        \draw[blue] (0)--(1)--(2)--(3);
        \node at (1,0.5) {$c_N$};
        \node at (2,1.5) {$c_{N-1}$};
        \node at (3,2.5) {$c_{N-2}$};
        \draw[blue] (c0)--(c1);
        \node at (-1,1.5) {$c_N$};
        \draw[red] (0)--(c0);
        \node at (-1,0.5) {$D_{N+1}$};
        \draw[red] (1)--(c1);
        \node at (1,1.7) {$D_{N+1}$};
        \node at (1.1,0) {\color{purple}\scriptsize$\mathrm{O}(3)-[C_N]$};
        \node at (2.2,1) {\color{purple}\scriptsize$\mathrm{O}(2)-[C_{N-1}]$};
        \node at (3.3,2) {\color{purple}\scriptsize$\mathrm{O}(1)-[C_{N-2}]$};
        \node at (3.3,3) {\color{purple}\scriptsize$\emptyset$};
        \node at (-2.1,1) {\color{purple}\scriptsize$\mathrm{O}(1)-[C_N]$};
        \node at (0,2.3) {\color{purple}\scriptsize$\emptyset$}; 
    \end{tikzpicture}}
\end{equation}

The mixed branch\footnote{Since the mixed branch includes the Coulomb branch it is also called enhanced Coulomb branch \cite{Argyres:2016xmc}.} is a direct product
\begin{equation}
    c_N\times D_{N+1}\qquad \rightarrow \qquad \raisebox{-.5\height}{\begin{tikzpicture}
        \node[hasse] (0) at (0,0) {};
        \node[hasse] (1) at (1,1) {};
        \node[hasse] (c0) at (-1,1) {};
        \node[hasse] (c1) at (0,2) {};
        \draw[blue] (0)--(1);
        \node at (1,0.5) {$c_N$};
        \draw[blue] (c0)--(c1);
        \node at (-1,1.5) {$c_N$};
        \draw[red] (0)--(c0);
        \node at (-1,0.5) {$D_{N+1}$};
        \draw[red] (1)--(c1);
        \node at (1,1.7) {$D_{N+1}$};
        \end{tikzpicture}}
\end{equation}

\paragraph{Special orthogonal gauge group.} Let us now consider the 3d $\mathcal{N}=4$ theory
\begin{equation}
   \raisebox{-.5\height}{ \begin{tikzpicture}
        \node[gauge,label=below:{SO$(3$)}] (1) at (0,0) {};
        \node[flavour,label=below:{$C_N$}] (2) at (1,0) {};
        \draw (1)--(2);
    \end{tikzpicture}}
\end{equation}
The Higgs branch Hasse diagram is \cite{Bourget:2019aer}:
\begin{equation}
  \raisebox{-.5\height}{  \begin{tikzpicture}
        \node[hasse] (0) at (0,0) {};
        \node[hasse] (1) at (1,1) {};
        \node[hasse] (3) at (3,3) {};
        \draw[blue] (0)--(1)--(3);
        \node at (1,0.5) {$c_N$};
        \node at (2.7,2) {$a_{2N-3}$};
        \node at (1.1,0) {\color{purple}\scriptsize$\mathrm{SO}(3)-[C_N]$};
        \node at (2.2,1) {\color{purple}\scriptsize$\mathrm{SO}(2)-[C_{N-1}]$};
        \node at (3.3,3) {\color{purple}\scriptsize$\emptyset$};
    \end{tikzpicture}}\;.
\end{equation}
On the most generic stratum the theory is completely higgsed. The Coulomb branch Hasse diagram of the $\mathrm{SO}(2)\sim \mathrm{U} (1)$ theory is
\begin{equation}
   \raisebox{-.5\height}{ \begin{tikzpicture}
        \node[hasse] (0) at (0,0) {};
        \node[hasse] (c1) at (-1,1) {};
        \draw[red] (0)--(c1);
        \node at (0.1,0.7) {$A_{2N-3}$};
        \node at (-1.2,0) {\color{purple}\scriptsize$\mathrm{SO}(2)-[C_{N-1}]$}; 
        \node at (-1.3,1) {\color{purple}\scriptsize$\emptyset$}; 
    \end{tikzpicture}}\;.
\end{equation}
The Coulomb branch Hasse diagram of the full SO$(3)$ theory is the same as that of the O$(3)$ theory
\begin{equation}
   \raisebox{-.5\height}{ \begin{tikzpicture}
        \node[hasse] (0) at (0,0) {}; 
        \node[hasse] (c1) at (-1,1) {};
        \draw[red] (0)--(c1);
        \node at (-1,0.5) {$D_{N+1}$};
        \node at (-1.1,0) {\color{purple}\scriptsize$\mathrm{SO}(3)-[C_N]$}; 
        \node at (-2.1,1) {\color{purple}\scriptsize$\mathrm{SO}(1)-[C_N]$};
    \end{tikzpicture}}\;.
\end{equation}
At this point it is important to note that
\begin{equation}
    A_{2N-3}/\mathbb{Z}_2=D_{N+1}\;,
\end{equation}
and that we can view $A_{2N-3}$ as $(A_{2N-3},\mathbb{Z}_2)$, where the $\mathbb{Z}_2$ is the global symmetry of $A_{2N-3}$ which we need to quotient to obtain $D_{N+1}$.

On the most general stratum of the Coulomb branch of the SO$(3)$ theory, there is a collection of $N$ free hypermultiplets, which we denote as a SO$(1)$ `gauge' theory. Its Higgs branch is $\mathbb{C}^{2N}$. We argue that we should view this Higgs branch as $(\mathbb{C}^{2N},\mathbb{Z}_2)$, where the $\mathbb{Z}_2$ is the symmetry of $\mathbb{C}^{2N}$ which we need to quotient to obtain $\mathbb{C}^{2n}/\mathbb{Z}_2=c_N$. The Hasse diagram of $(\mathbb{C}^{2N},\mathbb{Z}_2)$ is
\begin{equation}
   \raisebox{-.5\height}{ \begin{tikzpicture}
        \node[hasse] (0) at (0,0) {};
        \node[hasse] (1) at (1,1) {};
        \draw[blue,dashed] (0)--(1);
        \node at (1.4,0.5) {$(\mathbb{C}^{2N},\mathbb{Z}_2)$};
        \node at (1.1,0) {\color{purple}\scriptsize$\mathrm{SO}(1)-[C_N]$};
        \node at (1.3,1) {\color{purple}\scriptsize$\emptyset$};
    \end{tikzpicture}}\;.
\end{equation}
Putting this all together, we obtain the Hasse diagram of the full moduli space \cite{Grimminger:2020dmg}:
\begin{equation}
   \raisebox{-.5\height}{ \begin{tikzpicture}
        \node[hasse] (0) at (0,0) {};
        \node[hasse] (1) at (1,1) {};
        \node[hasse] (3) at (3,3) {};
        \node[hasse] (c0) at (-1,1) {};
        \node[hasse] (c1) at (0,2) {};
        \draw[blue] (0)--(1)--(3);
        \node at (1,0.5) {$c_N$};
        \node at (2.7,2) {$a_{2N-3}$};
        \draw[blue,dashed] (c0)--(c1);
        \node at (-1.4,1.7) {$(\mathbb{C}^{2N},\mathbb{Z}_2)$};
        \draw[red] (0)--(c0);
        \node at (-1,0.4) {$D_{N+1}$};
        \draw[red] (1)--(c1);
        \node at (1,1.7) {$A_{2N-3}$};
        \node at (1.1,0) {\color{purple}\scriptsize$\mathrm{SO}(3)-[C_N]$};
        \node at (2.2,1) {\color{purple}\scriptsize$\mathrm{SO}(2)-[C_{N-1}]$};
        \node at (3.3,3) {\color{purple}\scriptsize$\emptyset$};
        \node at (-2.1,1) {\color{purple}\scriptsize$\mathrm{SO}(1)-[C_N]$};
        \node at (0,2.3) {\color{purple}\scriptsize$\emptyset$}; 
    \end{tikzpicture}}
\end{equation}

The mixed branch is no longer a direct product, but it is a fibration with a global section, as was already discussed in detail in the 4d context in \cite[Sec.2.2.1]{Argyres:2016xmc}. Suggestively we can denote its Hasse diagram as
\begin{equation}
   \raisebox{-.5\height}{ \begin{tikzpicture}
        \node[hasse] (0) at (0,0) {};
        \node[hasse] (1) at (1,1) {};
        \node[hasse] (c0) at (-1,1) {};
        \node[hasse] (c1) at (0,2) {};
        \draw[blue] (0)--(1);
        \node at (1.5,0.5) {$\mathbb{C}^{2N}/\mathbb{Z}_2$};
        \draw[blue,dashed] (c0)--(c1);
        \node at (-1.5,1.5) {$(\mathbb{C}^{2N},\mathbb{Z}_2)$};
        \draw[red] (0)--(c0);
        \node at (-1.5,0.5) {$A_{N+1}/\mathbb{Z}_2$};
        \draw[red] (1)--(c1);
        \node at (1.6,1.5) {$(A_{N+1},\mathbb{Z}_2)$};
        \end{tikzpicture}}\;,
\end{equation} 
which mimics the description of the fibration in \cite{Argyres:2016xmc}.

\section{\texorpdfstring{Proof of \eqref{eq:countingLeavesSym}}{Proof}}

\label{app:proof}

In this Appendix, we prove formula \eqref{eq:countingLeavesSym} which gives the number of transitions of type $g$ in $\mathcal{M}_{G,k}$, and the number of transitions of type $A_{N-1}$ in $\mathrm{Sym}^k (\mathbb{C}^2 / \mathbb{Z}_N)$. 

To show this, it is enough to prove that when going from the diagram of $\mathcal{M}_{G,k-1}$ to that of $\mathcal{M}_{G,k}$ (respectively from $\mathrm{Sym}^{k-1}(\mathbb{C}^2/\Gamma_G)$ to $\mathrm{Sym}^k(\mathbb{C}^2/\Gamma_G)$), the number of added lines is $\sum\limits_{k'=0}^{k-1} p(k')$. 

This follows from the existence of two maps $f_1$ and $f_2$ from the set $P_{<k}$ of partitions of integers $<k$ to the set $P_k$ of partitions of $k$ defined by 
\begin{equation}
    f_1 ([ \lambda_1 , \dots , \lambda_r]) = [ \lambda_1 , \dots , \lambda_r , 1^{\lambda_1 + \dots + \lambda_r}]
\end{equation} 
and 
\begin{equation}
    f_2 ([ \lambda_1 , \dots , \lambda_r]) = [ \lambda_1 , \dots , \lambda_r , k-(\lambda_1 + \dots + \lambda_r)]
\end{equation}
The number of red lines (respectively green lines) in the diagram of $\mathcal{M}_{G,k}$ (resp. $\mathrm{Sym}^k(\mathbb{C}^2/\Gamma_G)$) connected to the partition $\lambda \in P_k$ is $|f_1^{-1} (\lambda)|$ (resp. $|f_2^{-1} (\lambda)|$), and we have 
\begin{equation}
    \sum\limits_{\lambda \in P_k} |f_1^{-1} (\lambda)| = |P_{<k}| =  \sum\limits_{\lambda \in P_k} |f_2^{-1} (\lambda)| \, . 
\end{equation}
This equality can be illustrated by the following table, which contains the partitions of $P_{<5}$ with the value of $f_1$ and $f_2$ labeling the columns and rows. 
\begin{center}
    \begin{tabular}{c|ccccccc}
 & $[1^5]$ & $[2,1^3]$ & $[2^2,1]$ & $[3,1^2]$ & $[3,2]$ & $[4,1]$ & $[5]$ \\ \hline 
$[1^5]$ & $[1^4]$ & $[1^3]$ & & $[1^2]$ & & $[1]$ & $\emptyset$  \\
$[2,1^3]$ &  &  $[2,1^2]$ & $[2,1]$ & &  $[2]$ & &   \\
$[2^2,1]$ &  & &  $[2^2]$ & & & &   \\
$[3,1^2]$ &  & & & $[3,1]$ & $[3]$ & &   \\
$[3,2]$ &  & & & & & &  \\
$[4,1]$ &  & & & & & $[4]$ &  \\
$[5]$ &  & & & & & &   \\ 
\end{tabular}
\end{center}

\providecommand{\href}[2]{#2}\begingroup\raggedright\endgroup

\end{document}